\documentclass[prd,superscriptaddress,nofootinbib]{revtex4}
\pdfoutput=1
\usepackage{amsmath,amssymb,epsfig}

\usepackage{graphicx}
\usepackage{bm}

\usepackage[normalem]{ulem} 
\usepackage{scalerel}
\usepackage{tikz}
\usetikzlibrary{svg.path}

\definecolor{orcidlogocol}{HTML}{A6CE39}
\tikzset{
	orcidlogo/.pic={
		\fill[orcidlogocol] svg{M256,128c0,70.7-57.3,128-128,128C57.3,256,0,198.7,0,128C0,57.3,57.3,0,128,0C198.7,0,256,57.3,256,128z};
		\fill[white] svg{M86.3,186.2H70.9V79.1h15.4v48.4V186.2z}
		svg{M108.9,79.1h41.6c39.6,0,57,28.3,57,53.6c0,27.5-21.5,53.6-56.8,53.6h-41.8V79.1z M124.3,172.4h24.5c34.9,0,42.9-26.5,42.9-39.7c0-21.5-13.7-39.7-43.7-39.7h-23.7V172.4z}
		svg{M88.7,56.8c0,5.5-4.5,10.1-10.1,10.1c-5.6,0-10.1-4.6-10.1-10.1c0-5.6,4.5-10.1,10.1-10.1C84.2,46.7,88.7,51.3,88.7,56.8z};
	}
}

\newcommand\orcidicon[1]{\href{https://orcid.org/#1}{\mbox{\scalerel*{
				\begin{tikzpicture}[yscale=-1,transform shape]
				\pic{orcidlogo};
				\end{tikzpicture}
      }{1}}}}

\usepackage{xcolor}

\usepackage{natbib,ifthen}
\usepackage[hidelinks]{hyperref}

\begin{document}

\newcommand{\fixme}[1]{{\textbf{Fixme: #1}}}
\newcommand{\ud}{{\rm d}}
\renewcommand{\eprint}[1]{\href{http://arxiv.org/abs/#1}{#1}}
\newcommand{\adsurl}[1]{\href{#1}{ADS}}
\newcommand{\ISBN}[1]{\href{http://cosmologist.info/ISBN/#1}{ISBN: #1}}
\newcommand{\jcap}{J.\ Cosmol.\ Astropart.\ Phys.}
\newcommand{\mnras}{Mon.\ Not.\ R.\ Astron.\ Soc.}
\newcommand{\progress}{Rep.\ Prog.\ Phys.}
\newcommand{\prlett}{Phys.\ Rev.\ Lett.}
\newcommand{\procspie}{Proc.\ SPIE}
\newcommand{\na}{New Astronomy}
\newcommand{\apjl}{ApJ.\ Lett.}
\newcommand{\physrep}{Physics Reports}
\newcommand{\aap}{A\&A}
\newcommand{\aapr}{A\&A Rev.}

\newcommand{\rstrike}[1]{{\color{red} \sout{#1}}}
\newcommand{\blue}[1]{{\color{blue} #1}}

\newcommand{\fpbh}{f_{{\rm PBH}}}
\newcommand{\erf}{\mathrm{erf}}


\title[Bayesian analysis of LIGO-Virgo PBH mergers]{Bayesian analysis of LIGO-Virgo mergers: Primordial vs.~astrophysical black hole populations}

\author{Alex Hall \orcidicon{0000-0002-3139-8651}}
\email{ahall@roe.ac.uk}
\affiliation{Institute for Astronomy, University of Edinburgh, Royal Observatory, Blackford Hill, \\ Edinburgh, EH9 3HJ, United Kingdom}

\author{Andrew D. Gow \orcidicon{0000-0002-4858-555X}}
\affiliation{Department of Physics and Astronomy, University of Sussex, \\ Brighton, BN1 9QH, United Kingdom}

\author{Christian T. Byrnes \orcidicon{0000-0003-2583-6536}}
\affiliation{Department of Physics and Astronomy, University of Sussex, \\ Brighton, BN1 9QH, United Kingdom}
   


\begin{abstract}
We conduct a thorough Bayesian analysis of the possibility that the black hole merger events seen in gravitational waves are primordial black hole (PBH) mergers. Using the latest merger rate models for PBH binaries drawn from a lognormal mass function we compute posterior parameter constraints and Bayesian evidences using data from the first two observing runs of LIGO-Virgo. We account for theoretical uncertainty due to possible disruption of the binary by surrounding PBHs, which can suppress the merger rate significantly. We also consider simple astrophysically motivated models and find that these are favoured decisively over the PBH scenario, quantified by the Bayesian evidence ratio. Paying careful attention to the influence of the parameter priors and the quality of the model fits, we show that the evidence ratios can be understood by comparing the predicted chirp mass distribution to that of the data. We identify the posterior predictive distribution of chirp mass as a vital tool for discriminating between models. A model in which all mergers are PBH binaries is strongly disfavoured compared with astrophysical models, in part due to the over-prediction of heavy systems having $\mathcal{M}_{{\rm chirp}} \gtrsim 40 \, M_\odot$ and positive skewness over the range of observed masses which does not match the observations. We find that the fit is not significantly improved by adding a maximum mass cut-off, a bimodal mass function, or imposing that PBH binaries form at late times. We argue that a successful PBH model must either modify the lognormal shape of the initial mass function significantly or abandon the hypothesis that all observed merging binaries are primordial. We develop and apply techniques for analysing PBH models with gravitational wave data which will be necessary for robust statistical inference as the gravitational wave source sample size increases.

\end{abstract}

\maketitle

\section{Introduction}
\label{sec:intro}

Primordial black holes  (PBHs,~\citep{1967SvA....10..602Z, 1971MNRAS.152...75H, 1974MNRAS.168..399C, 1975ApJ...201....1C}) have long been recognised as a unique dark matter candidate that does not require the existence of a new particle or modification to gravity (see Refs.~\citep{2016PhRvD..94h3504C, 2020ARNPS..7050520C, 2020arXiv200212778C, 2020arXiv200710722G} for recent reviews). Interest in PBHs has increased greatly due to the detection of black hole (BH) mergers emitting gravitational waves (GWs) by LIGO and Virgo~\citep{Abbott:2016blz}, since it is possible that the merging objects are primordial in origin~\citep{2016PhRvL.116t1301B, 2016PhRvL.117f1101S, 2017PDU....15..142C}.

Assuming that some fraction of the observed merger events are primordial binaries one can place bounds on the fraction of the dark matter that should be in PBHs of the relevant mass range to explain the observed merger rate (see Ref.~\citep{Sasaki:2018dmp} for a review). If all of the confirmed LIGO-Virgo events are PBH mergers then the fraction of dark matter in PBHs, $\fpbh$, is typically found to be a few $\times 10^{-3}$ depending on assumptions about the evolution and formation mechanism of the binary (see Ref.~\citep{2020JCAP...06..044D} for a comprehensive recent review), although $\fpbh \approx 1$ is still permitted in certain models (e.g.~Ref.~\citep{Jedamzik:2020ypm}).

The increase in sample size to ten events~\citep{2019PhRvX...9c1040A} since the first detection has allowed several groups to make fits of the PBH initial mass function to the LIGO data, typically concluding that a lognormal mass function with central mass $m_c\sim 10 \, M_\odot$  and a width of order unity is the best fit~\citep{2017JCAP...09..037R, 2018ApJ...864...61C, 2020PhRvD.101h3008W}. Connecting the empirical distribution of black hole source parameters to an initial mass function for PBHs is in general non-trivial and involves modelling the formation of the binary and its evolution through to the merger event~\citep{2017PhRvD..96l3523A, 2020JCAP...01..031G}, but the reward for this is a direct constraint on the conditions in the early Universe which gave rise to PBH formation. The initial mass function can be predicted from the spectrum of curvature fluctuations at the formation epoch, implying that constraints on the mass function can give unique information on  a host of poorly understood physics in the early Universe, including the small-scale power spectrum, non-Gaussianity, phase transitions, and inflation~\citep{Cai:2018tuh, Byrnes:2018clq, 2020PhRvD.101d3015V, 2020arXiv200803289G}. In general the calculation is more complicated for the extended mass functions required by the LIGO data if more than one event is primordial but, encouragingly, recent simplified models for PBH binary evolution tentatively give good agreement with the results of $N$-body simulations~\citep{2019JCAP...02..018R}.

What previous analyses have neglected however is whether the best-fit PBH model is a good fit to the data and, more specifically, whether the model is a good fit compared to simple astrophysical BH merger models such as those studied by the LIGO and Virgo collaborations. With the event rate in the recent LIGO-Virgo O3 observing run roughly double what it was in the O1 and O2 observing runs, the importance of a rigorous statistical analysis of the PBH formation scenario is becoming increasingly necessary. The required analysis can be compared with more conventional studies of stellar black hole binary populations using GW events~\citep{2016PhRvX...6d1015A, 2017ApJ...851L..25F, 2018ApJ...863L..41F, 2018PhRvD..98h4036G, 2019ApJ...882L..24A} where techniques such as Bayesian model comparison, tests of model consistency, and goodness-of-fit tests are becoming commonplace. These tests are in principle able to rule out whole classes of PBH mass functions for \emph{any} values of their parameters, if those mass functions predict merger populations which do not match the observations. The ability of Bayesian methods to quantify this is one of several advantages to pursuing this line of study.

In this paper we perform Bayesian tests of the PBH merger scenario using the binary black hole (BBH) merger events in the first two observing runs of LIGO-Virgo. We quantify how well the data fit the PBH merger scenario compared with simple astrophysically motivated models using the latest calculations for the formation of binaries during radiation domination and their subsequent evolution and possible disruption. The techniques we employ provide a link between the methodology of the LIGO-Virgo and GW community and that of the PBH community. We consider  individual source masses and redshifts in our analysis, which provides more constraining power than simply using the component spins (for which a Bayesian analysis in the spirit of ours was performed recently in Ref.~\citep{2019JCAP...08..022F}). By accounting for correlated parameter uncertainties, the non-uniform selection probability of LIGO-Virgo, and an accurate likelihood function for the population parameters, we provide a comprehensive statistical study of the PBH formation channel for merging black hole binaries. We pay particular attention to how the models are able to fit the data, and why certain models are favoured over others.

As is well known, Bayesian tests using the model posterior probability or evidence are sensitive to the priors assigned to the parameters of each model, which often lack a strong physical motivation when the models are phenomenological. In this case the best one can do is transparently present the chosen priors and check the sensitivity of the results to alternative choices. The sensitivity is typically only logarithmic, but we will be careful to account for uncertainty in the choice of prior when presenting our results.

Unless otherwise stated we adopt units where $G=c=1$. When computing background quantities we assume a flat $\Lambda$CDM cosmological model with parameters fixed to the best fitting values of Planck 2015~\citep{2016A&A...594A..13P}. We note that the sources considered in this work are all at sufficiently low redshift that our results are insensitive to the choice of cosmological model.

\section{Data}
\label{sec:data}

\begin{figure}
\centering
\includegraphics[width=\textwidth]{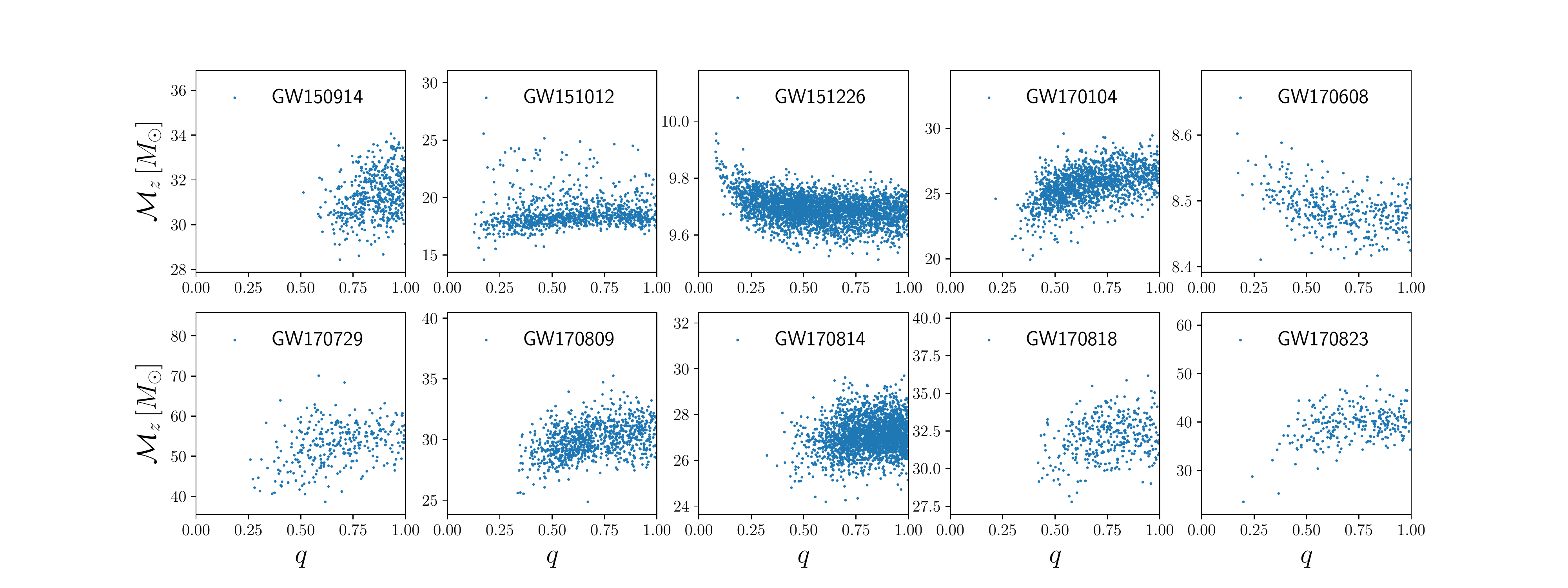}
\caption{Posterior samples from each of the sources in the O1O2 source catalogue, indicated in each panel. We show samples in the plane of redshifted chirp mass $\mathcal{M}_z$ and mass ratio $q$, marginalised over the other source parameters. Samples have been thinned by a factor of 16 for visual clarity.}
\label{fig:O1O2_data_Mcz_q}
\end{figure}

To test PBH models of binary mergers we use the ten BH-BH merger events in the Gravitational-Wave Transient Catalogue from the first two observing runs of LIGO-Virgo (GWTC-1,~\citep{2019PhRvX...9c1040A})\footnote{We discuss the implications of recent BBH detections in the O3a run in Section~\ref{sec:conclusions}.}. As we shall see, it is sufficient for our population-level analysis to use only samples from the posterior distribution of source parameters (masses, spins etc.) for each source. In Figure~\ref{fig:O1O2_data_Mcz_q} we show posterior samples in the plane of detector-frame (i.e.~redshifted) chirp mass $\mathcal{M}_z$ and mass ratio $q$, where $\mathcal{M}_z = (1+z)(m_1 m_2)^{3/5}(m_1 + m_2)^{-1/5}$ and $q=m_2/m_1$. Note that $m_2 < m_1$ has been enforced in the GWTC-1 posteriors.

The detector-frame chirp mass is a well-constrained parameter for each source, being constrained with typically $\lesssim 10\%$ precision. Being closely related to measurable aspects of the source waveforms (specifically the frequency evolution of the GW strain) it is also practically uncorrelated with other source parameters, as shown for the case of $q$ in Figure~\ref{fig:O1O2_data_Mcz_q}. This is in contrast with other descriptors of the absolute mass scale of the BBH; the heavier mass $m_1$ is typically only constrained at the 10\% -- 50\% level and is highly correlated with the lighter mass $m_2$, while the total mass is on average constrained with $\gtrsim 10\%$ precision and is often highly correlated with $q$. For this reason we will often present constraints in terms of chirp mass rather than heaviest mass or total mass\footnote{This point has also been recognised in the recent Ref.~\citep{2020arXiv200500892D}.}. Note that the redshifts of the sources in GWTC-1 are sufficiently low that there is only a small difference between source-frame and detector-frame masses.

Any inference of BBH populations must carefully account for the selection function of LIGO-Virgo. We adopt an accurate semi-analytic approach to computing $p_{\mathrm{det}}(\boldsymbol{\lambda})$, the probability of detecting a source given it has source parameters $\boldsymbol{\lambda}$, following the prescription in Refs.~\citep{2016PhRvX...6d1015A, 2019ApJ...882L..24A}. We use the public code \textsc{gwdet}~\citep{2017zndo....889966G} to compute $p_{\mathrm{det}}(m_1, m_2, z)$ on a grid of values for subsequent interpolation. Following the procedure described in Refs.~\citep{1993PhRvD..47.2198F, 2019PhRvD..99j3004G} and approximating detection as coming from a single interferometer, $p_{\mathrm{det}}$ is computed as
\begin{equation}
  p_{\mathrm{det}}(m_1, m_2, z) = \int^1_{\rho_*/\rho_{\mathrm{opt}}(m_1, m_2, z)} p(\omega) \ud \omega,
  \label{eq:pdet}
\end{equation}
where $\rho_{\mathrm{opt}}(m_1, m_2, z)$ is the signal-to-noise ($S/N$) for an optimally oriented source, face-on directly above the interferometer. The $S/N$ threshold for detection is approximated as $\rho_* = 8$, $\omega$ encodes all the angular dependence of the interferometer response, and the orientation and angular position of the source have been marginalised over assuming isotropy, encoded in the distribution $p(\omega)$. The signal-to-noise for an optimally-oriented source is computed using the routines in the \textsc{PyCBC} software package~\citep{2016CQGra..33u5004U}. To compute the noise power spectral density (PSD) we assume the IMRPhenomD waveform approximant assuming non-spinning black holes\footnote{Note that the assumption of non-spinning black holes should be a reasonable approximation in the case of PBH models, which are expected to have negligible spin at formation~\citep{2019JCAP...05..018D, 2020JCAP...03..017M}, although it is possible that subsequent accretion can lead to non-zero spin for the higher mass objects detected by LIGO~\citep{2020JCAP...04..052D}. In the case of astrophysical merger models we will see that this is a reasonable approximation to the $S/N$.}, and we approximate the PSD of each source in the GWTC-1 catalogue with the \textsc{PyCBC} analytic function \textsc{aLIGOEarlyHighSensitivityP1200087}~\citep{2016LRR....19....1A}, i.e.~we assume that each source is detected in a single aLIGO detector. This is a sufficiently good approximation for our purposes to the true PSD of each source, with the biggest difference arising at the lowest frequencies between $f = 10 \, \mathrm{Hz}$ and $f = 20 \, \mathrm{Hz}$, a frequency range to which our final results are insensitive.

\begin{figure}
\centering
\includegraphics[width=0.5\textwidth]{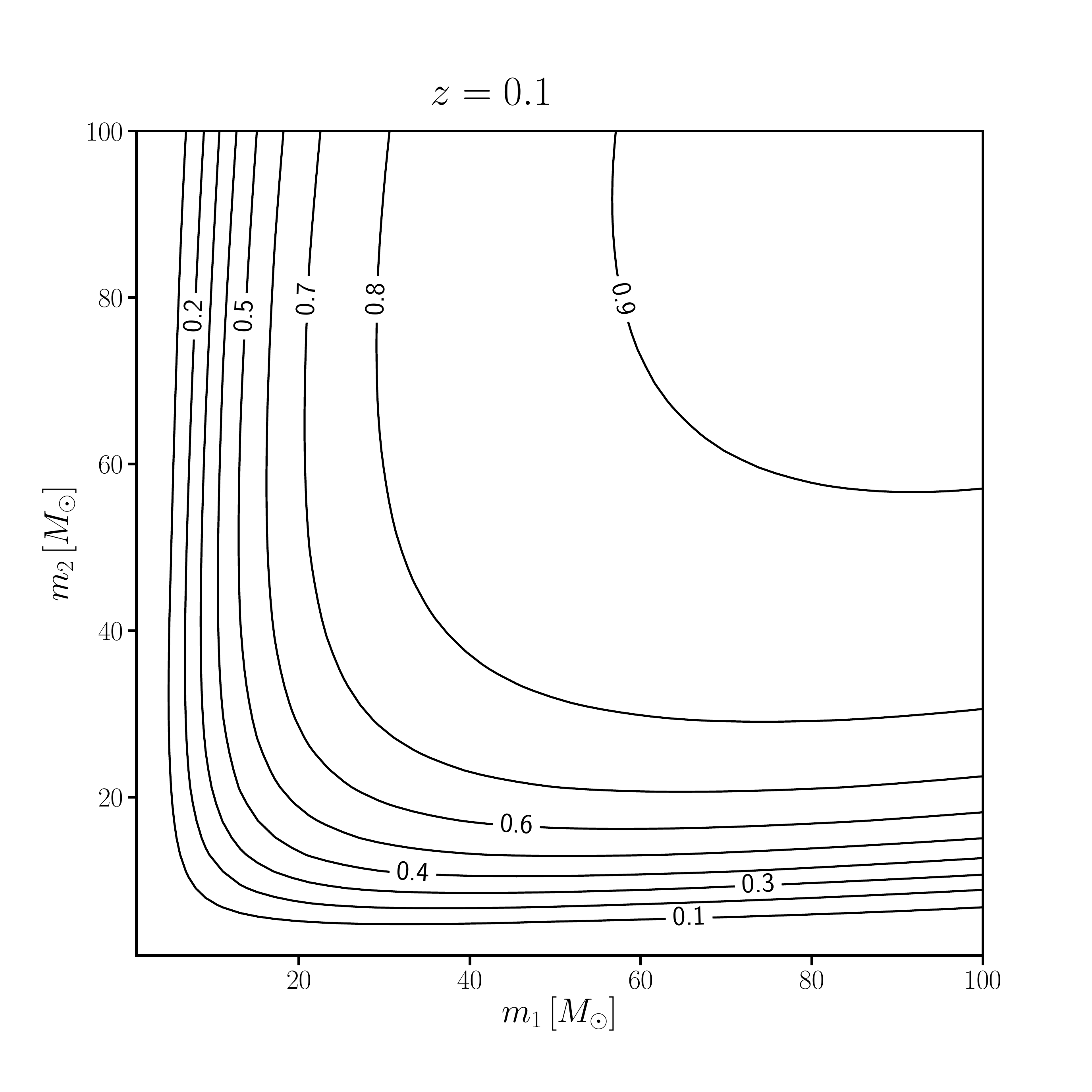}
\caption{Contours of constant detection probability at $z=0.1$ as a function of the source-frame component masses, assuming the O1O2 observing specifications defined in the text.}
\label{fig:pdet_z0p1}
\end{figure}

In Figure~\ref{fig:pdet_z0p1} we show contours of the detection probability for a source at $z=0.1$ as a function of the source-frame component masses (see, e.g. Ref.~\citep{2019PhRvD.100d3012W} for similar plots). The $S/N$ threshold results in a suppression of the detection probability below about $5 \, M_\odot$. The Figure makes it clear that LIGO in the O1O2 observing runs was sensitive to large mass ratios, with this increasingly true for the more sensitive O3 run which has yielded objects with $q \approx 0.3$~\citep{2020arXiv200408342T} and $q \approx 0.1$~\citep{Abbott:2020khf}. Note that the sensitivity falls to zero at very high (a few hundred solar) masses where the waveforms have no support above the minimum frequency $f_{{\rm low}}$. 

\begin{figure}
\centering
\includegraphics[width=0.7\textwidth]{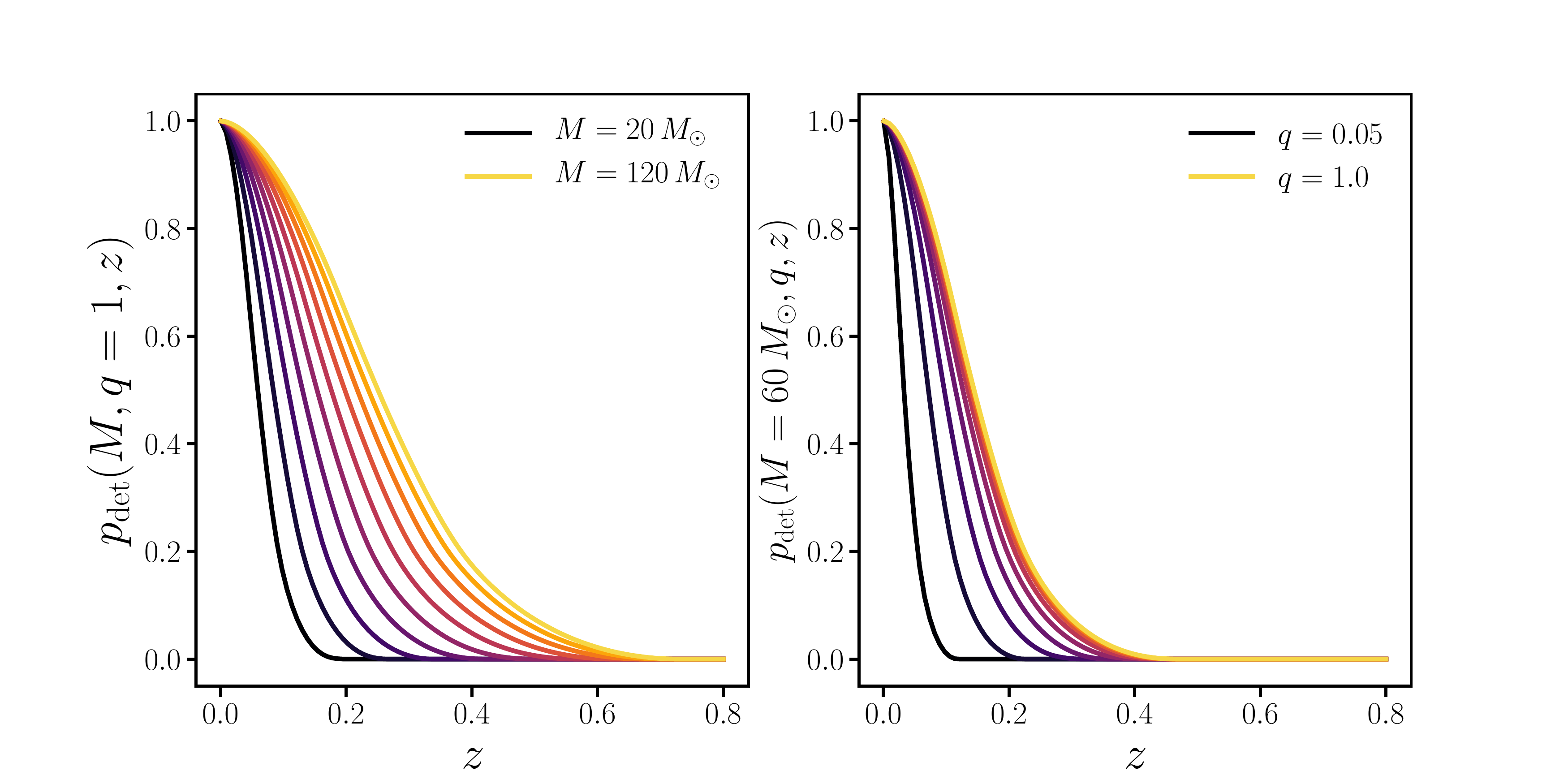}
\caption{Detection probability as a function of redshift for equal component masses and a range of total masses (left panel) and total mass $M=60 \, M_\odot$ and a range of mass ratios $q$ (right panel). Warmer colours (increasing from bottom to top in both panels) indicate greater values of $M$ or $q$.}
\label{fig:pdet_zs}
\end{figure}

In Figure~\ref{fig:pdet_zs} we show the redshift dependence of $p_{\mathrm{det}}$ for a range of total source frame masses and mass ratios. This dependence arises primarily from the $d_L^{-1}$ drop-off in the $S/N$, but there is also a dependence via the redshifted (detector-frame) masses $(1+z)m_{1,2}$ at fixed source mass. Sources with redshifts $z \gtrsim 0.7$ are undetectable in the O1O2 runs for any component masses, with this upper limit quickly dropping as the total mass is lowered. For a total mass $M = 20 \, M_\odot$ only sources with $z \lesssim 0.2$ are detectable, and then only for equal mass components. For comparison, the highest redshift in the GWTC-1 catalogue is $z \approx 0.49$ (GW170729, total mass $M \approx 84 \, M_\odot$, albeit with significantly non-zero spin), while the median redshift source is $z \approx 0.16$.

We close this section by noting that the methodology we have adopted for computing $p_{\mathrm{det}}$ is a commonly used approach in BBH population analyses (see, e.g.~Refs.~\citep{2018PhRvD..98h4036G, 2018ApJ...863L..41F, 2019MNRAS.484.4216R, 2019PhRvD..99j3004G}). Nonetheless, as discussed in Appendix A of Ref.~\citep{2019ApJ...882L..24A} our approximate method for computing the detection probability can overestimate the sensitive volume $\langle VT \rangle$ by up to a factor of 2 compared to the more accurate approach of injecting signals into the source detection pipeline (see their Figure 9). This means that our constraints on overall merger rates are expected to be underestimated by at most a factor of 2. This translates into an underestimation of $\fpbh$ of at most 20\% for the models we consider.

\section{Binary merger rate models}
\label{sec:rates}

In this section we present the models of merging black hole binaries which we confront with the GWTC-1 catalogue. The fundamental quantity we require for our statistical analysis is the differential merger rate density in the source frame as a function of masses and redshift.

\subsection{PBH merger models}
\label{subsec:PBHrates}

Many attempts have been made to model the merger rate of PBH binaries (e.g. Refs.~\citep{2016PhRvL.116t1301B, 2016PhRvL.117f1101S, 2017JCAP...09..037R, 2017PhRvD..95d3511N, 2017PhRvD..96l3523A, 2018PhRvD..98b3536K, 2018ApJ...864...61C}). As in Ref.~\citep{2020JCAP...01..031G} we use the formalism of Ref.~\citep{2019JCAP...02..018R}, which itself builds upon that in Ref.~\citep{2017PhRvD..96l3523A}. We will first give a brief sketch of the calculation before presenting the resulting PBH merger rate. Readers unconcerned with the derivation may skip to Equations~\eqref{eq:dR0} and~\eqref{eq:sfac} where the differential merger rate is presented.

The calculation of Ref.~\citep{2019JCAP...02..018R} follows a pair of PBHs drawn from a mass function $\psi(m)$ (normalised to unity), initially comoving with the cosmic expansion in radiation domination. The PBH pair decouples from the expansion and forms a high-eccentricity binary, with gravitational torquing from other surrounding PBHs and a smoothly distributed dark matter component with small Gaussian fluctuations. The PBH binary forms with dimensionless angular momentum $j \ll 1$, emits GWs and merges after a time~\citep{PhysRev.136.B1224}
\begin{equation}
  \tau = \frac{3}{85}\frac{r_a^4}{\eta M^3}j^7,
  \label{eq:tmerge}
\end{equation}
where $r_a$ is the semi-major axis of the binary, $M$ is the total mass, and $\eta$ is the symmetric mass ratio defined by $\eta = m_1 m_2/(m_1 + m_2)^2$. The ellipticity of the binary is given by $e = \sqrt{1 - j^2}$. The semi-major axis $r_a$ follows from the dynamics of the system prior to binary formation, and is given by $r_a \approx 0.1 a_{\mathrm{dc}} x_0$ where $a_{\mathrm{dc}}$ is the scale factor at decoupling and $x_0$ is an initial comoving separation. Decoupling takes place at roughly $a_{\mathrm{dc}} \approx a_{\mathrm{eq}}/\delta_b$ where $a_{\mathrm{eq}}$ is the scale factor at  matter-radiation equality and $\delta_b \gg 1$ is the effective density fluctuation generated by the PBH pair, given by $\delta_b = (M/2)/\rho_M V(x_0)$ with $\rho_M$ the background matter density and $V(x) = (4\pi/3)x^3$. We refer the reader to Ref.~\citep{2019JCAP...02..018R} for further discussion.

The time taken for a newly formed PBH binary to merge is a crucial factor in determining the rate of merging sources in the LIGO-Virgo sensitive volume, and Equation~\eqref{eq:tmerge} makes clear its high sensitivity to the angular momentum of the binary. At formation this angular momentum is imparted by gravitational torquing from other PBHs and fluctuations in the surrounding dark matter density. In the model of Refs.~\citep{2017PhRvD..96l3523A, 2019JCAP...02..018R} these dark matter fluctuations are modelled as Gaussian with variance $\langle \delta_M^2 \rangle$, with contributions from dark matter mass scales greater than $\sim 10^{-3} M \approx 10^{-2} \, M_\odot$ for PBHs in the LIGO mass range. Following Ref.~\citep{Eroshenko:2016hmn, 2017PhRvD..96l3523A} we assume a fixed value $\langle \delta_M^2 \rangle^{1/2} = 0.005$ extrapolating the linear adiabatic power spectrum measured on CMB scales, although one should note that models of PBH formation typically invoke enhanced small-scale power or non-Gaussianity in the dark matter distribution which could boost $\langle \delta_M^2 \rangle$ significantly. The variance of angular momentum fluctuations in the vicinity of the PBH binary is then
\begin{equation}
  \sigma^2_{j,M} = \frac{6}{5}j_0^2 \frac{\sigma_M^2}{\fpbh^2},
  \label{eq:sigj}
\end{equation}
where $j_0$ is a characteristic angular momentum (with $j_0 \ll 1$), $\fpbh \equiv \rho_{{\rm PBH}}/\rho_{{\rm DM}}$ is the ratio of the PBH energy density to the dark matter energy density, and $\sigma_M^2$ is defined in Ref.~\citep{2019JCAP...02..018R} as a `rescaled variance' given by $\sigma_M^2 \equiv (\Omega_{{\rm M}}/\Omega_{{\rm DM}})^2\langle \delta_M^2 \rangle$. Since the PBH binary is assumed to form deep in the radiation era when baryons are tightly coupled to photons, only the dark matter contributes to fluctuations in the local tidal field from the smooth matter component, and hence $\langle \delta_M^2 \rangle$ should appear on the right hand side of Equation~\eqref{eq:sigj}. The difference is negligible in comparison to the uncertainty on the variance on these small scales however, so for consistency with Ref.~\citep{2019JCAP...02..018R} we take $\sigma_M \approx 0.006$.

The total variance of the angular momentum imparted to the PBH binary consists of the dark matter fluctuations plus those of the surrounding PBHs, and is given by
\begin{equation}
  \sigma_j^2 = \sigma^2_{j,M} + \sigma^2_{j,\mathrm{PBH}} = \frac{6}{5}j_0^2\left(\frac{1 + \sigma^2_m/\langle m \rangle^2}{\bar{N}(y)} + \frac{\sigma^2_M}{\fpbh^2}\right),
\end{equation}
where $\sigma^2_m$ is the variance of the PBH mass function, $\langle m \rangle$ is the average PBH mass (angle brackets denote expectation values over $\psi/m$), and $\bar{N}(y)$ is the expected number of PBHs within a comoving radius $y$ of the binary. This latter quantity is needed because the model of Ref.~\citep{2019JCAP...02..018R} assumes that there is an \emph{exclusion zone} around the binary of radius $y$, inside of which no other PBH can reside lest its close proximity fatally disrupt the newly formed binary. The limit $\bar{N}(y) \rightarrow 0$ corresponds to no such exclusion.

The \emph{distribution} of the angular momentum imparted to the binary follows from assuming Gaussianity for the dark matter and Poisson statistics for the surrounding PBHs. As shown in Ref.~\citep{2017PhRvD..96l3523A} this results in a Holtsmark distribution for the latter. The resulting probability density $p_j$ for the angular momentum $j$ is
\begin{equation}
  j p_j(j) = \int_0^\infty \ud u \, u J_0(u) \exp \left[ -\bar{N}(y) \int \frac{\ud n(m)}{n} F\left(u \frac{m}{\langle m \rangle }\frac{1}{\bar{N}(y)}\frac{j_0}{j}\right) - u^2\frac{3}{10}\frac{\sigma^2_M}{\fpbh^2}\frac{j_0^2}{j^2}\right],
  \label{eq:pj}
\end{equation}
where the innermost integral is over the number density of PBHs with
\begin{equation}
  \frac{\ud n}{\ud m} = \rho_{\mathrm{PBH}} \frac{\psi(m)}{m},
  \label{eq:psi}
\end{equation}
and $F(x) = {}_1F_2(-1/2; 3/4, 5/4;-9x^2/16) - 1$ with ${}_1F_{2}$ a generalized hypergeometric function. In the limit $\bar{N}(y) \rightarrow 0$ and $\sigma_M \ll \fpbh$ we obtain the result of Ref.~\citep{2017PhRvD..96l3523A}, and in the limit $\bar{N}(y) \rightarrow \infty$ we obtain a Rayleigh distribution for $j$ with width $\sigma_j$, i.e.~torquing only by the Gaussian dark matter fluctuations.

As in Refs.~\citep{2019JCAP...02..018R, 2020JCAP...01..031G, 2020JCAP...06..044D} we take
\begin{equation}
  \bar{N}(y) = \frac{M}{\langle m \rangle} \frac{\fpbh}{\fpbh + \sigma_M},
  \label{eq:Eq3p5}
\end{equation}
which agrees well with the numerical simulations of Ref.~\citep{2019JCAP...02..018R} for $\fpbh \lesssim 10^{-1}$. To understand the form of this expression, first note the limiting case of a monochromatic mass function and $\sigma_M \ll \fpbh$. In this case, $\bar{N}(y) = 2$, i.e.~we expect two PBHs in the vicinity of the binary -- the two components black holes themselves. This agrees with the discussion in Ref.~\citep{1998PhRvD..58f3003I} that $y$ should be roughly the inter-particle distance of the PBH distribution. In the case that $\sigma_M \gg \fpbh$ we have $\bar{N}(y) \rightarrow 0$, i.e.~the exclusion region around the binary is expected to hold very few PBHs. Equation~\eqref{eq:Eq3p5} extends this simple picture to the case of a broad mass function, identifying the transition value of $\fpbh$ as $\sim \sigma_M$. We note that there is considerable uncertainty in the potential rate of disruption of newly formed binaries by surrounding PBHs, and deviations from Equations~\eqref{eq:pj} and~\eqref{eq:Eq3p5} may be expected in the case of broad mass functions.

The merger rate density at time $\tau$ of binaries in the model of Ref.~\citep{2019JCAP...02..018R} is given by $\ud R = S \times \ud R_0$, where
\begin{equation}
  \ud R_0 = \frac{1.6 \times 10^6}{\mathrm{Gpc^3 yr}} \fpbh^{\frac{53}{37}}\eta^{-\frac{34}{37}} \left(\frac{M}{M_\odot}\right)^{-\frac{32}{37}} \left(\frac{\tau}{\tau_0}\right)^{-\frac{34}{37}} \psi(m_1)\psi(m_2)\ud m_1 \ud m_2
  \label{eq:dR0}
\end{equation}
is the rate in the limit $\bar{N}(y) \rightarrow 0$ and $\sigma_M/\fpbh \rightarrow 0$, with $\tau_0 = 13.8 \times 10^9 \, \mathrm{yr}$, and $S$ is the \emph{suppression factor} given by
\begin{equation}
  S = \frac{e^{-\bar{N}(y)}}{\Gamma(21/37)}\int_0^{\infty} \ud v \, v^{-\frac{16}{37}} \exp{\left[-\bar{N}(y) \langle m \rangle \int_0^\infty \frac{\ud m}{m} \psi(m) F\left(\frac{m}{\langle m \rangle} \frac{v}{\bar{N}(y)}\right) - \frac{3\sigma^2_M v^2}{10 \fpbh^2} \right]}.
  \label{eq:sfac}
\end{equation}

The suppression factor $S$ quantifies the effect of demanding that no PBHs be present in a region of size $y$ around the binary (which would be expected to contain $\bar{N}(y)$ PBHs) \emph{and} the effect of dark matter density fluctuations imparting angular momentum to the binary. It is straightforward to show that $0 \leq S \leq 1$.

We will often present results for a PBH model having no suppression factor, i.e.~$S=1$, where the merger rate is given by Equation~\eqref{eq:dR0}. Some of the uncertainty in the precise formation mechanism is bracketed by the cases $S=1$ and the full expressions of Equations~\eqref{eq:dR0} and~\eqref{eq:sfac}. More precise numerical simulations of PBH binary formation and evolution will be required for a more quantitative investigation of the sensitivity of our results to the formation model~\citep{2019PhRvD.100h3528I,Jedamzik:2020ypm,Jedamzik:2020omx,Young:2020scc,Trashorras:2020mwn}.

It is important to note that the PBH mass functions we use are based on calculations of the primordial mass function $\psi$. It is possible that $\psi$ could evolve through mergers and/or accretion. So called second generation mergers, i.e.~those involving one or more BHs which have already undergone a previous merger, are expected to be very rare compared with primary mergers~\citep{2020PhRvD.101h3008W,2020JCAP...04..052D} (although also see Ref.~\citep{Liu:2019rnx}). Accretion is a highly non-linear process which is hard to model, but recently Ref.~\citep{2020JCAP...04..052D} have suggested this could play an important role on more massive PBHs due to the dark matter halo which forms around them at early times, and thereby acts as a significant additional gravitational attraction to nearby baryons (note that this only applies if $f_{\rm PBH}\ll 1$, but that is the case we consider in this paper). Ref.~\citep{2018PhRvD..98b3536K} showed that the DM halo which forms around PBHs~\citep{Adamek:2019gns,2019PhRvD.100h3528I} has minimal impact on the merger rate and the estimate of $f_{\rm PBH}$. Accretion onto a PBH can increase the initially negligible spin of a PBH, provided that the PBH mass grows significantly~\citep{2020JCAP...04..052D}. The most massive BH pair detected by LIGO is also the system with non-zero spin detected at highest significance, which may be consistent with a PBH model that includes a modest amount of accretion. However, the second-lightest BH merger event also has significant evidence for non-zero spin, suggesting that not all events are primordial. We will return to this issue in Section~\ref{sec:conclusions}. We note that there exists a window for which accretion has a non-negligible impact on the spin of the most massive PBHs but has very little impact on the PBH mass function~\citep{2020JCAP...06..044D}, but the impact of accretion is a highly non-linear process which deserves further study, see e.g.~Ref.~\citep{2020A&A...638A.132B}.

Recently Jedamzik has argued that the LIGO and Virgo results are consistent with $f_{\rm PBH}=1$ provided that PBHs follow a broad mass function with a large spike at $1\, M_\odot$~\citep{Jedamzik:2020ypm,Jedamzik:2020omx}, as motivated by the softening of the equation-of-state parameter during the Standard Model QCD phase transition~\citep{Byrnes:2018clq,Sobrinho:2020cco}. This result is based in numerical simulations of dense PBH clusters, and it does not appear to be in contradiction with the constraints we derive of $\fpbh \ll 1$ based on a relatively narrow mass function, see e.g.~Refs.~\citep{2019JCAP...02..018R,2020PhRvD.101d3015V} which showed that the analytic estimate for the merger rate which we used may be unreliable if $\fpbh \gtrsim 0.1$. See also Refs.~\citep{2019PhRvD.100h3528I,Young:2020scc,Trashorras:2020mwn} for further numerical studies of the PBH binary disruption rate.

Finally, our baseline results assume a lognormal mass function for the PBHs, given by
\begin{equation}
  \psi(m) = \frac{1}{m\sqrt{2\pi \sigma^2}}\exp{\left[-\frac{\ln^2{(m/m_c)}}{2\sigma^2}\right]},
\end{equation}
where $m_c$ is the peak of the function $m\psi(m)$ and $\sigma$ its logarithmic width. This is a good approximation to the PBH mass function in the case of formation from a smooth, symmetric peak in the power spectrum~\citep{PhysRevD.47.4244, PhysRevD.96.023514}, although deviations are expected in the case of particularly narrow power spectrum peaks~\citep{2020arXiv200903204G}. We consider the case of non-lognormal mass functions in Section~\ref{sec:extension}.

One consideration when using the model of Ref.~\citep{2019JCAP...02..018R} for lognormal mass functions is the over-suppression of the merger rate for very broad mass functions. Physically, one would expect a population of very light PBHs to have little effect on the merger rate in the LIGO mass range, since this light population does not contribute significantly to the gravitational attraction between two heavier PBHs, leaving negligible impact on the formation of the binary. However, in the calculation of~Ref.~\citep{2019JCAP...02..018R}, a large population of light black holes makes a large contribution to the expected number of PBHs in the vicinity of the binary, $\bar{N}(y)$, and are assumed to cause disruption to the heavier pair of PBHs \cite{2020JCAP...01..031G}. To ensure that this over-suppression does not affect the constraint presented in this paper, we quantify the value of $\sigma$ for which the lognormal distribution becomes broad enough that the suppression becomes significant. We do this by considering the differential merger rate for an equal mass merger in two cases: considering the full mass function, and a mass function with a low-mass cut-off, so that the population of light black holes is removed. For three different values of the PBH mass $m$ (the same for both PBHs in the binary), the inclusion of the low-mass population causes a significant suppression ($>10$ orders of magnitude) for $\sigma\gtrsim2$. As we shall see later, this is well above the range that the data favours, and so we assume that the model in Ref.~\citep{2019JCAP...02..018R} is valid for the constraints we present. We note that a thorough investigation of this effect will require the running of $N$-body simulations.

To compute the merger rate as a function of mass and redshift we use a fast and accurate approximation to the suppression factor valid for lognormal mass functions, described in Appendix~\ref{app:numerics}.

In Figure~\ref{fig:merger_rate_shape} we show the dependence of the source-frame differential merger rate density $\ud R/\ud m_1\ud m_2$ on the component masses, for several choices of the lognormal mass-function parameters. The lognormal distribution has a characteristic skewness towards large masses, giving rise to a skewness towards large total mass. This also gives rise to a broad range of mass ratios, as seen by the off-diagonal extent of the merger rates, which increases with $\sigma$.
\begin{figure}
\centering
\includegraphics[width=0.7\textwidth]{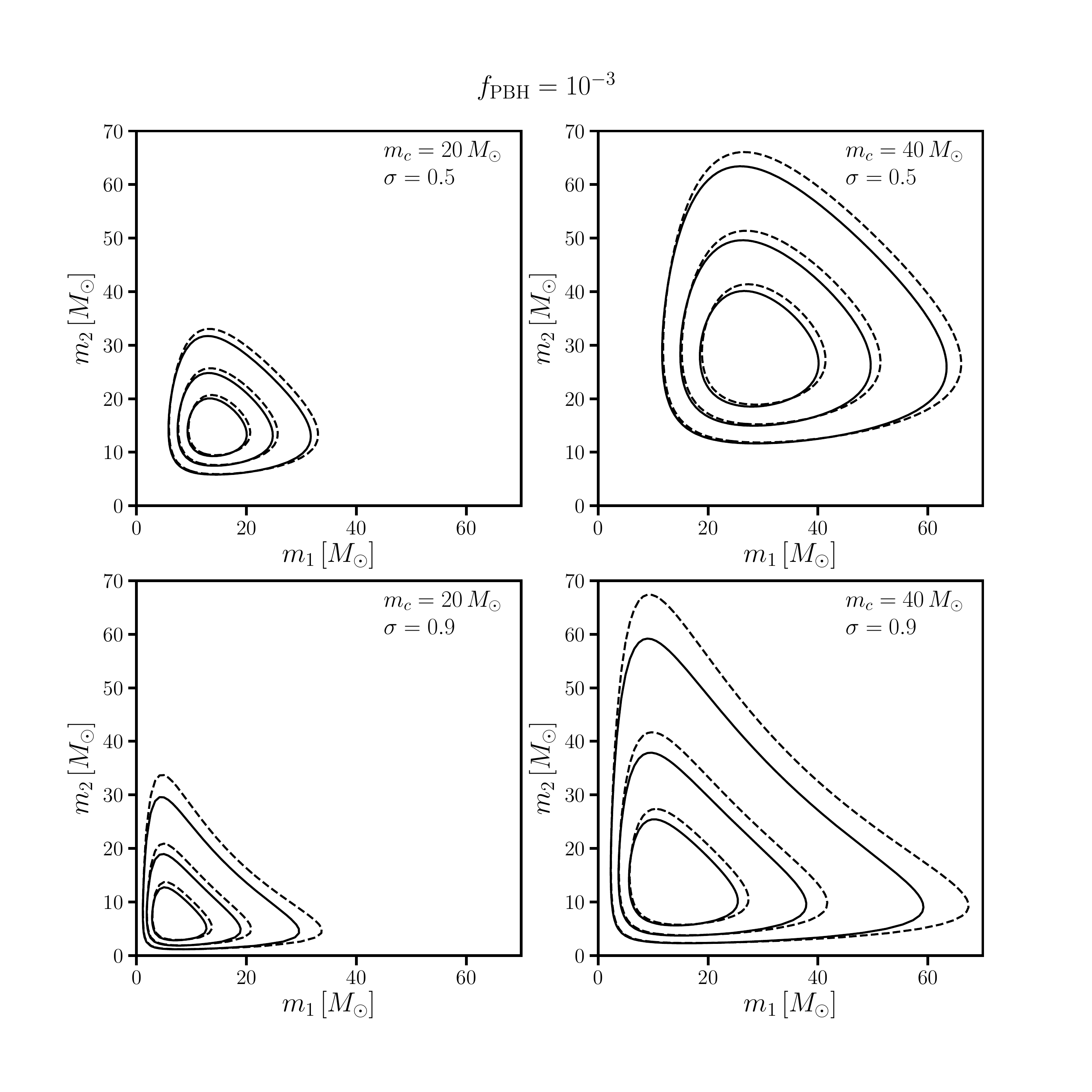}
\caption{Source-frame differential merger rate density $\mathrm{d}R/\mathrm{d}m_1\mathrm{d}m_2$ for lognormal PBH mass functions at $z=0.1$ and $\fpbh = 10^{-3}$, as a function of the individual source-frame masses. Solid and dashed contours are predictions with and without the suppression factor accounting for 3-body interactions respectively, and contours are drawn at 25\%, 50\% and 75\% of the peak value. The lognormal mass-function parameters are indicated in the top-right corner of each panel.}
\label{fig:merger_rate_shape}
\end{figure}

Importantly, for $\sigma \lesssim 1$, the shape of the merger rate distribution is primarily controlled by the mass function terms $\psi(m_1)\psi(m_2)$, with only limited sensitivity to the factors of $M$ and $\eta$ which multiply these terms in Equation~\eqref{eq:dR0}. These factors primarily control the shape of the tails of the distribution. When $\sigma \gtrsim 1$ the mass function is broad and $\psi(m_1)\psi(m_2)$ varies more slowly over a fixed mass range, such that factors of $M$ and $\eta$ can be relatively more important.

This behaviour is also seen in the importance of the suppression factor $S$, indicated by the difference between the solid and dashed lines in Figure~\ref{fig:merger_rate_shape}. The influence of $S$ on the shape of the merger rate distribution is weak for the mass function parameters plotted (which we will see correspond to those favoured by the data), increasing in importance for larger $\sigma$. The suppression factor depends only on the total mass, via Equation~\eqref{eq:Eq3p5}, with this dependence weakening for small $\fpbh/\sigma_M$.

Note also that the peak of the lognormal mass function is at $m_c e^{-\sigma^2}$, its mean is at $m_c e^{\sigma^2/2}$, and its median is at $m_c$, i.e.~for large $\sigma$ the distribution is significantly skewed.

The rate of merger events observable today can be found by integrating $\ud R$ over mass and volume, and is given by
\begin{equation}
  \frac{\beta}{T_{\mathrm{obs}}} = \int \ud z \ud m_1 \ud m_2 \,\frac{1}{(1+z)}\frac{\ud V_c}{\ud z} \frac{\ud R}{\ud m_1 \ud m_2}  p_{\mathrm{det}}(m_1, m_2, z),
  \label{eq:dN}
\end{equation}
where $\ud R = S \times \ud R_0$ with $\ud R_0$ and $S$ given by Equations~\eqref{eq:dR0} and~\eqref{eq:sfac}, and $\ud V_c$ is the comoving volume of a thin spherical shell of width $\ud z$. The factor of $(1+z)^{-1}$ accounts for the difference between proper (source frame) rate and observed (detector frame) rate. The total number of detectable mergers is $\beta$, and $T_{\mathrm{obs}}$ is the observation time. Note that Equation~\eqref{eq:dN} assumes that $p_{\mathrm{det}}$, and hence the interferometer PSD, is independent of time. We make this approximation throughout our analysis.

\begin{figure}
\centering
\includegraphics[width=0.7\textwidth]{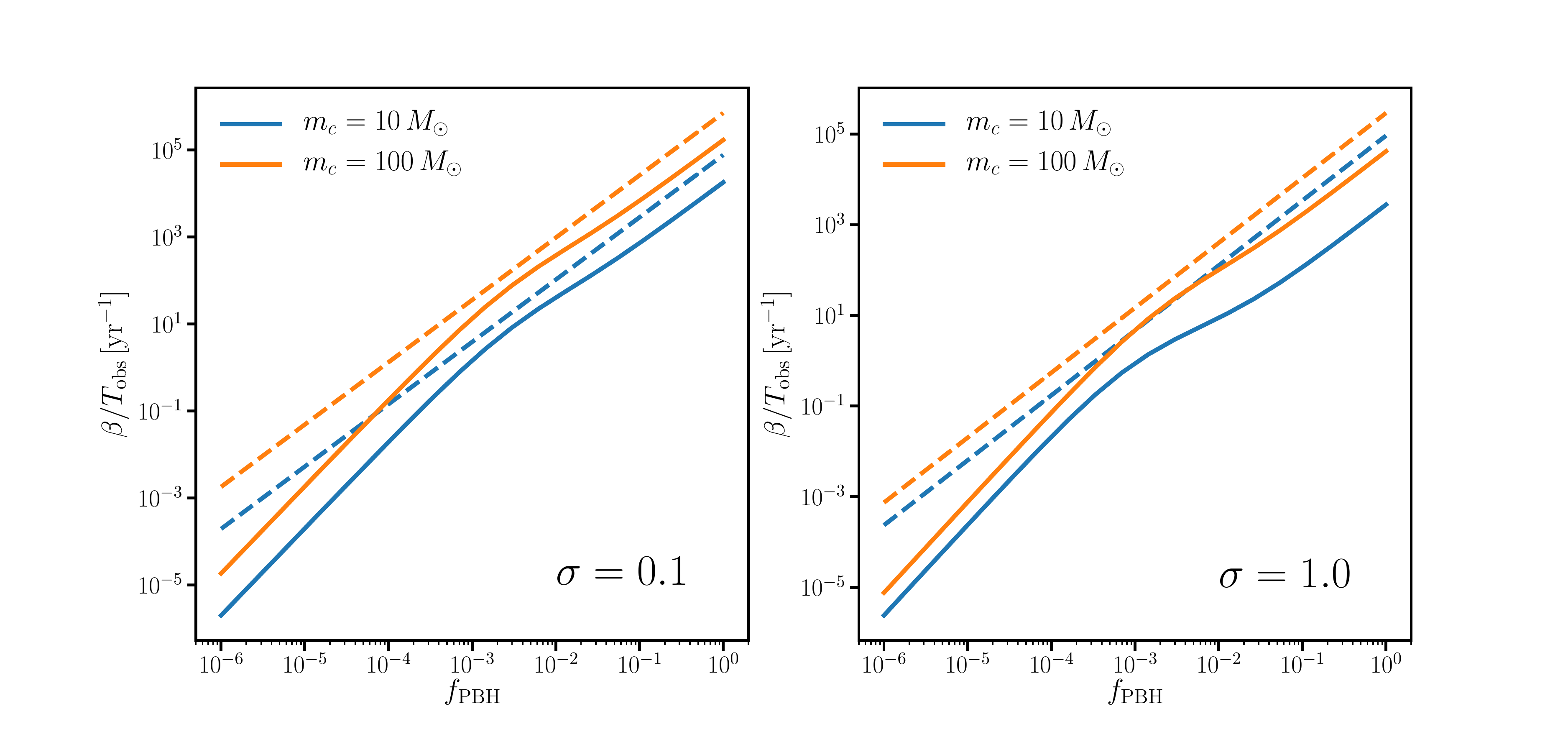}
\caption{Rate of detectable PBH mergers, per year as a function of $\fpbh$ with (solid) and without (dashed) the suppression factor, for $m_c = 10 \, M_\odot$ (blue) and $m_c = 100 \, M_\odot$ (orange), and for $\sigma=0.1$ (left panel) and $\sigma=1.0$ (right panel).}
\label{fig:beta}
\end{figure}

In Figure~\ref{fig:beta} we plot the detectable merger rate $\beta/T_{\mathrm{obs}}$ as a function of $\fpbh$, for representative values of $m_c$ and $\sigma$, and using the $p_{\mathrm{det}}$ described in Section~\ref{sec:data}. For a six-month observation period, the figure shows that 10 binary merger events would be expected for $\fpbh \sim$ a few $\times 10^{-3}$, with $\fpbh \sim 10^{-2}$ possible for large $\sigma$ and $m_c \sim 10 \, M_\odot$. Note that $\fpbh$ primarily controls the \emph{amplitude} of the merger rate, with only a minor impact on its mass dependence. This will be important when testing these models against the LIGO data.
%

\subsection{LIGO empirical merger models}
\label{subsec:ABrates}

As well as PBH models of binary mergers, we also consider two empirical distributions often used to model BBH populations. These models, termed Model A and Model B, were introduced in Refs.~\citep{2016PhRvX...6d1015A, 2017PhRvL.118v1101A} and extended in Refs.~\citep{2017ApJ...851L..25F, 2017PhRvD..95j3010K, 2018ApJ...856..173T, 2019PhRvD.100d3012W}. We use the forms as presented in Ref.~\citep{2019ApJ...882L..24A}.

Both models can be described by an intrinsic merger rate given by
\begin{equation}
  \frac{\ud R}{\ud m_1 \ud m_2} =
  \begin{cases}
    R_0 \, C(m_1) \, m_1^{-\alpha} q^{\beta_q} &\quad \mathrm{if} \quad m_{\mathrm{min}} \leq m_2 \leq m_1 \leq m_{\mathrm{max}} \\
    0 &\quad \mathrm{otherwise}
  \end{cases}
  \label{eq:dRAB}
\end{equation}
where $R_0$ is a constant amplitude, $q = m_2/m_1$, and $C(m_1)$ is such that the marginal distribution for the heavier mass is $m_1^{-\alpha}$. Model A fixes $m_{\mathrm{min}} = 5 \, M_\odot$ and $\beta_q = 0$, allowing $m_{\mathrm{max}}$, $R_0$, and $\alpha$ to vary. Model B allows all five parameters to vary. When $\beta_q = 0$ we have $C(m_1) \propto 1/(m_1 - m_{\mathrm{min}})$. In the default formulation of these models the rate $R_0$ is assumed to be independent of redshift.

The two models considered here are not intended to be detailed physical models of the merger rate of stellar black holes, rather they are empirical parameterisations that allow for straightforward computation for comparison with data. They do however have two important features, motivated by astrophysics, which will prove crucial when comparing with PBH models; the upper and lower cut-off in mass. The lower mass cut-off is motivated by observations of X-ray binaries~\citep{2010ApJ...725.1918O} and appears to be roughly $5 \, M_\odot$, with a mass gap expected between this and the predicted upper limit for a neutron star of roughly $2 \, M_\odot$ (see however Ref.~\citep{Abbott:2020khf} for a recent detection of a compact object in this mass gap). The upper limit is more uncertain, and is partly motivated a posteriori from the GWTC-1 catalogue. There is some astrophysical motivation for an upper limit of roughly $50 \, M_\odot$ from pulsational pair-instability supernovae, and potentially a mass gap between $50$--$150 \, M_\odot$ due to the combined effect of this with pair instability (see Ref.~\citep{2019PhRvX...9c1040A} for relevant references). Importantly the PBH models do not require upper or lower limits for the black hole mass, which will prove crucial for discriminating models. The detailed power-law behaviour of Models A and B is less physically motivated, so the LIGO models may be seen as a combination of astrophysical and empirical considerations.

The models considered in Ref.~\citep{2019ApJ...882L..24A} also specify distributions for the spin parameters~\citep{2017PhRvD..96b3012T}. To facilitate model comparison with the PBH merger scenario, where there is considerable uncertainty in the form of the spin distribution, we will neglect information from spin by marginalising over it as discussed further in Section~\ref{sec:stats}. We note in passing that all but two of the sources (GW151226 and GW170729) in the GWTC-1 catalogue are consistent with zero spin at 90\% confidence. We refer the reader to Ref.~\citep{2019JCAP...08..022F, 2020arXiv201013811G} for recent Bayesian analyses using spin in the PBH context and Ref.~\citep{2020arXiv200500023K} for a Bayesian study including a zero-spin black hole population, which could be considered a simplified proxy for a PBH model.

\begin{figure}
\centering
\includegraphics[width=0.7\textwidth]{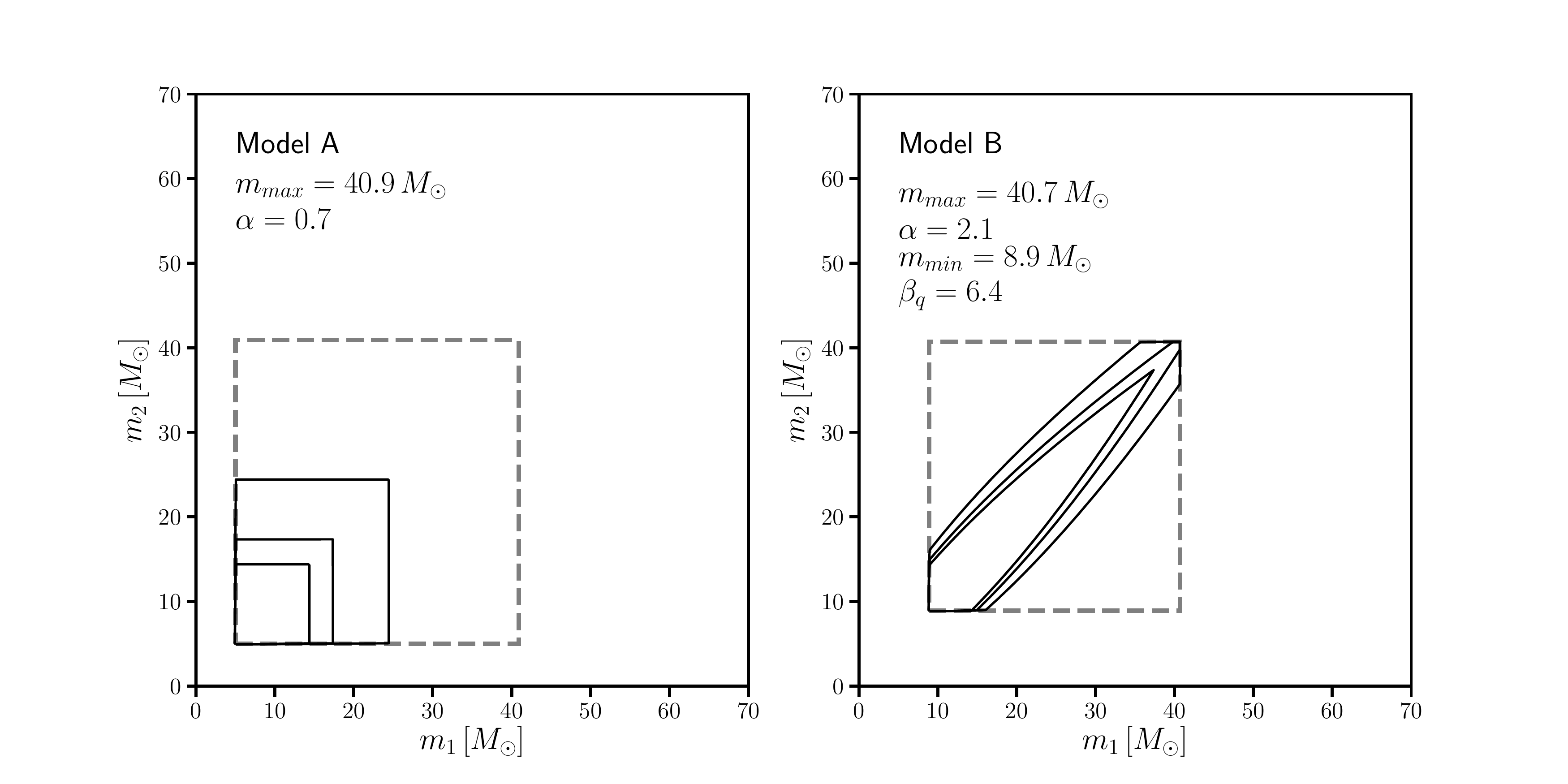}
\caption{Source-frame differential detectable merger rate for LIGO Model A (left panel) and Model B (right panel), as a function of the individual source-frame masses. The population parameters are indicated in the top left of each panel, and correspond to the values maximising the posterior of the O1O2 data. Contours are drawn at 25\%, 50\% and 75\% of the peak value, and the dashed lines indicate the boundaries of the distribution where the rates go to zero.}
\label{fig:AB_merger_rate_shape}
\end{figure}

In Figure~\ref{fig:AB_merger_rate_shape} we show the mass-dependence of the merger rate for Models A and B, analogous to Figure~\ref{fig:merger_rate_shape} in the case of PBHs, for representative values of the models' parameters. Note that for consistency with the PBH models we have extended the definition of $m_1$ and $m_2$ in Equation~\eqref{eq:dRAB} to the full mass plane, removing the requirement that $m_2 \leq m_1$. In the case of Model A the contours are perfect squares, being independent of the lighter mass for fixed heavier mass. For Model B with positive $\beta_q$ the merger rate is strongest for $q\approx 1$ and the contours are hence more concentrated around the diagonal.
That the distributions in Figure~\ref{fig:AB_merger_rate_shape} appear somewhat unphysical is a consequence of the explicit symmetry breaking between $m_1$ and $m_2$ in Equation~\eqref{eq:dRAB}. We emphasise that these distributions are not intended to be realistic models of astrophysical black hole binary formation, but rather capture broad features in the source population distributions.

A more complex model, dubbed Model C, is also analysed in Ref.~\citep{2019ApJ...882L..24A} and found to be superior fit to the GWTC-1 sources, although it is not significantly preferred over Model B. Both Model B and Model C were found to be better fits than Model A, so for simplicity we only consider the two models A and B, noting that Bayesian evidence ratios against Model C can be easily deduced from Table 3 or Ref.~\citep{2019ApJ...882L..24A}. A future extension of this work would be to compare PBH source distributions with more realistic models incorporating relevant astrophysical effects, in the manner of Refs.~\citep{2018ApJ...856..173T, 2019PhRvD..99j3004G}.

\section{Statistical framework}
\label{sec:stats}

To assess the merger rate models described in Section~\ref{sec:rates} against data from the BBH sources in the GWTC-1 catalogue, we must specify a likelihood function. As described in, e.g., Refs.~\citep{2004AIPC..735..195L, 2018ApJ...863L..41F, 2019MNRAS.486.1086M}, the likelihood of the observed data $\mathbf{d} = \{\mathbf{d}_i \}_{i=1 \dots N_{\mathrm{obs}}}$ from $N_{\mathrm{obs}}$ sources given population hyperparameters $\boldsymbol{\theta}$ and a merger rate model $M$ is
\begin{equation}
  p(N_{\mathrm{obs}}, \mathbf{d} \lvert \boldsymbol{\theta}, M) \propto \left[\prod_{i=1}^{N_{\mathrm{obs}}} \int  \ud m_1 \ud m_2 \ud z \, p(\mathbf{d}_i \lvert m_1, m_2, z) \frac{\ud N}{\ud m_1 \ud m_2 \ud z}(m_1, m_2, z \vert \boldsymbol{\theta}, M) \right] e^{-\beta(\boldsymbol{\theta}, M)},
  \label{eq:like}
\end{equation}
where $p(\mathbf{d}_i \lvert m_1, m_2, z)$ is the likelihood of each observed data set (i.e.~GW strain data) given the source masses and redshift (marginalising over all other parameters), $\ud N$ is the number of mergers in a mass and redshift interval, and $\beta(\boldsymbol{\theta}, M)$ is defined in Equation~\eqref{eq:dN}, i.e.
\begin{equation}
  \frac{\ud N}{\ud m_1 \ud m_2 \ud z}(m_1, m_2, z \vert \boldsymbol{\theta}, M) = T_{\mathrm{obs}} \frac{1}{(1+z)}\frac{\ud V_c}{\ud z} \frac{\ud R}{\ud m_1 \ud m_2}(m_1, m_2, z \vert  \boldsymbol{\theta}, M).
\end{equation}
This likelihood correctly accounts for the interferometer selection function via $p_{\mathrm{det}}$, which enters via the expected number of detected mergers $\beta$, and the uncertainties on the parameters of each source, which may be correlated. It assumes each source is independent. Since our data covers the O1 and O2 observing runs of LIGO, we take $T_{\mathrm{obs}} = 169.5 \, \mathrm{days}$.

As shown in Ref.~\citep{2019MNRAS.486.1086M}, the likelihood in Equation~\eqref{eq:like} is the product of a Poisson likelihood for the observed number of detections $N_{\mathrm{obs}}$ when $\beta(\boldsymbol{\theta}, M)$ were expected with a likelihood for observable data $\mathbf{d}_i$, the result being an inhomogeneous Poisson likelihood. Note that the selection function $p_{\mathrm{det}}$ only enters via $\beta$, since the observed data is observable by definition. The likelihood thus consistently accounts for information on population models coming from the observed distribution of source parameters $\emph{and}$ the overall number of detections -- these were considered separately for PBH models in a frequentist approach in Ref.~\citep{2020JCAP...01..031G}.

We note that several works constraining PBH merger models with GW data have used likelihoods differing from Equation~\eqref{eq:like}, for example Refs.~\citep{2019JCAP...02..018R, 2020JCAP...06..044D}. We emphasise that Equation~\eqref{eq:like} correctly accounts for the source parameter correlations and selection effects, and its use ensures that posteriors on the PBH model parameters are unbiased\footnote{Our approach is more similar to the recent Ref.~\citep{2020PhRvD.101h3008W}, differing in our self-consistent treatment of $\fpbh$ in the merger rate amplitude and our use of a suppression factor accounting for three-body effects.}.

We approximate the integral over the source likelihoods in Equation~\eqref{eq:like} with a sum over Monte Carlo samples from the source posteriors available from the GWTC-1 catalogue~\citep{2019PhRvX...9c1040A}. To do this we need to first divide out the source prior which was used in the LIGO inference, i.e.~we have
\begin{equation}
  p(N_{\mathrm{obs}}, \mathbf{d} \lvert \boldsymbol{\theta}, M) \propto \left[\prod_{i=1}^{N_{\mathrm{obs}}} \left \langle \frac{1}{\pi(z^i, m_1^i, m_2^i)}\frac{\ud N}{\ud m_1 \ud m_2 \ud z}(m_1^i, m_2^i, z^i \lvert \boldsymbol{\theta}, M) \right \rangle_{\{ z^i, m_1^i, m_2^i\}} \right]e^{-\beta(\boldsymbol{\theta}, M)},
  \label{eq:likeMC}
\end{equation}
where $\pi(z^i, m_1^i, m_2^i)$ is the prior on parameters for source $i$, and angle brackets denote an expectation value over MCMC samples from the source posterior. The prior is uniform in the detector-frame masses and scales as $d_L^2$ in the space of luminosity distance $d_L$, i.e.
\begin{equation}
  \pi(z, m_1, m_2) \propto (1+z)^2 d_L^2(z) \frac{\ud d_L(z)}{\ud z}.
\end{equation}

We ignore any information coming from the spin of the black holes by marginalising over spin parameters when averaging over the MCMC samples -- this is equivalent to assuming that the merger rates are independent of spin. Similarly we assume all intrinsic merger rates are independent of angular position and orientation, and marginalise over these parameters. This is also implemented in our treatment of the detection probability $p_{\mathrm{det}}$ in Equation~\eqref{eq:pdet}, which makes the implicit assumption that the signal-to-noise is not significantly impacted by the component spins\footnote{Alternatively, our neglect of spin can be phrased as the imposition of zero spin in all source components plus the assumption that the spin parameters are uncorrelated with the inferred masses and redshifts. This latter assumption allows us to include all the posterior samples when computing Equation~\eqref{eq:likeMC} and not just those lying in the zero-spin hypersurface. This is a reasonable approximation for the GWTC-1 sources.}.

We use Bayes' theorem to compute the posterior of the population hyperparameters, $p(\boldsymbol{\theta} \lvert \mathbf{d}, N_{\mathrm{obs}}, M) \propto p(\boldsymbol{\theta} \vert M)p( \mathbf{d}, N_{\mathrm{obs}} \lvert \boldsymbol{\theta}, M)$. For our lognormal PBH mass function we have $\boldsymbol{\theta} = [\fpbh, m_c, \sigma]$. This requires us to specify priors on the population hyperparameters. As is common in Bayesian inference problems the choice of these priors is somewhat arbitrary. We will see that the data is sufficiently constraining that the priors have negligible impact on posterior parameter constraints, but can significantly impact Bayesian evidences. We will see later that this latter prior dependence can be unpicked using a suitable approximation to the evidence.

In Table~\ref{tab:priors} we show the priors adopted in our inference runs. In the case of the LIGO Models A and B we use the priors adopted in Ref.~\citep{2019ApJ...882L..24A}. In the case of the PBH models we take a uniform prior on $\log \fpbh$ motivated by the fact that $\fpbh$ primarily controls the amplitude of the merger rate and its order of magnitude is unknown. We assume uniform priors on $\log m_c$ and $\log \sigma$ since their orders of magnitude are similarly unconstrained a priori, and for the reason that these would be the Jeffreys' priors on these parameters if the likelihood were proportional to $\psi(m)$\footnote{Note that the Jeffreys' prior for a Poisson distribution with rate parameter $\lambda$ is $\pi(\lambda) \propto \lambda^{-\frac{1}{2}}$, i.e.~uniform in $\sqrt{\lambda}$. In the PBH model the amplitude of the merger rate scales roughly as $\fpbh^{\frac{53}{37}}$, so an uninformative prior might be expected to scale roughly as $\fpbh^{-\frac{53}{74}} \approx \fpbh^{-0.7}$, i.e.~uniform in $\fpbh^{\frac{21}{74}}$. Our prior, scaling like $\fpbh^{-1}$, is therefore approximately uninformative.}.
\begin{table}
  \begin{tabular}{c c}
    \hline
    \hline
    Parameter & Prior \\
    \hline
    $\log_{10}{\fpbh}$ & $[-6, 0]$ \\
    $\log_{10}{m_c \, [M_\odot]}$ & $[0, 4]$ \\
    $\log_{10}{\sigma}$ & $[-1, 0.7]$ \\
    \hline
    $\log_{10}{R_0}$ & $[-1, 3]$ \\
    $m_{\mathrm{max}}\,[M_\odot]$ & $[30, 100]$ \\
    $m_{\mathrm{min}}\,[M_\odot]$ & $[5, 10]$ \\
    $\alpha$ & $[-4, 12]$ \\
    $\beta_q$ & $[-4, 12]$ \\
    \hline
    $\lambda$ & $[0, 0.5]$ \\
    $\log_{10}{m_{c,1} \, [M_\odot]}$ & $[-1, 3]$ \\
    $\log_{10}{m_{c,2} \, [M_\odot]}$ & $[-1, 3]$ \\
    \hline
   \end{tabular}
  \caption{Priors used in this work. All priors are uniform within the limits given in the right-hand column. Models sharing parameters which vary have the same priors on those parameters. See the main text for the definition of these parameters.}
  \label{tab:priors}
\end{table}

Finally, we assume that all the BBH sources in the catalogue are primordial in origin when performing inference under a PBH model, and that all sources are astrophysical when using Model A or Model B. In principle we should account for the possibility that some binaries consist of PBH pairs and some are astrophysical pairs (the merger rate of mixed PBH-astrophysical black hole binaries~\citep{Vattis:2020iuz, Tsai:2020hpi} is expected to be small compared with that of PBH-PBH binaries for the values of $\fpbh$ we consider). This could be implemented by introducing an extra parameter controlling the proportion of sources in each formation channel~\citep{2017CQGra..34cLT01V, 2019JCAP...08..022F}. For simplicity we do not take this approach, and instead treat all ten sources as either primordial or astrophysical, using Bayesian model selection to compare how well the respective models fit the data. Our constraints on $\fpbh$ should thus be interpreted as upper limits.

\section{Bayesian inference from the GWTC-1 catalogue}
\label{sec:bayes}

We use the likelihood in Equation~\eqref{eq:likeMC} with the models described in Section~\ref{sec:rates} and priors listed in Table~\ref{tab:priors} to draw samples from the posterior distribution of each model's parameters. We use the nested sampling algorithm~\citep{2004AIPC..735..395S, skilling2006} with multi-ellipsoidal bounded sampling~\citep{2009MNRAS.398.1601F} as implemented in \textsc{dynesty}~\citep{2020MNRAS.493.3132S} to draw samples from the posterior.

The nested sampling algorithm also computes the evidence for each model $M$, integrating over the prior as
\begin{equation}
  p(M \vert \mathbf{d}, N_{\mathrm{obs}}) \propto p(M) \int \ud \boldsymbol{\theta} \, p( \mathbf{d}, N_{\mathrm{obs}} \vert \boldsymbol{\theta}, M)p(\boldsymbol{\theta} \vert M),
\end{equation}
where $p(M)$ is the prior on the model. We assume that $p(M)$ is uniform, such that Bayes factors are equivalent to evidence ratios
\begin{equation}
  \frac{Z_{M_1}}{Z_{M_2}} \equiv \frac{\int \ud \boldsymbol{\theta} \, p( \mathbf{d}, N_{\mathrm{obs}} \vert \boldsymbol{\theta}, M_1)p(\boldsymbol{\theta} \vert M_1)}{\int \ud \boldsymbol{\theta} \, p( \mathbf{d}, N_{\mathrm{obs}} \vert \boldsymbol{\theta}, M_2)p(\boldsymbol{\theta} \vert M_2)}.
\end{equation}
We express all evidences relative to that of Model B, and quote errors on the evidence using the default first-order approximation produced by \textsc{dynesty}\footnote{We verify that this approximation to the evidence agrees with the more accurate \texttt{simulate\_run} approximation in \textsc{dynesty} to within 10\% in all cases, and also agrees well with resampled and jittered approximations to the evidence.}.

In Table~\ref{tab:baseline_results} we present the marginalised parameter constraints on the parameters of each model, as well as the evidence relative to Model B. We will first discuss the parameter constraints on the baseline PBH models and the empirical LIGO Models A and B, before discussing the evidences and model consistency tests. We will then introduce the extensions to the baseline models, the results of which are also listed in Table~\ref{tab:baseline_results} for completeness.
\begin{table}
  \begin{tabular}{p{30mm} p{20mm} p{20mm} p{15mm} p{15mm} p{25mm} p{25mm}}
      \hline
      \hline       
      Parameter & \multicolumn{6}{c}{Model}\\
      \hline
                                         & PBH   & PBH, $S=1$ & Model A & Model B & PBH, $S$=1, \hphantom{text} $m_{\mathrm{max}}= 50 \, M_\odot$ & PBH, $S$=1, skew-bimodal \\
          \hline
          $\log_{10}{\fpbh}$                      & $-2.30^{+1.16}_{-0.35}$    & $-2.76^{+0.25}_{-0.24}$  &    --                 &   --                   & $-2.72^{+0.25}_{-0.25}$    & $-2.74^{+0.23}_{-0.23}$   \\
          $\log_{10}{m_c \, [M_\odot]}$    & $1.38^{+1.36}_{-0.13}$    &  $1.26^{+0.12}_{-0.22}$  &     --                 &    --                  &  $1.91^{+1.91}_{-0.76}$    &    --          \\
          $\log_{10}{\sigma}$                      & $-0.09^{+0.49}_{-0.24}$    & $-0.21^{+0.24}_{-0.16}$  &    --                 &    --                  &  $0.27^{+0.23}_{-0.47}$    &    --           \\
          \hline
          $m_c \, [M_\odot]$              &  $24.23^{+528.62}_{-6.31}$ & $18.06^{+5.72}_{-7.10}$  &    --                  &   --                  &  $81.28^{+6525.7}_{-67.15}$ &     --          \\
          $\sigma$                                &   $0.82^{+1.71}_{-0.35}$   &  $0.61^{+0.45}_{-0.19}$  &    --                 &   --                   &  $1.86^{+1.30}_{-1.23}$    &      --          \\
          \hline
          $\log_{10}{R_0}$                         &    --                   &      --              &  $1.63^{+0.50}_{-0.45}$  &  $1.55^{+0.41}_{-0.43}$  &      --                &        --            \\
          $m_{\mathrm{max}}\,[M_\odot]$     &    --                   &     --                & $42.65^{+18.96}_{-5.99}$ & $42.73^{+35.11}_{-6.31}$ & $50.0$                 &        --            \\
          $m_{\mathrm{min}}\,[M_\odot]$     &     --                  &     --                &  $5.00$                &  $7.88^{+1.30}_{-2.64}$  &      --                 &       --           \\
          $\alpha$                                &     --                  &     --                &  $0.94^{+1.59}_{-2.38}$  &  $1.93^{+1.70}_{-1.96}$  &      --                 &        --           \\
          $\beta_q$                               &     --                  &     --                &  $0.00$                &  $6.62^{+5.04}_{-6.62}$  &      --                 &        --           \\
          \hline
          $\lambda$                               &      --              &     --                   &   --                 &   --                   &            --      &   $0.35^{+0.14}_{-0.27}$   \\
          $\log_{10}{m_{c,1} \, [M_\odot]}$&    --                &      --                  &    --                &    --                  &            --      &   $1.08^{+0.57}_{-0.38}$    \\
          $\log_{10}{m_{c,2} \, [M_\odot]}$&    --                &       --                 &    --                &    --                  &            --      &   $1.57^{+0.08}_{-0.62}$    \\
          $m_{c,1} \, [M_\odot]$          &    --                &      --                  &    --                &    --                  &            --      &    $12.02^{+32.65}_{-10.82}$ \\
          $m_{c,2} \, [M_\odot]$          &    --               &       --                 &     --               &    --                  &             --      &    $37.15^{+7.52}_{-28.24}$  \\
          \hline
          $\ln{L^*/L^*_{\mathrm{B}}}$               &   $-6.99$             &  $-7.14$                 &   $-2.51$           &  $0.00$                 &    $-5.44$                &    $-3.53$         \\
          $\ln{\mathrm{Occam}}$            &    $-6.13$           &    $-8.21$               &    $-5.71$            &  $-6.74$               &     $-5.46$              & $-7.73$              \\
          $\ln{Z_{\mathrm{Lap}}/Z_{NS}}$             & $1.60^{+0.16}_{-0.16}$  & $0.26^{+0.17}_{-0.17}$    & $0.77^{+0.15}_{-0.15}$   & $0.63^{+0.16}_{-0.16}$    &  $0.54^{+0.13}_{-0.13}$     & $1.92^{+0.18}_{-0.18}$  \\
          $\ln{Z_{NS}/Z_{NS,\mathrm{B}}}$            & $-7.35^{+0.23}_{-0.23}$  & $-8.25^{+0.23}_{-0.23}$  & $-1.62^{+0.22}_{-0.22}$ &  $0.00$                  &    $-4.01^{+0.21}_{-0.21}$  & $-5.79^{+0.24}_{-0.24}$ \\
           \hline
  \end{tabular}
  \caption{Median and 95\% credible intervals for the parameters of each model considered. The bottom four rows display difference in best-fit log-likelihood between each model and LIGO Model B, the log of the Occam factor defined in the text, the difference in log-evidence between the \textsc{dynesty} nested sampling estimate and the Laplace approximation defined in the text, and the Bayesian evidence ratios computed from nested sampling along with uncertainties.}
  \label{tab:baseline_results}
\end{table}

\subsection{Parameter constraints}
\label{subsec:parameters}

\subsubsection{PBH models}
\label{subsubsec:PBHpars}

Our baseline PBH model uses the merger rate model of Equation~\eqref{eq:dR0} with a suppression factor given in Equation~\eqref{eq:sfac}. To study the influence of the suppression factor we also consider a model with $S=1$. Differences between these two models can be roughly interpreted as encapsulating the uncertainty associated with PBH binary disruption.

\begin{figure}
\centering
\includegraphics[width=0.7\textwidth]{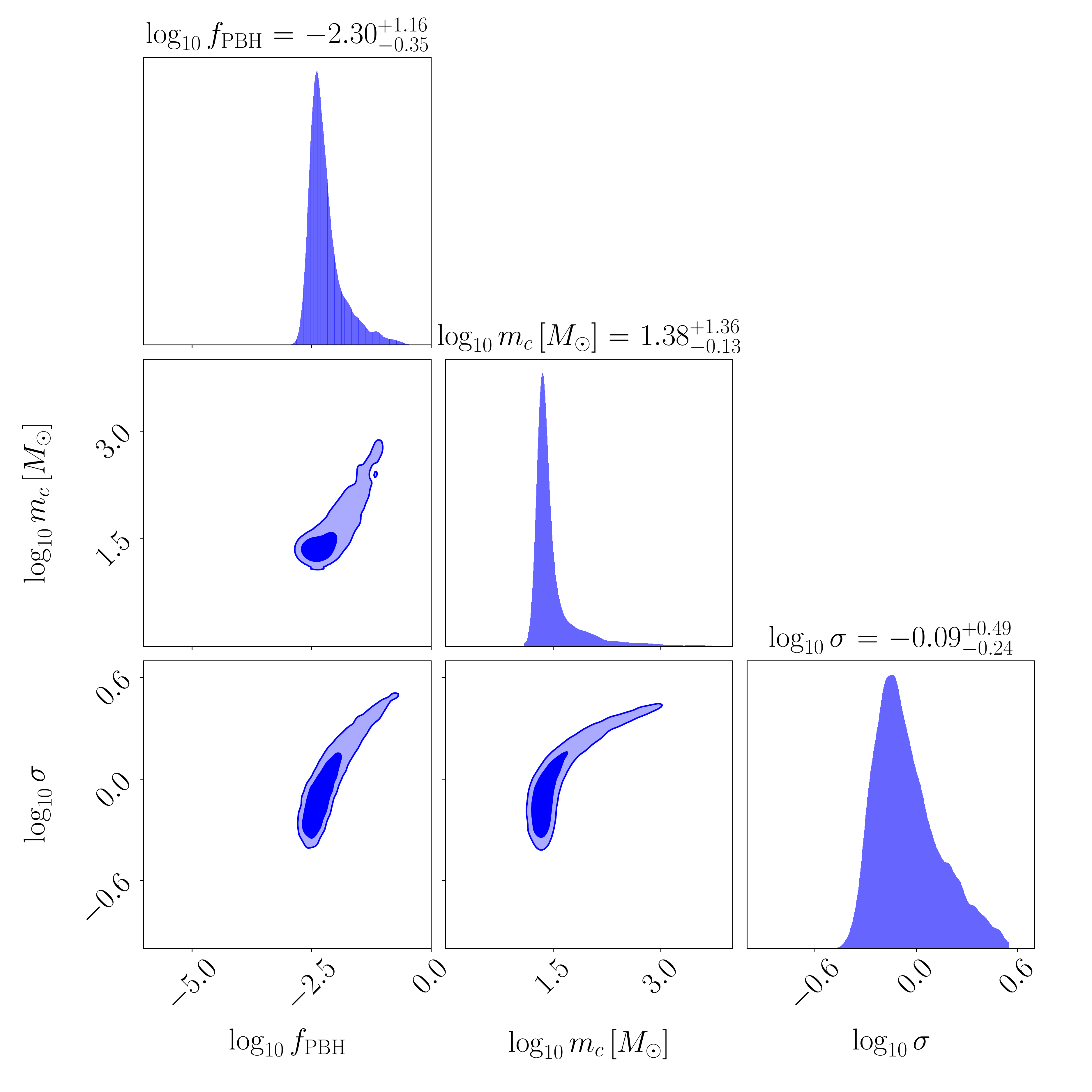}
\caption{\emph{Off-diagonal panels}: Two-dimensional 68\% and 95\% marginal posterior quantiles for the parameters of the lognormal PBH model including the 3-body suppression factor, given the GWTC-1 data. The plot boundaries correspond to the extent of the (uniform) priors on the parameters shown. \emph{Diagonal panels}: One-dimensional marginal posterior densities for the parameters. Above each panel are the marginalised posterior median and 95\% posterior quantiles for each parameter.}
\label{fig:PBH_posterior}
\end{figure}

In Figure~\ref{fig:PBH_posterior} we plot two-dimensional Bayesian credibility intervals (68\% and 95\% weighted posterior quantiles) for $\log_{10} \fpbh$ and the lognormal mass function parameters $\log_{10}m_c$ and $\log_{10}\sigma$, along with the marginalised one-dimensional posteriors. Note that the priors on these parameters are uniform, with limits given in Table~\ref{tab:priors}. Constraints on these, and the derived parameters $m_c$ and $\sigma$, are given in Table~\ref{tab:baseline_results}.

The posterior constraints (median and 95\% credible intervals) on these parameters are
\begin{align}
  \log_{10} \fpbh &= -2.30^{+1.16}_{-0.35} \nonumber \\
  m_c &= 24.23^{+528.62}_{-6.31} \, M_\odot \nonumber \\
  \sigma &= 0.82^{+1.71}_{-0.35}.
  \label{eq:PBHresults}
\end{align}
Figure~\ref{fig:PBH_posterior} demonstrates that the posterior is highly non-Gaussian, with a pronounced curving degeneracy between all three parameters. There is however a clear peak around the median values quoted above, with a preferred value of $\fpbh \approx 0.005$, assuming all the BBHs in GWTC-1 are primordial. As hinted at in Section~\ref{sec:rates} $\fpbh = 1$ is strongly disfavoured, with this model drastically overproducing BBH mergers. The preferred mass function parameters roughly correspond to the average mass of the components in the catalogue and the approximate spread in values.

The degeneracy tail in Figure~\ref{fig:PBH_posterior} skews the one-dimensional posteriors to large values of $m_c$, $\sigma$, and $\fpbh$. This tail (also visible in the likelihood plots in Ref.~\citep{2019JCAP...02..018R}) is a three-parameter degeneracy caused by the suppression factor, Equation~\eqref{eq:sfac}. We investigate its origin in detail in Appendix~\ref{app:degs}. Briefly, the suppression factor can allow for enhanced $\fpbh$ without overproducing mergers by increasing $\bar{N}(y)$, since $S \approx e^{-\bar{N}(y)}$. We compute this using Equation~\eqref{eq:Eq3p5}, which depends on the lognormal mass function parameters as $\bar{N}(y) \propto M/\langle m \rangle = (M/m_c)e^{\sigma^2/2}$. Large-$\sigma$ mass functions are highly skewed; the total mass $M$ is typically $\sim 2m_c$, meaning a high proportion of masses in the integral contributing to $\beta$, Equation~\eqref{eq:dN}, have $M \gg \langle m \rangle $ when $\sigma \gtrsim 1$, giving large suppression factors. Models with high $\sigma$ and high $\fpbh$ also need high $m_c$ in order to give an acceptable fit to the $\sim 10$ solar mass region occupied by the LIGO sources. Fixing $M$ to the LIGO mass scale implies that $m_c$ must be increased when $\sigma$ is increased to keep $m_ce^{-\sigma^2/2}$ fixed in order to keep the suppression factor constant in the observed mass range. This results in a three-parameter degeneracy allowing for $\fpbh$ as high as 0.07. We note that this partly arises due to the ambiguity of defining a `typical' mass scale in models with highly skewed and broad mass functions, which raises concerns about the validity of Equation~\eqref{eq:Eq3p5}. We note that the peak of the posterior is reasonably robust to the degeneracy tail, and that more accurate simulations will be needed to investigate the formation and evolution of PBH binaries with these extreme mass functions.

\begin{figure}
\centering
\includegraphics[width=0.45\textwidth]{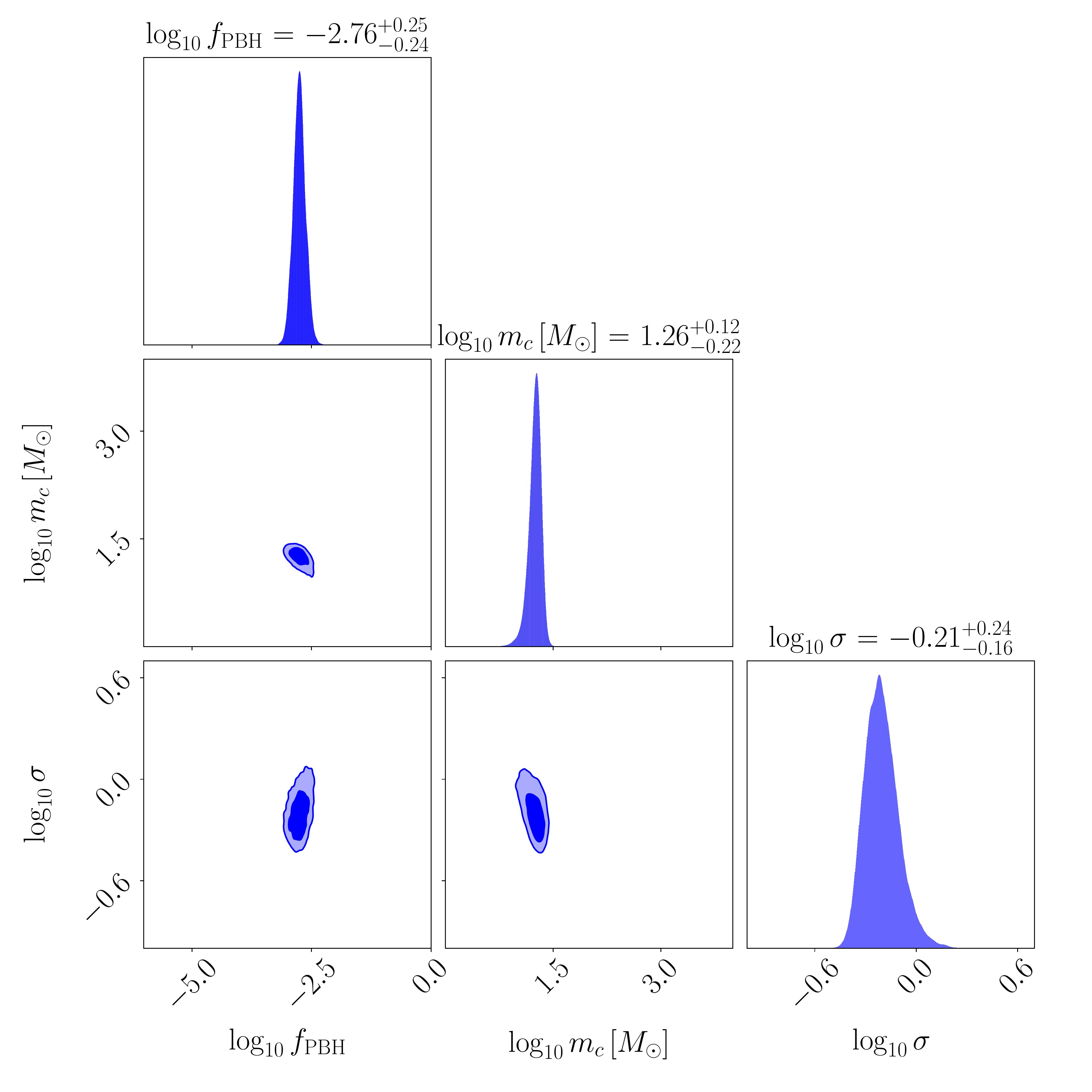}
\includegraphics[width=0.45\textwidth]{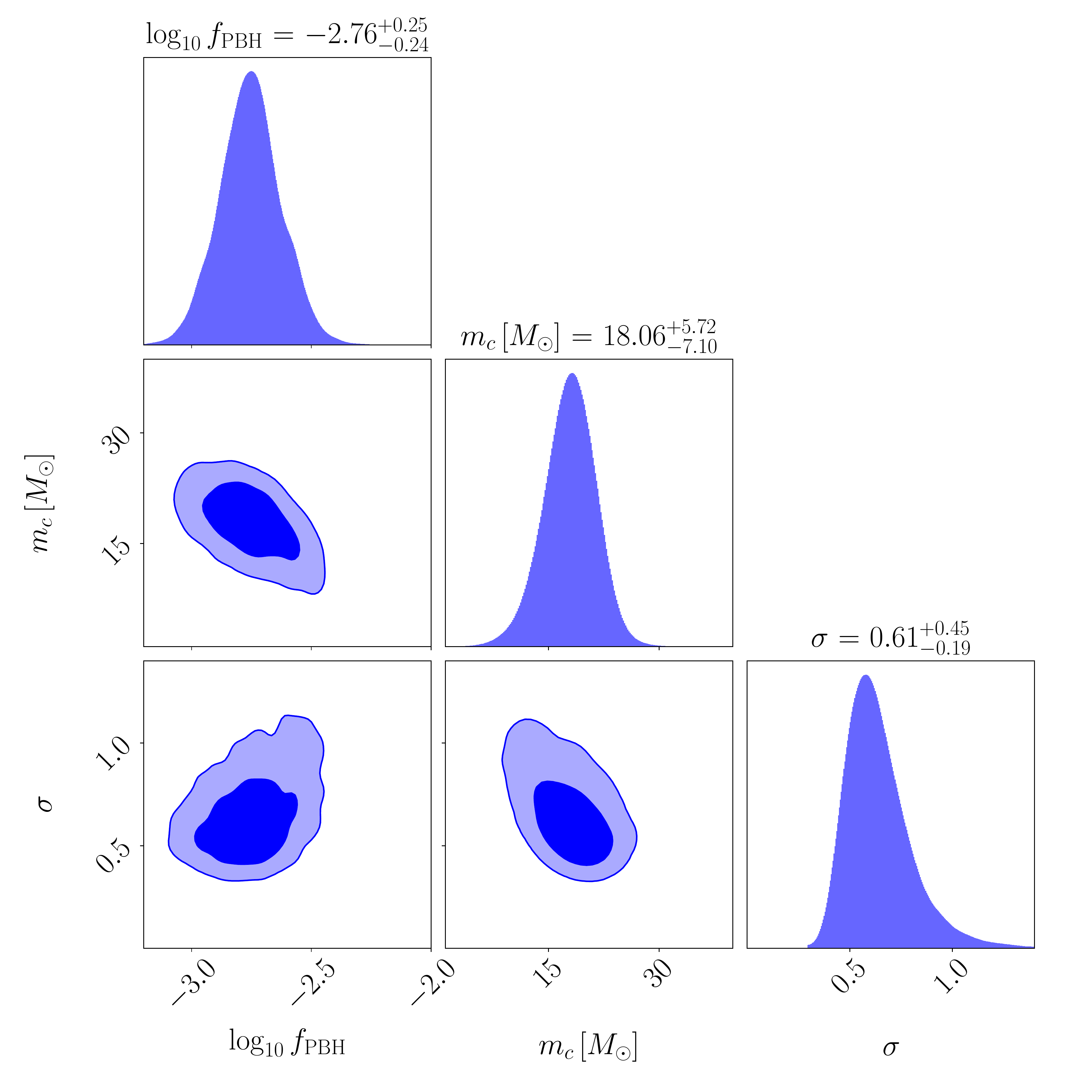}
\caption{\emph{Left, off-diagonal panels}: Two-dimensional 68\% and 95\% marginal posterior quantiles for the parameters of the lognormal PBH model without the 3-body suppression factor, given the GWTC-1 data. The plot boundaries correspond to the extent of the (uniform) priors on the parameters shown. \emph{Left, diagonal panels}: One-dimensional marginal posterior densities for the parameters. Above each panel are the marginalised posterior median and 95\% posterior quantiles for each parameter. \emph{Right}: Same as left panel for the parameters $(\log_{10}{\fpbh}, m_c, \sigma)$.}
\label{fig:PBH_posterior_nosup}
\end{figure}

This explanation for the degeneracy tail is supported by Figure~\ref{fig:PBH_posterior_nosup}, which shows the posterior for the PBH model with the suppression factor set to unity. In this case there is no mechanism available to suppress the merger rate when $\fpbh$ is high, and the constraints are much more Gaussian and confined in parameter space.

The constraints on the parameters for this $S=1$ model are (median and 95\% credible interval)
\begin{align}
  \log_{10} \fpbh &= -2.76^{+0.25}_{-0.24} \nonumber \\
  m_c &= 18.06^{+5.72}_{-7.10} \, M_\odot \nonumber \\
  \sigma &= 0.61^{+0.45}_{-0.19}.
\end{align}
These constraints arise from fitting the observed mass scale and spread in observed masses (which effectively fix $m_c$ and $\sigma$) and fitting $\fpbh$ to match the observed rate of mergers. There is a slight tendency for the data to simultaneously prefer low values of $m_c$ and high values $\sigma$, as seen in the right-hand panel of Figure~\ref{fig:PBH_posterior_nosup}. This combination keeps the merger rate roughly constant in the observed mass region, although the degeneracy is weak.

The median values of the parameters are fairly stable to switching on the suppression factor, with smaller values of $m_c$ and $\sigma$ now preferred due to the absence of the degeneracy tail and the median $\fpbh$ now $\approx 0.0017$, i.e.~almost a factor of three smaller. Once again, $\fpbh \approx 1$ is highly disfavoured.

We note that for both PBH models the priors chosen are sufficiently broad that they do not influence the posterior constraints on the parameters, as shown by Figure~\ref{fig:PBH_posterior} and the left panel of Figure~\ref{fig:PBH_posterior_nosup}.

\subsubsection{LIGO Model A and Model B}
\label{subsubsec:ABpars}

\begin{figure}
\centering
\includegraphics[width=0.45\textwidth]{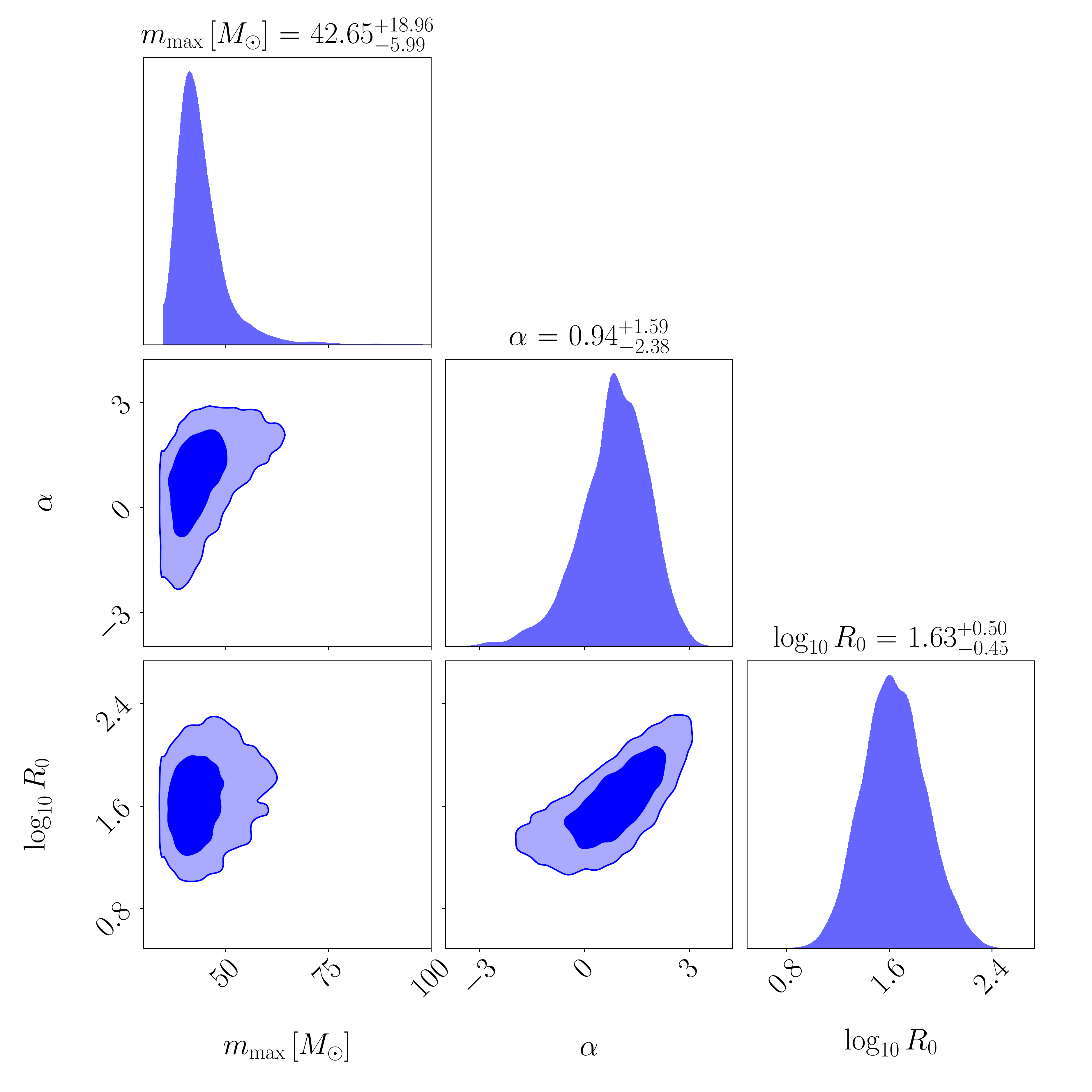}
\includegraphics[width=0.45\textwidth]{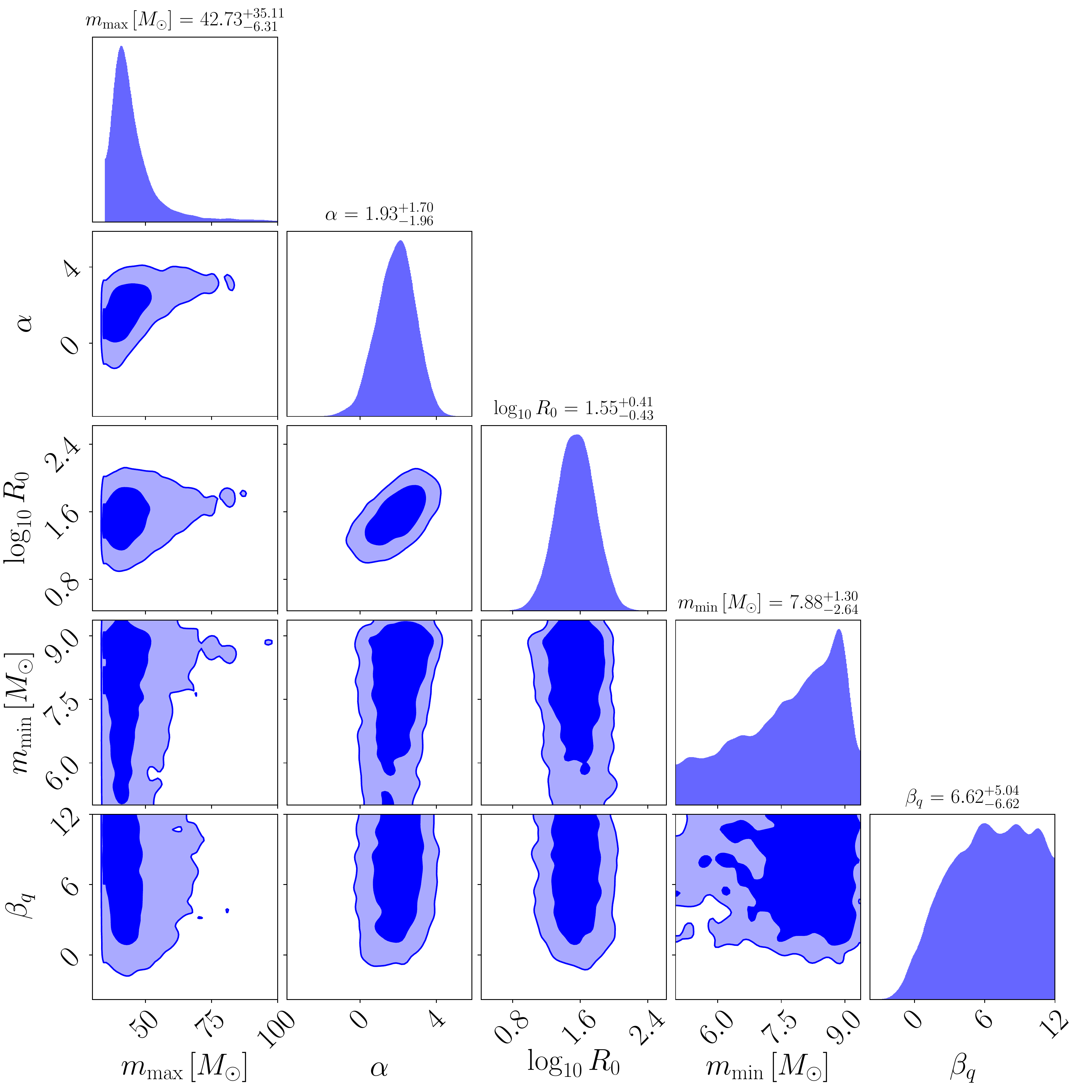}
\caption{Posteriors of the parameters of LIGO Model A (left panel) and Model B (right panel) given the GWTC-1 data. The meaning of the contours and quoted error significance are the same as in Figure~\ref{fig:PBH_posterior}.}
\label{fig:posterior_AB}
\end{figure}

In Figure~\ref{fig:posterior_AB} we show the posterior constraints on the parameters of the empirical LIGO models A and B, using the priors listed in Table~\ref{tab:priors}. These posteriors are fully consistent with those presented in Ref.~\citep{2019ApJ...882L..24A}, with only weak constraints provided on the Model B parameters $m_{\mathrm{min}}$ and $\beta_q$. In contrast the constraints on the upper component mass limit $m_{\mathrm{max}}$ and the power law slope in the distribution of the heavier mass $\alpha$ are reasonably well-constrained, being determined by the maximum component mass in the catalogue and the typical spread in masses respectively (c.f.~the $m_c$ and $\sigma$ parameters of the PBH model). The amplitude of the merger rate simply fits the observed number of mergers, analogous to $\fpbh$ in the PBH model.
%
%

The parameter constraints (median and 95\% credible intervals) for Model A are
\begin{align}
  m_{\mathrm{max}} &= 42.65^{+18.96}_{-5.99} \, M_\odot \nonumber \\
  \alpha &= 0.94^{+1.59}_{-2.38} \nonumber \\
  R_0 &= 42.66^{+92.24}_{-27.52} \, \mathrm{Gpc}^{-3} \mathrm{yr}^{-1},
\end{align}
while those for Model B are
\begin{align}
  m_{\mathrm{max}} &= 42.73^{+35.11}_{-6.31} \, M_\odot \nonumber \\
  \alpha &= 1.93^{+1.70}_{-1.96} \nonumber \\
  R_0 &= 35.48^{+55.72}_{-22.30} \, \mathrm{Gpc}^{-3} \mathrm{yr}^{-1} \nonumber \\
  m_{\mathrm{min}} &= 7.88^{+1.30}_{-2.64} \, M_\odot \nonumber \\
  \beta_q &= 6.62^{+5.04}_{-6.62}.
\end{align}
These constraints are consistent with those presented in Ref.~\citep{2019ApJ...882L..24A}, with the exception of $R_0$ which we find to be typically smaller with $R_0^{\mathrm{LIGO}}/R_0^{\mathrm{here}} \approx 1.50$ for both models. This can be explained by the difference in $p_{\mathrm{det}}$ arising from using the semi-analytic approximation described in Section~\ref{sec:data} vs.~a more accurate method using pipeline injections, as discussed in Appendix A of Ref.~\citep{2019ApJ...882L..24A}. Our approximation overestimates the LIGO sensitive volume by between a factor of 1.4 and 1.9 depending on $m_{\mathrm{max}}$ and $\alpha$ (top left panel of Figure 11 in Ref.~\citep{2019ApJ...882L..24A}), leading to an \emph{underestimate} of $R_0$ by roughly the same factor in order to keep the total number of observed events fixed.

We close this section by noting that the weak constraints on the Model B parameter $\beta_q$ reflect the weak constraints on mass ratios in the GWTC-1 catalogue (see Figure~\ref{fig:O1O2_data_Mcz_q}). Recently the LIGO-Virgo Collaboration reported detections of BBH mergers with significantly asymmetric masses having $q \approx 0.3$~\citep{2020arXiv200408342T} and $q\approx 0.1$~\citep{Abbott:2020khf}. Consequently the constraints on $\beta_q$ tighten significantly when these sources are included. We only make use of the sources detected in the O1 and O2 observing runs in this work, but discuss the implications of reported O3 detections in Section~\ref{sec:conclusions}. In a future work we intend to repeat the analysis of this work with the $\mathcal{O}(100)$ detections expected in the final O3 catalogue.

\subsection{Evidences, goodness-of-fit tests and model consistency}
\label{sec:fits}

Having presented constraints on the parameters of the two PBH models and the two empirical LIGO models, we now examine the quality of model fits and compare the models using the Bayesian evidence.

\subsubsection{Posterior merger rate distributions}
\label{subsubsec:dbetas}

\begin{figure}
\centering
\includegraphics[width=\textwidth]{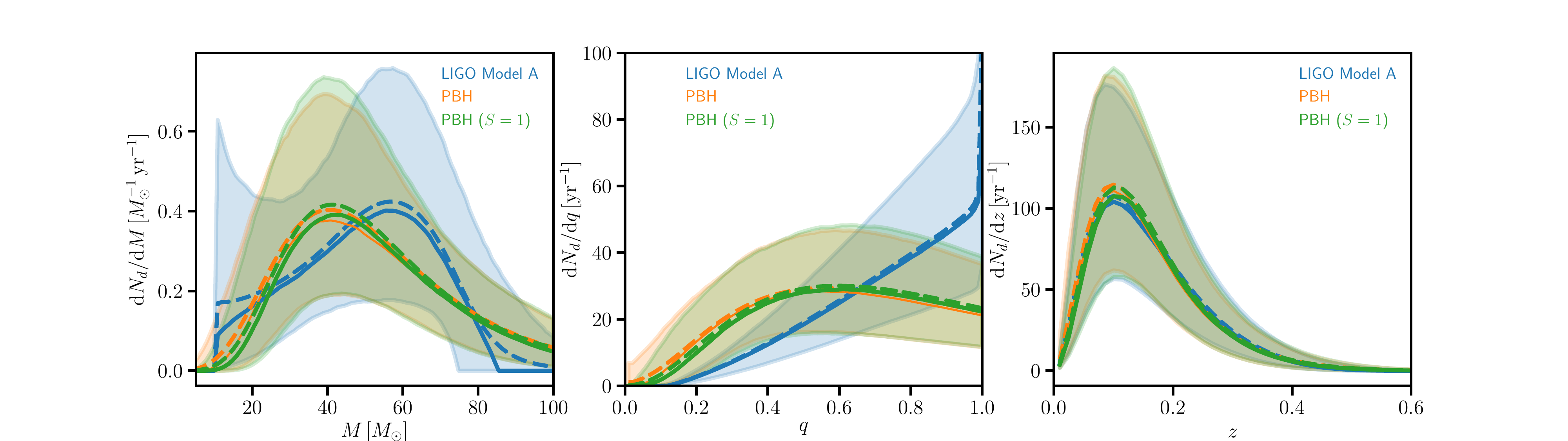}
\caption{Differential detector-frame merger rates with respect to total mass (left panel), mass ratio (middle panel) and redshift (right panel) for LIGO Model A (blue), the lognormal PBH model (orange) and the lognormal PBH model with suppression factor set to unity (green). In each case we plot the median and 90\% quantiles over the posterior samples for each model given the GWTC-1 data (solid lines and shaded bands), and the (weighted) mean over the samples (dashed lines).}
\label{fig:dbetas}
\end{figure}

We first examine the preferred distributions of source parameters in each model by computing the allowed values of the differential detector-frame merger rate, plotted in Figure~\ref{fig:dbetas}. These figures show the derivative of $N_d \equiv \beta/T_{\mathrm{obs}}$ with respect to total mass $M$, mass ratio $q$, and redshift $z$ for each model averaged over the posterior distributions of the population hyperparameters. We show results for the PBH model, the PBH model with $S=1$, and Model A. The area under each curve in Figure~\ref{fig:dbetas} is fixed at roughly $10/T_{\mathrm{obs}}$, since most of the posterior mass lies in a region where $\beta \approx 10$, matching the 10 observed sources. 

The preferred merger rates are similar between the three models, with the suppression factor making very little difference to the results. This is also evident from the parameter posteriors of the two models which have most of their mass in a similar region of parameter space, the degeneracy tail in the suppression factor model having little influence on the preferred differential merger rates. The redshift dependence in all three cases is determined from that of the selection probability $p_{\mathrm{det}}$ and the comoving volume element, c.f. Figure~\ref{fig:pdet_zs}.

The distributions in total mass for the PBH models inherit the lognormal shape of the mass function $\psi(m)$, with a peak at roughly $40 \, M_\odot$ and a long tail to high masses. Model A in contrast has both a minimum and maximum cut-off in mass. Note that the median Model A merger rate is monotonically decreasing in its heavier mass ($\alpha \approx 0.4$), but this is counteracted by the detection probability $p_{\mathrm{det}}$ which increases with mass (see Figure~\ref{fig:pdet_z0p1}). These combined effects produce a peak in the merger rate around $M \approx 60 \, M_\odot$, and a smaller peak around $m_{\mathrm{min}}$ where the differential merger rate formally diverges (visible in the 90\% confidence region). In the case of the PBH models the exponential fall-off at high masses dominates over $p_{\mathrm{det}}$ giving a single peak.

The most pronounced differences occur in the dependence on mass ratio $q$ (middle panel of Figure~\ref{fig:dbetas}). As pointed out in Ref.~\citep{2020JCAP...01..031G}, the broad PBH mass function allows for mass ratios significantly different from unity. In contrast, and as is evident from comparing Figure~\ref{fig:AB_merger_rate_shape} and Figure~\ref{fig:merger_rate_shape}, Model A merger rates typically have more symmetric masses. Despite the visible difference in the $q$ distribution in Figure~\ref{fig:dbetas}, the errors on $q$ in the GWTC-1 catalogue are typically large, and the mass ratio has little discriminatory power between the PBH and LIGO models\footnote{We note that a source with $q \approx 0.3$ or $q\approx0.1$, as recently detected in the O3 run, is predicted to be significantly more likely in the PBH models than under Model A.}.

\subsubsection{Bayesian evidence ratios}
\label{subsubsec:evidences}

In Table~\ref{tab:baseline_results} we present the Bayesian evidence for each model relative to that of Model B. These quantities indicate the posterior preference for each model after marginalising over each of its parameters.

For the PBH model we find a (natural log) evidence ratio of $-7.35 \pm 0.23$ compared to Model B (errors here are approximately standard deviations). On the Jeffreys' scale (e.g., ~\citep{2007MNRAS.378...72T}) this corresponds to `decisive' evidence in favour of Model B compared with the PBH model. For the PBH model with $S=1$ the log-Bayes factor is $-8.25 \pm 0.23$, i.e.~this model is even more disfavoured compared with Model B. The evidence ratio between the PBH models is $0.90 \pm 0.23$, i.e.~the data do not show evidence for a suppression factor given our choice of priors.

For Model A we find a log-evidence ratio of $-1.62 \pm 0.22$, i.e.~positive or substantial but not strong evidence in favour of Model B. This is consistent with the result reported in Ref.~\citep{2019ApJ...882L..24A} of $-1.42$. We note that Table 3 of Ref.~\citep{2019ApJ...882L..24A} implies that the LIGO Model C is slightly (but not significantly) preferred over Model B. We find Model A is strongly preferred over the PBH models with log-evidences of $-5.73 \pm 0.23$ and $-6.63 \pm 0.23$ for the cases with and without the suppression factor respectively.

Taken at face value these evidences suggest that both PBH models are strongly disfavoured compared with the simple empirical models A and B. However it is well known that Bayesian evidences can be strongly influenced by the choice of priors, so it is beneficial to delve a bit deeper into the evidence ratios. We can make progress by employing the \emph{Laplace approximation} for the evidence, discussed in Ref.~\citep{10.5555/971143}. This assumes that the posterior is approximately Gaussian around its peak (which occurs at the point $\boldsymbol{\theta}_{{\rm BF}})$, such that the integral over parameters can be approximated (for a uniform prior) as
\begin{equation}
  p(\mathbf{d} \vert M) \approx p(\mathbf{d} \vert \boldsymbol{\theta}_{{\rm BF}}, M) \times \frac{\sqrt{\mathrm{det}(2\pi \mathbf{C})}}{\mathrm{Vol}_{\pi(\boldsymbol{\theta})}},
  \label{eq:Laplace}
\end{equation}
where $\mathbf{C}$ is the covariance matrix of the posterior and $\mathrm{Vol}_{\pi(\boldsymbol{\theta})}$ is the prior volume (i.e.~the volume of the cube defining our uniform priors). The first term on the right-hand side of Equation~\eqref{eq:Laplace} is the likelihood value at of the best-fitting model, a quantifier of model fit quality well known from classical statistics. The second term is the `Occam factor' expressing the ratio of the posterior volume to the prior volume. The Occam factor quantifies the degree to which the region of acceptable parameter values shrinks upon arrival of the data, and penalises models for which this shrinkage is large i.e.~models which require finely tuned parameter values amongst those which were allowed a priori.

Since the posterior of both Model B and the PBH model with suppression are significantly non-Gaussian, the Laplace approximation is expected to be only a coarse model for the evidence. In Table~\ref{tab:baseline_results} we give the differences between the Laplace-approximated evidence $Z_{\mathrm{Lap}}$ and the nested sampling estimate $Z_{NS}$. We find that $Z_{\mathrm{Lap}}$ provides a remarkably good approximation to the true evidences, with log-ratios ranging from roughly 0.26 for the $S=1$ model (which has the most Gaussian posterior) to 1.60 for the full PBH model (which has a strongly non-Gaussian posterior). In all cases the discrepancy between $Z_{\mathrm{Lap}}$ and $Z_{NS}$ is significantly smaller than the difference from the evidence for Model B.

With the Laplace-approximated evidences we can start to understand why some models are favoured over others. In Table~\ref{tab:baseline_results} we show the ratios of the terms in Equation~\eqref{eq:Laplace} with those of Model B. The ratio of first terms is just $L^*/L_B^*$, the likelihood ratio of the best-fit model compared with that of Model B. In the case of the PBH models we find that this term dominates the evidence ratio. The Occam factor is similar between the PBH model with suppression factor and Model B, but is more penalising for the $S=1$ PBH model since the shrinkage in prior volume is much greater, as evident from Figure~\ref{fig:PBH_posterior_nosup}.

The log-Bayes factors depend on the prior volume, via the Occam factor, as $\ln Z \sim -\ln \mathrm{Vol}_{\pi(\boldsymbol{\theta})}$. The evidence ratios of the PBH models compared with the LIGO models are thus sensitive to the prior range on $\fpbh$ and the mass function parameters. If, for example, we reduced the prior lower limit on $\fpbh$ from $10^{-6}$ to $10^{-16}$ we would reduce the log-Bayes factor compared to Model B by roughly one\footnote{In reality our posterior limits on $\fpbh$ are upper limits due to our assumption that every source is a PBH-PBH merger, so the Occam factor is probably not as penalising as this example suggests.}. This prior range could be easily exceeded if instead we placed a uniform-in-log prior on the primordial power spectrum amplitude, to which $\fpbh$ is exponentially sensitive, a point we discuss further in Section~\ref{sec:conclusions}. Similarly, increasing the ranges of $m_c$ and $\sigma$ would also increase the evidence against the PBH models compared with the LIGO models.

Note that there is no freedom to \emph{reduce} the prior range of the PBH model parameters without being overly informative, i.e.~we cannot attempt to boost the evidence of PBH models by making the Occam factor less penalising, unless some other prior information or physical insight demands it. Could we instead try to penalise the LIGO models to restore the prior ambivalence between models? In the case of Model A, we would need to increase the prior volume by a factor of roughly 300 to give an evidence ratio of unity with the PBH model. This could be achieved by expanding each side of the prior cube by a factor of roughly 6.7, i.e.~with priors $m_{\mathrm{max}} [M_\odot] = [30, 500]$, $\log_{10} R_0 = [-12, 14]$, $\alpha = [-50, 58]$. A priori this seems an extreme prior range which is likely to be unphysical. Rather than change the limits of the prior we could also change its density such that the prior volume contained more prior mass (note this would require a modification to Equation~\eqref{eq:Laplace}). In the absence of a more fundamental astrophysical theory there is no obviously preferred choice of parameter combination on which to impose a uniform prior. There is thus no well-motivated way to make the data favour the PBH model over the LIGO models by simply changing parameter priors.

We note that the prior volumes cancel in the evidence ratio of the PBH model with and without a suppression factor. We can therefore make the robust statement that the GWTC-1 data are not sensitive to the suppression factor, and inference of PBH models may proceed with $S=1$ with negligible loss of accuracy.

\subsubsection{Posterior predictive distributions}
\label{subsubsec:ppds}

Having seen that the likelihood ratio is primarily responsible for the evidence against a lognormal PBH model, and having argued that changing the prior on parameters to restore model parity is challenging, we now investigate the cause of the likelihood differences in detail. Equation~\eqref{eq:Laplace} tell us that it is sufficient to consider only the likelihood at the best-fit model, but the same conclusions can be reached by considering the likelihood averaged over the model space allowed by the data -- the \emph{posterior predictive distribution} (PPD), defined as
\begin{equation}
  p(D| \mathbf{d}, N_{\mathrm{obs}}, M) = \int \ud \boldsymbol{\theta} \, p(D \vert \boldsymbol{\theta}, M) p(\boldsymbol{\theta} | \mathbf{d}, N_{\mathrm{obs}}, M),
  \label{eq:ppd}
\end{equation}
where $D$ is \emph{unseen} data. Equation~\eqref{eq:ppd} is similar to the Bayesian evidence, except it is now an integral of the likelihood of new data over model parameters allowed by the old data $\mathbf{d}$ and $N_{\mathrm{obs}}$.

The PPD is a useful quantity to compute since it can be used to approximate the part of the evidence ratio coming from the likelihood ratio in the Laplace approximation Equation~\eqref{eq:Laplace}. It can also be used to assess the absolute quality of the model fit in a more `Bayesian' way than a classical $\chi^2$ test~\citep{Gelman96posteriorpredictive}.

We approximate the integral in Equation~\eqref{eq:ppd} with an average over posterior samples from our nested sampling runs. The challenge in implementing the PPD is finding an approximation for the likelihood of unseen data $p(D \vert \boldsymbol{\theta}, M)$, as so far we have only needed the likelihood of the GWTC-1 data as a function of parameters, which we extracted indirectly via the GWTC-1 posterior samples. In Appendix~\ref{app:ppd} we present a detailed derivation of this approximate likelihood. This results in a PPD for the redshifted chirp mass given by

\begin{equation}
  p(\{ \widehat{\mathcal{M}}_z \} \vert \{ \mathbf{d} \}) = \prod_{i=1}^{N_{\mathrm{obs}}} \int \ud \mathcal{M}_z \, p(\widehat{\mathcal{M}}^i_z \vert  \mathcal{M}_z) p(\mathcal{M}_z \vert \{ \mathbf{d} \}),
  \label{eq:convlike_text}
\end{equation}
where
\begin{equation}
  p(\mathcal{M}_z \vert \{ \mathbf{d} \}) = \int \ud \boldsymbol{\theta} \, p(\mathcal{M}_z \vert \boldsymbol{\theta}) p(\boldsymbol{\theta} \vert \{ \mathbf{d} \}).
  \label{eq:PPD_text}
\end{equation}
In these expressions $\{ \widehat{\mathcal{M}}_z \}$ denotes the set of `measured' chirp masses, with members of the set denoted by $\widehat{\mathcal{M}}^i_z $.

The PPD in Equation~\eqref{eq:PPD_text} can be convolved with the individual constraints on chirp mass from each source to give a likelihood for new unseen chirp masses averaged over population parameters allowed by existing data. When evaluated at the actual chirp mass values in the GWTC-1 catalogue, this gives the likelihood function marginalised over the absolute merger rate and the population parameters -- this is approximately equivalent to the likelihood evaluated at the best-fitting population model, which is the key quantity in determining whether the Bayesian evidence favours PBH over the LIGO models.

\begin{figure}
\centering
\includegraphics[width=0.8\textwidth]{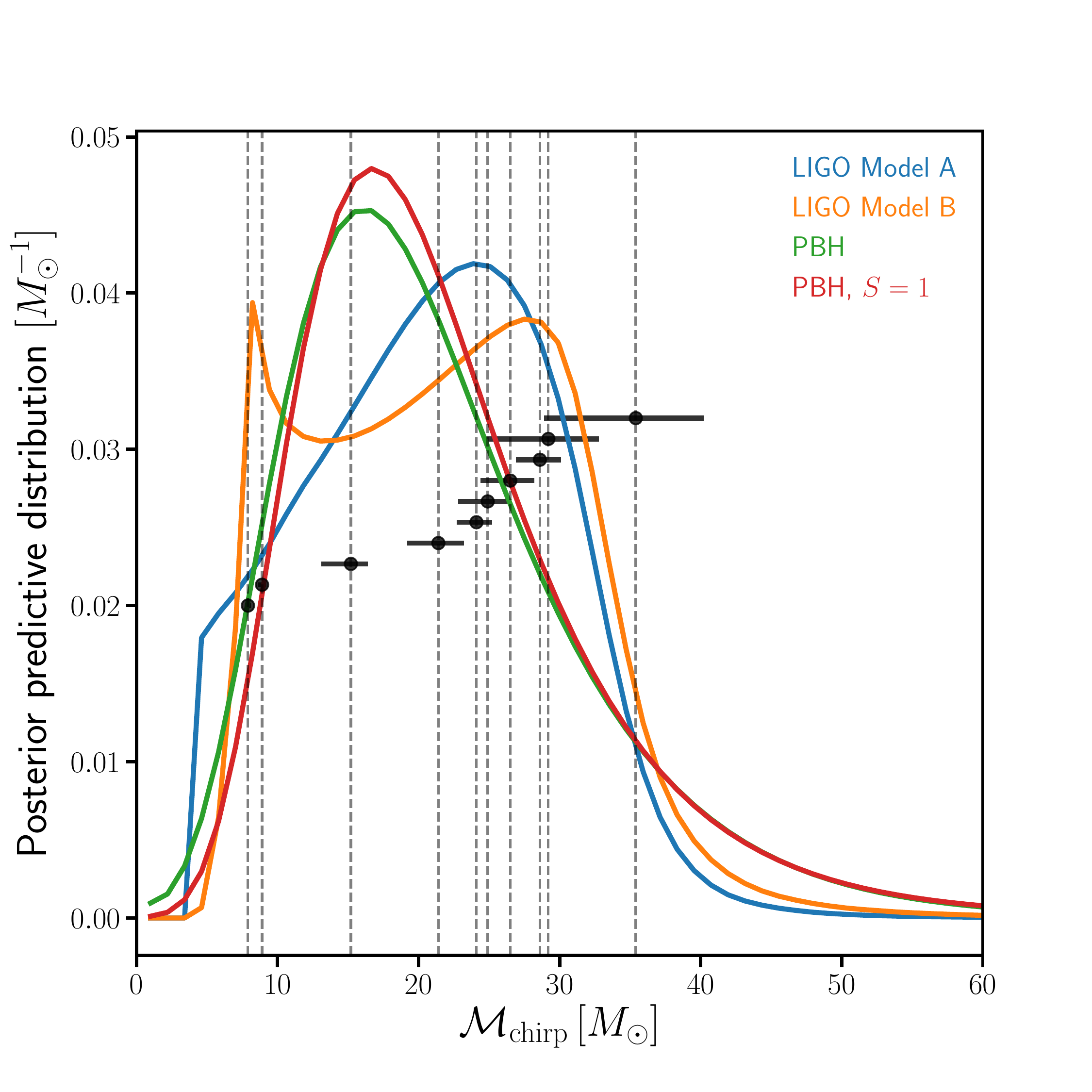}
\caption{Posterior predictive distribution (PPD) of the source-frame chirp mass, given the data, for LIGO Model A (blue), LIGO Model B (orange), the lognormal PBH model (green), and the lognormal PBH model with suppression factor set to unity (red). Note that there is a thin spike with width $\sim 10^{-3} \, M_\odot$ at $\mathcal{M}_{{\rm chirp}} \approx 4.35 \, M_\odot$ where the PPD diverges in the case of Model A, not visible on this plot due to the resolution. This spike corresponds to $m_1 \approx m_2 \approx m_{{\rm min}}$, and gives negligible contribution to the integrated PPD. We also show the median (black vertical lines and points) and 90\% credible intervals (black horizontal bars) of the source-frame chirp masses of the GWTC-1 BBH sources~\citep{2019PhRvX...9c1040A} (with an arbitrary vertical offset for visual clarity).}
\label{fig:PPD}
\end{figure}

In Figure~\ref{fig:PPD} we show the PPD on source-frame chirp mass\footnote{The detector-frame chirp mass PPD is very similar due to the low redshifts of the sources, but is significantly more computationally expensive to generate due to the suppression factor of the PBH model.} given by Equation~\eqref{eq:PPD_text}, along with the central values and 90\% confidence intervals for the sources in the GWTC-1 catalogue -- the best-fit likelihood for each model is approximately the PPD plotted in the figure convolved with each of the source posteriors and then evaluated at their central values. Ignoring the uncertainties in observed chirp mass, this simply amounts to recording the height of the PPD curves where they intersect each of the observed values. A model with a peak in its PPD located in the vicinity of a large number of measured chirp masses will have a higher likelihood than a model peaking away from where the observations are. This directly translates into a higher Bayesian evidence via the Laplace approximation Equation~\eqref{eq:Laplace}. Put even more coarsely, the likelihood ratio is effectively comparing the coherence of the empirical histogram of chirp masses with the predicted distribution for each model.

The key features of the curves in Figure~\ref{fig:PPD} are similar to those in the left panel of Figure~\ref{fig:dbetas} where we plotted the equivalent distribution for total mass. The suppression factor makes little difference to the PBH model, which demonstrates a lognormal shape in the chirp mass distribution inherited from the mass function $\psi(m)$. Model A and Model B both have sharp peaks around $m_1 \approx m_2 \approx m_{\mathrm{min}}$, with the peak in Model A narrower than the resolution of the plot, having width $\approx 10^{-3} \, M_\odot$. These peaks are due to a formal divergence in the merger rate caused by the $C(m_1)$ term in Equation~\eqref{eq:dRAB}, i.e.~the requirement that the marginal distribution in the heavier mass be a power law. The secondary peak around $\mathcal{M}_{{\rm chirp}} \approx 30 \, M_\odot$ is due to the detection probability. No such peak is seen in the PBH model due to its more extreme fall-off with increasing chirp mass.

The combined effect of the merger rate and the detection probability is that the PPD of Model B is able to peak sharply at the location of the two well-measured light binaries with $\mathcal{M}_{{\rm chirp}} \lesssim 10 \, M_\odot$, predicting fewer sources in the range $10$ -- $20 \, M_\odot$ in agreement with observations, before peaking again in the region $20$ -- $30 \, M_\odot$ just where the majority of the measurements are. Model A can also do this to a lesser extent, but is disfavoured compared with Model B because it gives less likelihood to the two light sources. This is simply a reflection of the fact that Model B has the freedom to fit the minimum component mass $m_{\mathrm{min}}$. Since there will always be a sharp peak in the chirp mass distribution at the minimum chirp mass, it is always advantageous for a model to place $m_{\mathrm{min}}$ as close to the actual minimum mass as possible. The penalty incurred from the Occam factor in this fine tuning process is substantially outweighed by the increase in likelihood. Model A in contrast has a fixed $m_{\mathrm{min}}$.

Turning now to the PBH models in Figure~\ref{fig:PPD}, it is clear that a lognormal distribution will struggle to fit the observed distribution of chirp masses compared with Models A and B. The mass function parameters $m_c$ and $\sigma$ are fit to ensure the lognormal peaks in the correct mass range and has a width encompassing the observed range of values, but the detailed shape is a poor fit to the data even with only 10 points. The LIGO Models are able to fit the key features of the empirical distribution, namely the high density of chirp masses in the $20$ -- $30 \, M_\odot$ region, the relative dearth in sources between $10$ -- $20 \, M_\odot$, and, in the case of Model B, the two light sources with $\mathcal{M}_{{\rm chirp}} \lesssim 10 \, M_\odot$.

The actual likelihoods at the observed data points are given approximately by the values of the curves where they intersect the vertical dashed lines in Figure~\ref{fig:PPD}. While the PBH models intersect the sources at $\sim 15 \, M_\odot$ and $\sim 35 \, M_\odot$ at higher values than both LIGO models, they both fail to capture the cluster of sources in the $20$ -- $30 \, M_\odot$ region. PBH models having a peak in this region are not as favoured as those having a peak around $\sim 15 \, M_\odot$, since they typically over-predict sources at heavier mass compared with lighter mass. The likelihood ratio (and hence the evidence ratio under the Laplace approximation) penalises the PBH models precisely for this reason. If new data populated Figure~\ref{fig:PPD} with many binaries having chirp masses greater than $35 \, M_\odot$, we would expect the lognormal PBH models to perform relatively better since they naturally predict a long positive tail in the distribution. The absence of sources above $35 \, M_\odot$, readily detectable in O1 and O2, penalises PBH models which predict they should be there.

\begin{figure}
\centering
\includegraphics[width=0.6\textwidth]{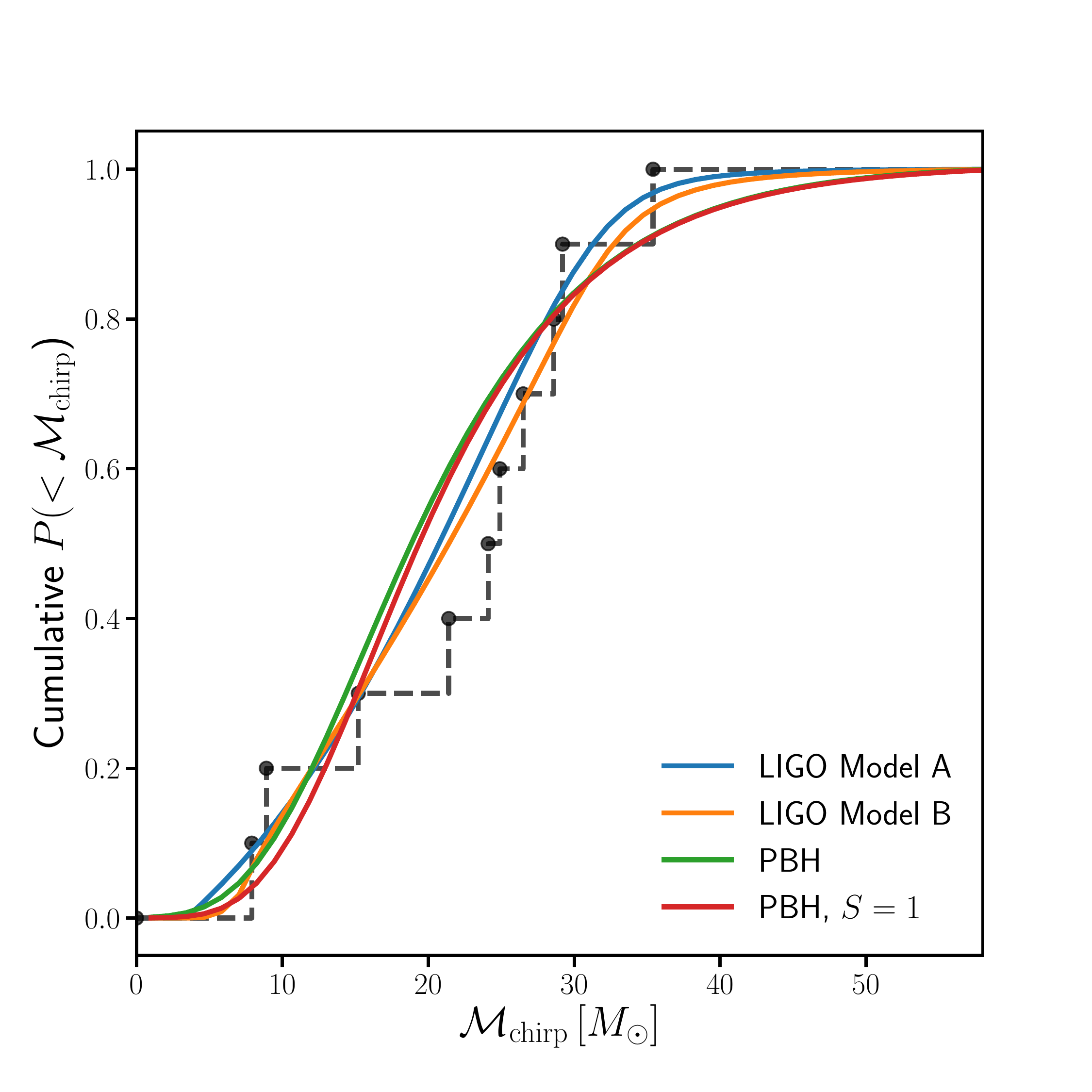}
\caption{Posterior cumulative distribution function (CDF) of source-frame chirp mass, given the data, along with the empirical CDF of the sources, for LIGO Model A (blue), LIGO Model B (orange), the lognormal PBH model (green) and the lognormal PBH model with suppression factor set to unity (red).}
\label{fig:CDF}
\end{figure}

An alternative way of looking at the differences between the PPD and the measured chirp mass is in the cumulative version of the PPD (posterior cumulative distribution function -- CDF) found by integrating from zero mass up to some specified value. The CDF for the LIGO models in terms of heavier mass was also studied in the recent Ref.~\citep{2020ApJ...891L..31F}. In Figure~\ref{fig:CDF} we show this quantity for the PBH models along with Models A and B.

The CDF is constrained to lie between zero and unity, and the model curves show significant overlap when plotted in this way. All models over-predict the number of mergers with chirp mass $\lesssim 20 \, M_\odot$, which can be seen as the relative dearth of sources in the $10$ -- $20 \, M_\odot$ region of Figure~\ref{fig:PPD}, with Model B clearly performing the best. Note that this behaviour is not evident in the CDF of the heavier mass plotted in Ref.~\citep{2020ApJ...891L..31F}. We have argued that the chirp mass is the more appropriate parameter to use since it is less correlated with other parameters, is the best constrained source parameter to leading order, and has a PPD most directly related to the likelihood and the evidence.

The CDF permits the use of a Kolmogorov-Smirnov (KS) test using the empirical CDF (shown as the grey lines and points in Figure~\ref{fig:CDF}). However, this test looks only at the maximum deviation of the predictions from the data (occurring around $20 \, M_\odot$), and the $p$-values from a one-sample KS test are all between $0.3$ and $0.8$, indicating that all models are acceptable fits to the data. This is not as powerful a test as the likelihood ratio or Bayesian evidence however, which uses the detailed shape of the chirp mass distribution to assess its ability to fit the data.

To summarise the results of this section, we have seen that the lognormal PBH models are significantly disfavoured compared with the empirical LIGO models, quantified by the Bayesian evidence ratio. We have shown that the evidence ratios between each model can be well approximated by the product of a likelihood ratio and an Occam factor. The Occam factor is sensitive to the prior volumes and the evidence ratio can be made to restore the prior ambivalence towards all models by broadening the priors on the LIGO models, but extreme values must be imposed to achieve this. The likelihood ratios are the dominant source of evidence against the PBH models, and we have shown how this can be reduced to the ability of models to predict the empirical distribution of chirp masses in the GWTC-1 catalogue. One of the main results of this work, Figure~\ref{fig:PPD}, demonstrates that the lognormal mass function struggles to match the detailed distribution of observed chirp mass, predicting positive skewness when the data appear to prefer negative skewness. In contrast Model A can predict negatively skewed chirp mass distributions and has consequently higher likelihood. Model B can additionally predict the two low mass events and the relative dearth of objects at intermediate masses, and is favoured over Model A. We caution that the LIGO models have some features which lack strong physical motivation and hence we do not advocate that the lognormal PBH model should be abandoned in favour of these models. We have instead shown why models in which every source is a PBH-PBH merger struggle in comparison, and identified the features of the data that need to be explained if the lognormal mass function is to become favourable. LIGO Model C is both more physical and a better fit to the data than Models A and B, and therefore by extension is significantly preferred (in terms of the Bayes factor) over the PBH models. It thus seems almost certain that successful PBH models will either need to abandon the prediction that every merger detected by LIGO and Virgo is a PBH merger or introduce a physical mechanism that significantly modifies the primordial lognormal mass function.

\section{Extensions to the lognormal PBH mass function}
\label{sec:extension}

We have seen that PBH models with a lognormal mass function do not provide as good a fit to the LIGO data as simple empirical models. Since this family of mass functions is highly constrained, having only two free parameters in addition to an overall normalisation, we now study simple extensions to $\psi(m)$ to investigate whether a better fit might be achieved with minimal modification.

We set the suppression factor equal to unity for all extended models considered in this section. The results of Section~\ref{sec:bayes} showed that the suppression factor has only a modest influence on the preferred models while greatly increasing the run-time of the likelihood calculation, so for simplicity we set $S=1$ henceforth.

\subsection{Lognormal with a high mass cut-off}
\label{subsec:mhicut}

We argued in Section~\ref{sec:bayes} that one of the reasons a lognormal struggles to fit the observations is its long positive tail to high chirp masses, not seen in the data. By comparison, both the LIGO models we consider have explicit cut-offs at high component masses. We therefore now consider a new mass function $\psi_{\mathrm{cut}}$, defined by
\begin{equation}
  \psi_{\mathrm{cut}}(m) \propto
  \begin{cases}
    \frac{1}{m\sqrt{2\pi \sigma^2}}\exp{\left[ -\frac{\ln^2(m/mc)}{2\sigma^2}\right]} & \quad m \leq m_{\mathrm{max}}\\
    0 & \quad m > m_{\mathrm{max}}
  \end{cases}
  \label{eq:psimhicut}
\end{equation}
where the normalisation is chosen such that $\psi_{\mathrm{cut}}$ integrates to unity. We note that such an upper mass cut-off is difficult to construct in PBH formation models.

For simplicity, and in order to approximately maximise the evidence for the PBH models without adding a penalty for adding a new free parameter, we fix $m_{\mathrm{max}} = 50 \, M_\odot$, such that the maximum source-frame chirp mass is $\mathcal{M}_{\mathrm{chirp}} \approx 44 \, M_\odot$. Figure~\ref{fig:PPD} shows that this lies just beyond the 90\% upper chirp mass of the heaviest source, which implies that this choice of cut-off is not too restrictive.

\begin{figure}
\centering
\includegraphics[width=0.7\textwidth]{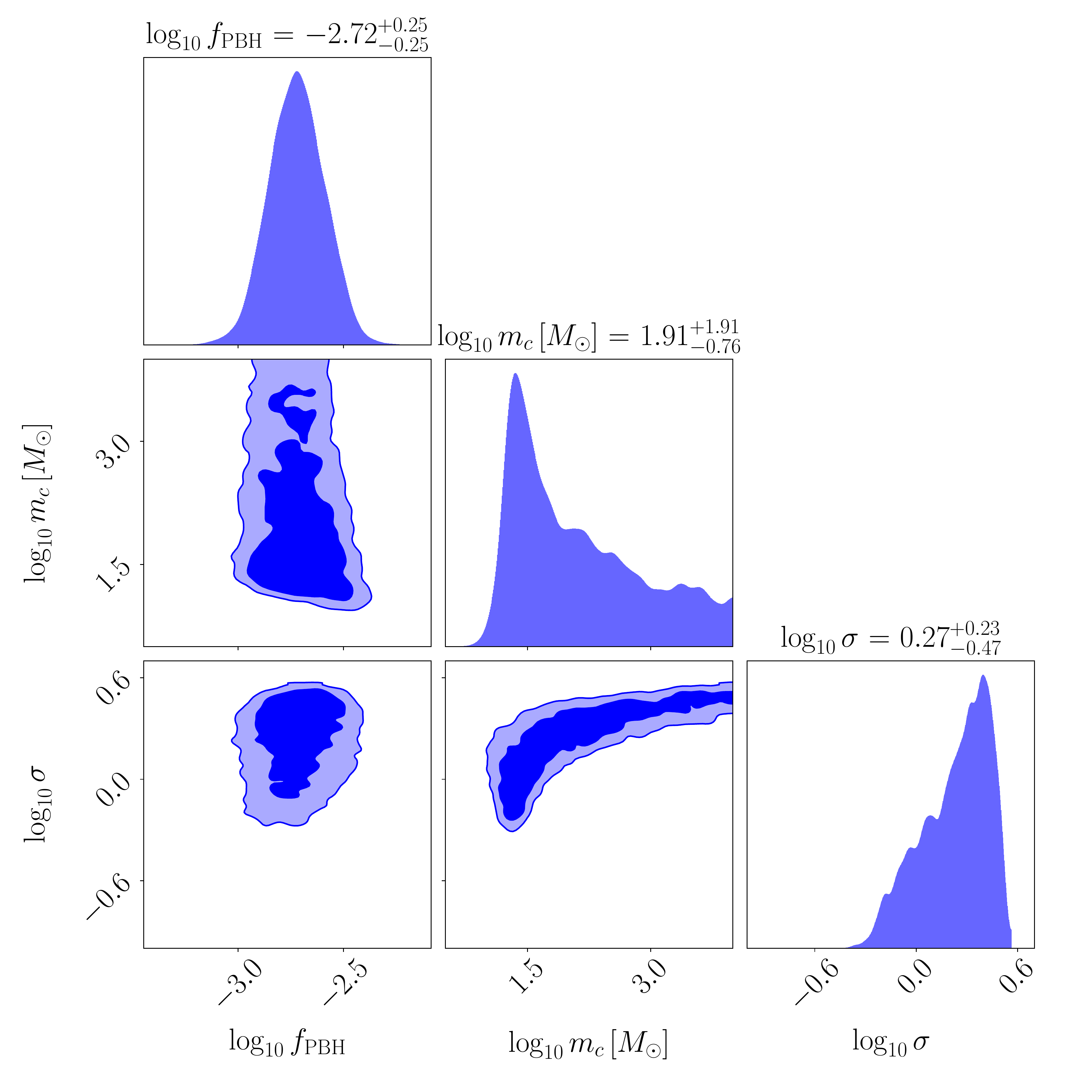}
\caption{Posteriors of the lognormal model without the suppression factor and imposing a maximum cut-off mass (source frame) of $m_{\mathrm{max}} = 50 \, M_\odot$. The meaning of the contours and quoted error significance are the same as in Figure~\ref{fig:PBH_posterior}.}
\label{fig:PBH_posterior_nosup_mhicut}
\end{figure}

We re-run the parameter inference and evidence calculation on the GWTC-1 catalogue data using this model, with the suppression factor set to unity and the same priors on $\fpbh$, $m_c$, and $\sigma$ as the lognormal model (i.e.~uniform in the log of each parameter). The resulting parameter posteriors are shown in Figure~\ref{fig:PBH_posterior_nosup_mhicut}.

The constraints on $\fpbh$ in this model are almost the same as the no cut-off case, with values preferred which give $\beta \approx 10$ events. The main difference comes in the preferred values of $m_c$ and $\sigma$, with Figure~\ref{fig:PBH_posterior_nosup_mhicut} revealing a long degeneracy tail stretching to high values of $m_c$ and $\sigma$. This is caused by the high-mass cut-off permitting values of $m_c \gg m_{\mathrm{max}}$ if $\sigma$ is sufficiently large that there is still a high likelihood of mergers happening in the observed mass range. When $\sigma$ is small enough, $m_c$ corresponds to this observed mass range and is constrained to similar values as in the absence of a cut-off.

The log-evidence of the cut-off model compared to Model B is given in Table~\ref{tab:baseline_results}, and is $-4.01 \pm 0.21$. This model is thus strongly preferred over the models without a cut-off. Model A and Model B are both still strongly preferred over this cut-off model. The Laplace-approximated evidence ratio differs from the nested sampling estimate by only $0.54 \pm 0.13$, but due to the highly non-Gaussian posterior we do not expect this to be accurate. Nevertheless, this approximation implies that the evidence ratio is dominated by the likelihood ratio against Model B, with the Occam factor now less penalising due to the larger posterior volume permitted by the data (compare Figure~\ref{fig:PBH_posterior_nosup_mhicut} with Figure~\ref{fig:PBH_posterior_nosup}). We note that had we allowed $m_{\mathrm{max}}$ to vary and be constrained by the data this conclusion might change, due to the large prior volume that could be assigned to $m_{\mathrm{max}}$. We note however that with fixed $m_{\mathrm{max}}$ the evidence ratios compared to the no-cut-off models are independent of the priors, which are the same amongst these PBH models.

\begin{figure}
\centering
\includegraphics[width=0.7\textwidth]{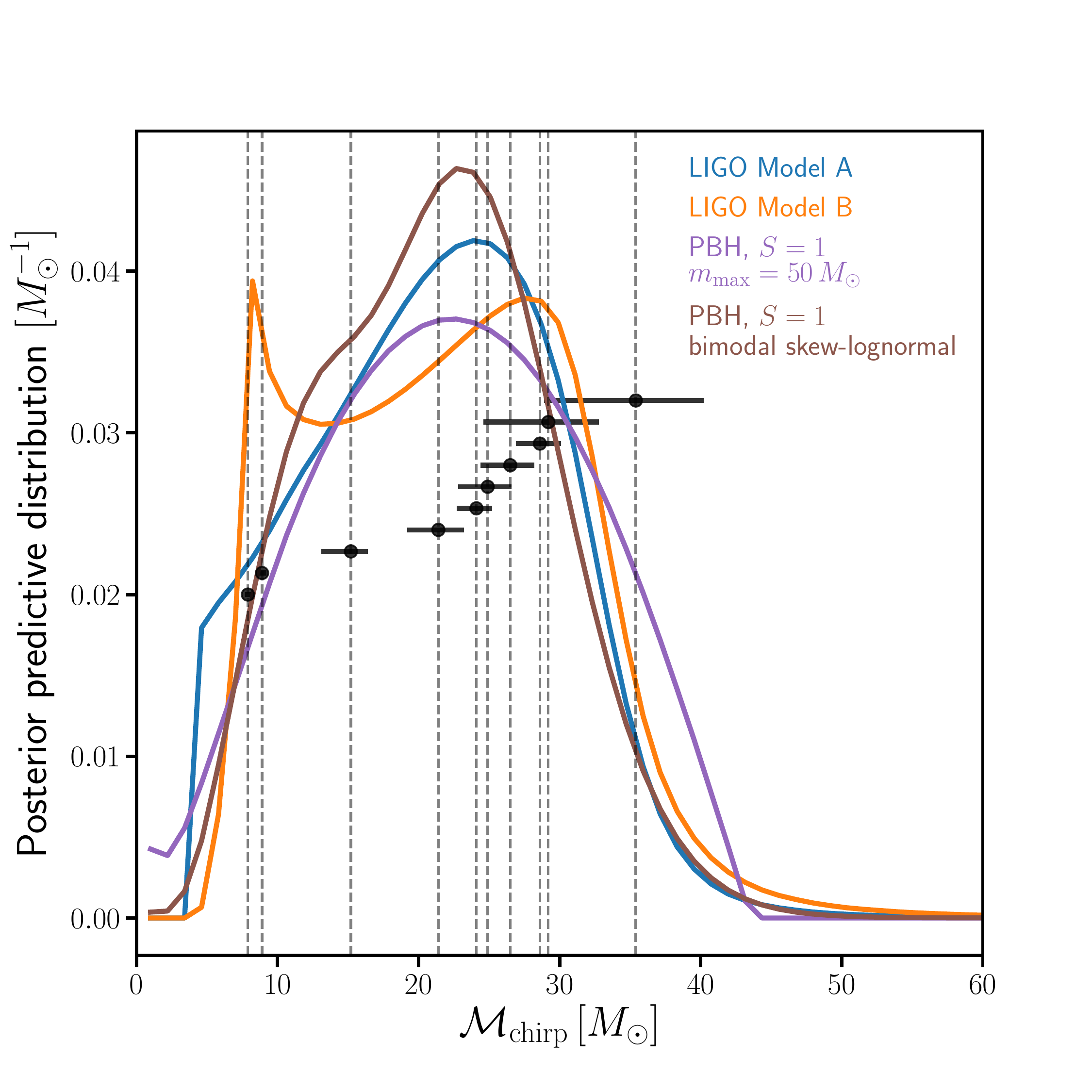}
\caption{PPD for two extensions to the lognormal PBH mass function: the $S=1$ lognormal model with a cut-off $m_{{\rm max}} = 50 \, M_\odot$ (purple), and the bimodal skew-lognormal model with $S=1$ (brown). We also show LIGO Model A (blue) and Model B (orange), which are the same as in Figure~\ref{fig:PPD}. }
\label{fig:PPD_ext}
\end{figure}

In Figure~\ref{fig:PPD_ext} we show the PPD of the source-frame chirp mass for this model (purple curve) along with that of LIGO models A and B. With the preferred values of $m_c$ skewed to values $\gg m_{\mathrm{max}}$, the shape of the posterior-averaged likelihood now looks quite different to the lognormal case. In particular, the large positive skewness has been suppressed by the cut-off, and the distribution is approximately symmetric about its peak at roughly $23 \, M_\odot$. As in the lognormal case, the preferred values of the parameters $m_c$ and $\sigma$ are such that the overall distribution has roughly the correct absolute mass scale and a width incorporating the observed chirp masses. It is clear from Figure~\ref{fig:PPD_ext} that the broader distribution allowed by the cut-off mass function is a better fit to the data, and accounts for the increased evidence for this model. Despite this, there is thus no additional freedom to fit the detailed distribution of the data beyond $m_c$ and $\sigma$, and Model B still provides a better fit. Model A is also preferred to the cut-off lognormal model, due in part to its low-mass cut-off which allows the likelihood of the two well-measured low-mass sources to be higher.

\subsection{Skew-Lognormal}
\label{subsec:skewln}

While a lognormal mass function for PBHs may be shown to be an excellent approximation to a wide range of peak-like features in the primordial power spectrum, for very narrow peaks a negatively-skewed lognormal is a better approximation~\citep{2020arXiv200903204G}. We thus consider a skewed lognormal mass function given by
\begin{equation}
  \psi_{\mathrm{skew}}(m) = [1 + \erf(\alpha \Delta)]\psi(m),
\end{equation}
where $\psi(m)$ is a lognormal mass function and $\Delta \equiv \ln(m/m_c)/(\sqrt{2} \sigma)$ is the logarithmic mass deviation. The skewness is parameterised by a parameter $\alpha$ controlling the argument to the error function $\erf(x)$. As shown in Ref.~\citep{2020arXiv200903204G}, a delta function in the power spectrum corresponds to $\alpha \approx -2.6$ and $\sigma \approx 0.56$, which we impose here. This leaves $\fpbh$ and $m_c$ as the only free parameters of this model.

Running the nested sampling inference with the skew-lognormal model, we find the best-fitting values of the parameters are $\fpbh = 1.4 \times 10^{-3}$ and $m_c = 34.0 \, M_\odot$, comparable with the results found for the lognormal model. The (log) likelihood ratio at the best-fit compared to Model B is $-8.3$, i.e.~very similar to the full lognormal ($S=1$) case. This is due to the relatively weak skewness of the model and the fact that $\sigma = 0.56$ is actually quite close to the value preferred by the data.

The evidence for the model relative to Model B is $-7.80 \pm 0.23$, i.e.~slightly preferred compared with the non-skewed PBH model with $S=1$ but still strongly disfavoured compared to both Model A and Model B. The model provides a fit to the data comparable with the unskewed lognormal mass function, i.e.~not competitive with the LIGO empirical models. A Laplace approximation to the evidence is very accurate, and shows that the increased evidence for the model results from a less penalising Occam factor due to its reduced dimensionality compared with the non-skewed model. However, this change is not enough to overcome the big difference in likelihood ratio which gives rise to strong evidence against the model compared with the LIGO models.

%
%
%
%
\subsection{Bimodal Skew-Lognormal}
\label{subsec:bisk}

The distribution of measured chirp masses in the GWTC-1 catalogue has a cluster of sources with $\mathcal{M}_{\mathrm{chirp}} \approx 30 \, M_\odot$, a relative dearth of sources between $10 \, M_\odot$ and $20 \, M_\odot$, and two well-measured light sources with $\mathcal{M}_{\mathrm{chirp}} < 10 \, M_\odot$. Motivated by this, we consider a mixture of two skew-lognormal mass functions for PBHs given by
\begin{equation}
  \psi_{\mathrm{skew,bi}}(m; m_{c,1}, m_{c,2}) = \lambda \psi_{\mathrm{skew}}(m; m_{c,1}) + (1 - \lambda) \psi_{\mathrm{skew}}(m; m_{c,2}),
  \label{eq:skewbi}
\end{equation}
where $\psi_{\mathrm{skew}}$ is the skew-lognormal distribution introduced in Section~\ref{subsec:skewln}, i.e.~each component has fixed skewness parameter $\alpha = -2.6$ and scale parameter $\sigma=0.56$. Such a mass function could arise from two distinct narrow peaks in the primordial power spectrum whose amplitudes must be tuned if a comparable number of PBHs are to be generated by each peak~\citep{Cai:2018tuh,Carr:2018poi}. However, we note that very close peaks will not produce the distribution Equation~\eqref{eq:skewbi} in detail due to the two peaks `smearing' into each other.

We choose log-uniform priors in location parameters $m_{c,1}$ and $m_{c,2}$ with limits given in Table~\ref{tab:priors}. To avoid redundant likelihood calculations implied by the symmetry of Equation~\eqref{eq:skewbi} we impose a uniform prior on $\lambda$ in the range $[0, 0.5]$, such that $m_{c,2}$ is defined as the location parameter of the dominant component in the mixture.

\begin{figure}
\centering
\includegraphics[width=0.7\textwidth]{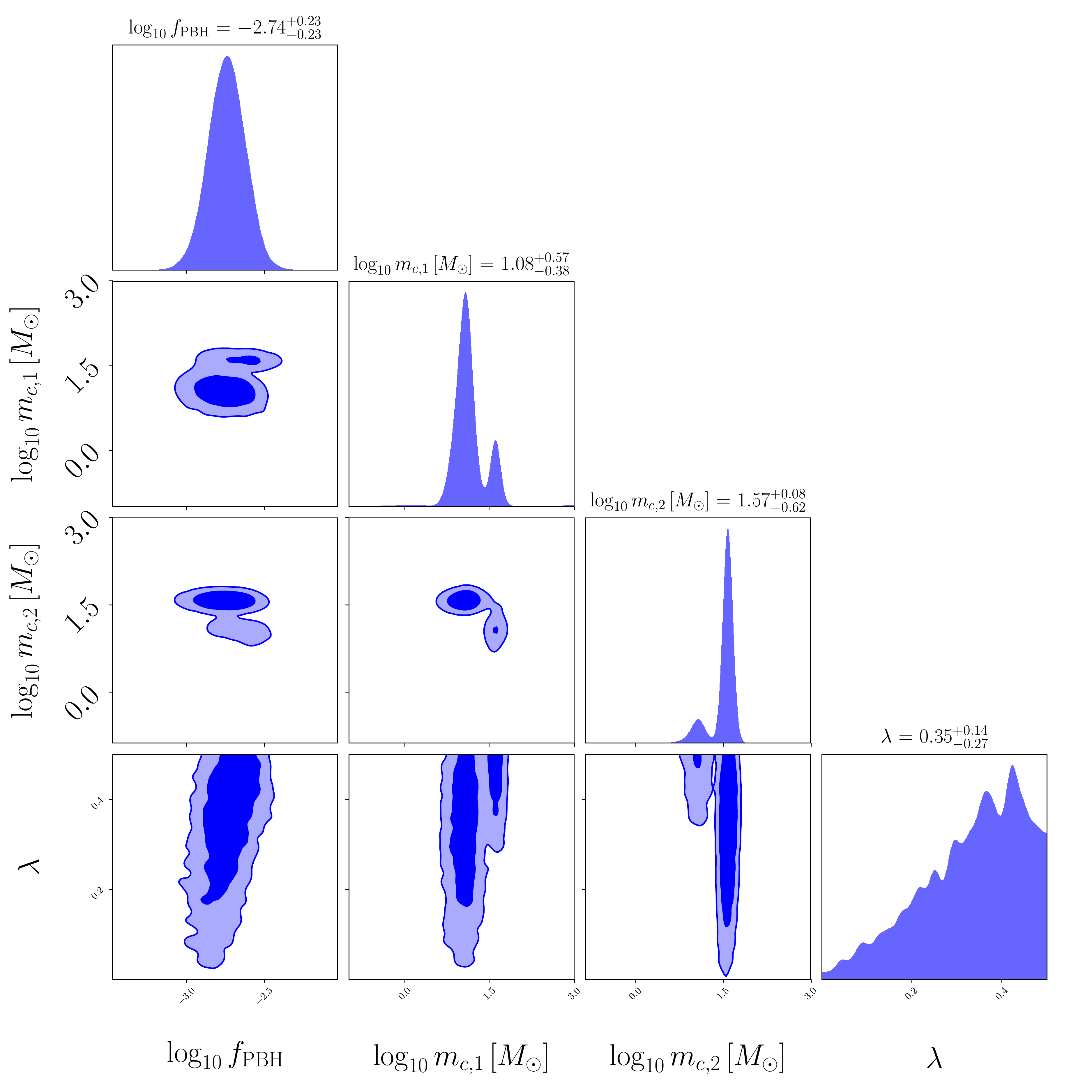}
\caption{Posteriors of the bimodal skew-lognormal model without the suppression factor, fixing the shape parameter of each component to $\sigma=0.56$ and the skewness parameter to $\alpha=-2.6$, roughly corresponding to delta functions in the power spectrum. The meaning of the contours and quoted error significance are the same as in Figure~\ref{fig:PBH_posterior}.}
\label{fig:PBH_posterior_nosup_skewbi}
\end{figure}

In Figure~\ref{fig:PBH_posterior_nosup_skewbi} we show the posterior constraints on the parameters of the skew-bimodal model. As for the other model extensions studied in this section, the preferred values of $\fpbh$ are such that the total number of events is roughly $10$, the median value being $\fpbh=0.002$ in this model. The distribution in the $[m_{c,1}, m_{c,2}]$ plane is bimodal, with a dominant peak at $m_{c,1} \approx 10 \, M_\odot$ and $m_{c,2} \approx 35 \, M_\odot$, which matches our expectation given the observed chirp masses; for $q=1$ this corresponds to a dominant component in the mass function at $\mathcal{M}_{\mathrm{chirp}} \approx 30 \, M_\odot$ and a sub-dominant component at $\mathcal{M}_{\mathrm{chirp}} \approx 9 \, M_\odot$. The sub-dominant peak in the mass posterior corresponds approximately to swapping which of these mass function peaks is dominant, preserving their location.

The mixture parameter $\lambda$ is poorly constrained, with values around $\lambda \approx 0.4$ typically preferred. Bimodality in the posterior appears when $\lambda \gtrsim 0.35$, which is to be expected since the data is not constraining enough to distinguish which mass function peak is dominant when the difference is sufficiently small, i.e.~when $\lambda$ is sufficiently close to the point of symmetry at $\lambda = 0.5$.

The evidence for this model is reported in Table~\ref{tab:baseline_results}, and is $-5.79 \pm 0.24$ compare to Model B. There is thus `substantial' evidence (on the Jeffreys' scale) for this model compared with both the PBH lognormal models considered previously, although the Bayesian evidence in favour of Models A and B is still comparatively strong. The likelihood ratio to Model B is much more favourable for the skew-bimodal model compared with the other PBH models, and is only marginally smaller than Model A. The Occam factor is comparatively more penalising than both the LIGO models, although the highly non-Gaussian parameter posterior makes the Laplace approximation a poor estimate of the true evidence.

In Figure~\ref{fig:PPD_ext} we show the PPD for the source-frame chirp mass in this model (brown curve). The peak at small masses has been skewed to heavier masses by the detection probability, but the two components are still clearly distinguishable in this plot. The comparable likelihood ratio between the skew-bimodal mass function and Model A (blue curve) is clear from Figure~\ref{fig:PPD_ext}, with the additional low-mass component matching Model A's ability to assign high likelihood to the two well-measured light BBHs. The PPD suggests that the skew-bimodal model performs less well compared with Model B due to its inability to predict a sharp peak at low chirp masses. This could potentially be remedied by allowing one or both of the variance parameters in the skew-lognormal components to vary, although this would come at the price of a more penalising Occam factor (i.e.~overfitting the data).

The skew-bimodal model is thus successful at matching the fit of the observed distribution of chirp masses provided by the LIGO models, and is the most successful of the PBH models we consider. We note that the two preferred central values of the components are reasonably close, such that this model might not be expected to be an accurate approximation to two delta-function peaks in the primordial power spectrum. We also note that constructing a physical mechanism that could produce two such peaks of comparable amplitude is not straightforward and the a priori motivation for this is weak.

It is perfectly possible of course that the apparent peak at low chirp masses is not a `real' feature of the population but a consequence of the low-number statistics. The Bayesian evidence accounts for this, but the likelihood ratio does not, so we caution against attempting to construct models to fit the detailed empirical distribution of chirp masses in general. Nonetheless, we have seen that Bayesian evidence favours the skew-bimodal model over a single lognormal component.

\subsection{Late-time PBH capture model}
\label{subsec:late}

The final merger rate model we consider is a `late-time capture' model in which PBH binaries form in the late Universe via two-body encounters. We adopt the model of Ref.~\citep{2017PDU....15..142C, 2020arXiv200706481C} in which the differential merger rate is given by
\begin{equation}
  \frac{\ud R}{\ud m_1 \ud m_2} \propto \frac{(m_1 + m_2)^{10/7}}{(m_1 m_2)^{5/7}} \psi(m_1)\psi(m_2).
  \label{eq:latetime}
\end{equation}
This model follows from the analytical calculations of Refs.~\citep{1989ApJ...343..725Q, 2002ApJ...566L..17M} which models two compact objects on an initially parabolic or hyperbolic trajectory which become bound due to the radiation of gravitational waves, using an accurate quasi-Newtonian approximation. Ref.~\citep{2017PDU....15..142C} additionally assumes that the relative velocity of the two objects is independent of their masses, an assumption which may break down in detail due to mass segregation in halos. The pre-factor of Equation~\eqref{eq:latetime} is a free parameter in this model which can be high enough to give the observed merger rate even for low $\fpbh$, due to enhancements in the merger rate within dense halos. We assume that the mass functions in Equation~\eqref{eq:latetime} are lognormals with parameters $m_c$ and $\sigma$, having the same priors as in our baseline PBH model.

The best-fit values for $m_c$ and $\sigma$ were obtained with a numerical optimization routine and are $16.5 \, M_\odot$ and $0.56$ respectively, i.e.~almost identical to the baseline $S=1$ PBH model. The best-fit mass function has a shape in the $(m_1, m_2)$ plane almost identical to that of the baseline early-time formation model, differing only in the tails. This is due to the low value of $\sigma$ preferred by the data, which keeps the mass function compact and suppresses the influences of mass-dependent pre-factors multiplying the lognormal $\psi(m_1)\psi(m_2)$ term in the merger rate. Given the similarity with the early-time model we chose not to run the full nested sampling inference on the late-time capture model.

This model thus provides a very similar fit to the early-time model, with the quality of the fit dominated by the lognormal shape of the mass function. This may be compared to the result that the suppression factor makes little difference to the best-fitting PBH model and its maximum likelihood, and suggests that radical mass-dependent evolution of the initial PBH mass function is required to give a good fit to the data when the mass function is lognormal.

\section{Conclusions}
\label{sec:conclusions}

In this work we have confronted the latest PBH binary merger models with the catalogue of gravitational wave merger events from the first two observing runs of LIGO-Virgo. We have adopted a Bayesian formalism throughout, which has allowed us to place posterior probabilities on the parameters of the PBH model given the data, accounting for the source parameter correlations and the interferometer selection function. Assuming all the observed black hole mergers are primordial and marginalising over the mass function parameters we find $\fpbh = (5.0^{+67.4}_{-2.8}) \times 10^{-3}$ (median and 95\% confidence), with a long positive tail allowed by the data due to a suppression in the merger rate from demanding the binary is not disrupted by other PBHs. Relaxing this requirement gives the smaller value $(1.7^{+1.4}_{-0.7}) \times 10^{-3}$. The preferred mass function parameters are such that the observed black hole mass scale and spread in masses is correctly predicted. A lognormal fit gives a central value of $\approx 20 \, M_\odot$ and a logarithmic width of order unity, consistent with previous results~\citep{2019JCAP...02..018R, 2020JCAP...06..044D, 2020arXiv200500892D}.

Going beyond parameter constraints, we have studied the quality of the model fits using several Bayesian tests. We computed the Bayesian evidence for the PBH models and popular astrophysically motivated models, finding in all cases that the astrophysical models were favoured decisively. By making a Laplace approximation we decomposed the evidence ratio into a likelihood ratio (well known from classical frequentist statistics) and an Occam factor quantifying the sensitivity to the parameter priors. This exercise showed that the evidence ratio was dominated by the likelihood ratio, i.e.~by the relative goodness of fit of the best-fitting models in the PBH and astrophysical scenarios. We were able to show that this may be understood by comparing the predicted distributions of the chirp mass with the observed values, and identified the posterior predictive distribution as a crucial descriptor of the relative quality of the model fits.

Using the posterior predictive distribution we were able to show that PBH models struggle because they predict a chirp mass distribution with a close-to-lognormal shape, in marked contrast to the observations. The LIGO interferometers were sensitive enough in their first two observing runs to detect black holes with chirp masses well beyond the $40 \, M_\odot$ upper limit of the sources which were actually detected. A lognormal distribution has a long positive tail to high masses, over-predicting the abundance of high-mass binaries. Likewise, the detailed distribution of the observed sources is not well predicted compared with the empirical astrophysical models. These empirical models are parameterised in terms of the heavier mass and the mass ratio, which gives rise to a detectable chirp mass distribution much preferred by the data over a lognormal PBH model. An explicit lower and upper mass cut-off in these models also boosts their evidences significantly over the PBH scenario. While such high- and low-mass cut-offs might be expected in stellar-origin black holes they are generally not expected in PBH models, which are consequently disfavoured.

These statements hold true for almost any choices of the black hole mass function parameters and $\fpbh$, although our modelling is expected to be inaccurate for high $\fpbh$ and for very broad mass functions. We also studied simple extensions to the lognormal model in an attempt to better fit the data, finding that a bimodal mass function with small negative skewness (as expected from narrow peaks in the primordial power spectrum) provides a marginally improved fit, although at the cost of a posteriori reasoning and little physical motivation. At face value our results strongly disfavour the possibility that all the sources seen in the first two LIGO-Virgo observing runs are merging PBH binaries forming from a smooth, symmetric peak in the primordial power spectrum. It is therefore worth discussing how these conclusions might be relaxed or challenged.

Firstly, we caution that the empirical astrophysical models we have considered are really parameterisations and that we limit our analysis to the case that all the black holes are either astrophysical or primordial. A more sophisticated analysis would consider a mixed model with both forms of black holes. Our results are therefore not evidence against the possibility that LIGO-Virgo have detected PBHs, but evidence against all of the detected black holes being PBHs. 

To test a concrete example of a mixed stellar-PBH population of binary mergers, we computed the likelihoods of the PBH ($S=1$) model and the LIGO models excluding the two sources possessing significant non-zero spin, GW151226 and GW170729. Since PBHs formed during radiation domination are expected to have zero spin at formation, these two sources are the most well motivated for exclusion from the analysis. Restricting to the reduced catalogue of 8 binary mergers, we find a slightly reduced best-fit $\fpbh \approx 1.4 \times 10^{-3}$ and a smaller $\sigma \approx 0.48$, reflecting the fact that the two spinning sources lie near the extremes of the chirp mass distribution (GW151226 is the second-lightest system, GW170729 is the heaviest). The maximum likelihood value of the PBH model compared to Model B is slightly increased (-6.7, up from -7.1) with that of Model A approximately unchanged. It thus appears that even without the two spinning sources we can expect the LIGO models to have significantly higher evidence compared with the PBH models, due to the PBH model struggling to fit the truncated and negatively skewed distribution of chirp masses.

Secondly, we have assumed that PBH binaries evolve from formation through to merger without modification to their dynamics from external astrophysical processes. In particular we have neglected the possibility that matter accretes onto the binary, a potentially important effect influencing its angular momentum and mass~\citep{2007ApJ...662...53R, 2020JCAP...04..052D}. Naively one would expect accretion of material onto the binary to effectively skew the mass function towards heavier masses. As we have seen that the lognormal mass function already has too much positive skewness, this is likely to make the fit to the LIGO-Virgo data worse. It thus appears that the possibility that all sources are PBH binaries is even less likely with the inclusion of accretion.

Thirdly, our results have sensitivity to the priors placed upon the model parameters. We have argued that there is little strong motivation for tightening the priors on the PBH model, but alternative choices could broaden them significantly. The abundance parameter $\fpbh$ is exponentially sensitive to the amplitude of the peak in the primordial power spectrum which produced the PBHs. If we chose to impose a uniform prior on the order of magnitude of the peak amplitude this would translate to a prior uniform in $\log \log \fpbh$, which would potentially increase the prior volume of the PBH model by many orders of magnitude. However, this would only increase the degree to which the Bayes factor disfavours the PBH model.

Alternatively, we could boost the Bayes factor in favour of the PBH models by making alternative choices for the priors of the astrophysical models. We have argued that, if the functional form of this prior is left unmodified, this cannot result in the PBH model being favoured over the astrophysical models without unphysical choices for the prior range. Alternatively one could make alternative choices for the functional form of the astrophysical parameter priors. One can make the Bayes factor arbitrarily favourable towards the PBH model this way. It is possible that future models will find relations between the model parameters and more fundamental quantities related to the physics of stellar black hole binaries, in which case well-motivated choices for the priors of these fundamental parameters could result in a significantly broadened prior volume for the empirical parameters. It will therefore be necessary to re-run our analysis if such model refinements become available.

Is there a way to test the quality of the PBH model fit without reference to an empirical astrophysical model? Ref.~\citep{2020JCAP...01..031G} found, using frequentist measures such as the $\chi^2$ and KS test, that the LIGO-Virgo data are not an unlikely realisation for a reasonable range of PBH model parameters. Our results are not in conflict with this conclusion, since we have focused largely on the \emph{relative} quality of the model fits compared to astrophysically motivated models. We found that the detailed distribution of the chirp mass was the key discriminator in our tests. Since the $\chi^2$ and KS tests do not make full use of this distribution but instead compress it down to test statistics we do not expect these to be particularly powerful in quantifying the quality of the PBH model fit. The Bayesian methodology has the advantage of using all the information available, which is one reason why we have focused on Bayesian model evidence ratios rather than frequentist statistical tests.

We have also neglected information coming from the spins of the merging black holes in our model comparisons. In reality we expect the spin distributions of PBH and astrophysical mergers to be different, and including spin could impact our conclusions. A typical PBH spin distribution would have more weight in non-spinning objects, due to the negligible spin that PBHs are expected to have at formation. Given that all but two of the LIGO-Virgo sources we consider are consistent with zero spin we might expect that including spin information boosts the relative probability of PBH models. Ref.~\citep{2019JCAP...08..022F} performed a Bayesian comparison of PBH and astrophysical models using their differing predictions for black hole spin, finding that the data are not currently constraining enough to discriminate between the models. There is also theoretical uncertainty in how the initial spin distribution of PBHs evolves in the presence of accretion~\citep{2020JCAP...04..052D}. Furthermore, we have seen that excluding the two objects with non-zero spin has little effect on our conclusions. These considerations suggest that including spin in our analysis would not change our results significantly; the chirp mass, being a well-measured parameter for each system whose distribution is sensitive to model parameters, will likely remain the discriminating observable. A future extension of this work will be to include spin, with realistic astrophysical and PBH population distributions.

In this work we have tested a specific model for the formation and subsequent evolution of a PBH binary. There exist models with dramatically different behaviour allowing much larger $\fpbh$, such as those of Ref.~\citep{Jedamzik:2020ypm, Jedamzik:2020omx}, which could yield quite different Bayesian evidences. However, there is no reason to expect that such models will provide good fits to the full set of GW events evidenced by the fact that the late time capture model we studied does not provide a significantly different fit to the data. The subject of PBH binary evolution is an active and rapidly evolving field of study and testing a broader range of merger models is a valuable extension of our formalism, which we defer to a future work.

It therefore appears that our conclusions are robust to including these added complications; both the full LIGO-Virgo sample of merging black holes and the subset consistent with zero spin are not well fit by PBH-PBH binaries compared with simple astrophysically motivated models.

Finally, it is interesting to consider how our results might change with the inclusion of recent new detections in the third observing run of LIGO-Virgo (O3). This has so far yielded a system, GW190814, with low mass ratio $q \approx 0.1$ having one component in the lower mass gap with $m \approx 2.6 \, M_\odot$~\citep{Abbott:2020khf}, the black hole binary merger GW190412 with low reported mass ratio $q \approx 0.3$~\citep{2020arXiv200408342T}, the black hole merger GW190521 with total mass $150 \, M_{\odot}$ and upper-mass gap component black holes~\citep{Abbott:2020tfl}, as well as roughly 50 new binary black hole detections. We caution that these three named sources have been singled out for publication by virtue of being `unusual', and hence the conclusions we can draw about population models are limited without including the full unbiased sample. A repeat of our analysis on the full sample of O3 events is forthcoming so for now we focus on these three unusual systems. Firstly we note that two low mass-ratio systems are more probable in PBH models than astrophysical models due to the extended mass function and lack of any mass correlation which might arise from mass transfer. As pointed out in Ref.~\citep{2020arXiv200812320B}, the constraint on the mass ratio of GW190412 is strongly dependent on the priors assumed for the source parameters, with $q \approx 1$ an equally good fit to the data when a low-spin prior (as might be expected for a PBH binary) is imposed. Focusing on chirp mass, which we have argued is where most of the constraining power comes from, GW190814 has $\mathcal{M}_{\mathrm{chirp}} = 6.09 \pm 0.06 \, M_\odot$ and GW190412 has $\mathcal{M}_{\mathrm{chirp}} = 13.3 \pm 0.4 \, M_\odot$, both in the source frame. Comparing the posterior predictive distribution given the O1 and O2 samples shows that both sources lie at the lighter end of the distribution, with GW190412 lying close to the peak of the PBH distribution. GW190814 is in the light tail of all the distributions we considered, although the constraining power of this object comes more from the implications of its low-mass component in the context of stellar black hole formation models. We note that GW190412 has a non-negligible spin parameter $\chi_{\mathrm{eff}} = 0.25^{+0.08}_{-0.11}$, while GW190814 is consistent with zero spin, raising the possibility that GW190412 is problematic for both PBH-PBH and stellar-stellar merger channel (although see the point above and Ref.~\citep{2020arXiv200812320B}). GW190814 on the other hand appears consistent with both, with high probability in the PBH model due to its low mass ratio. The source GW190521 has a chirp mass of roughly $65 \, M_\odot$, with both components having non-negligible spin. Comparison with Figure~\ref{fig:PPD} shows that this source lies far in the high-mass tail of the distribution implied by the O1O2 sample. It is possible that the inclusion of this source in the sample could reduce some of the negative skewness in the chirp mass distribution and improve the PBH fit, although its chirp mass is so large that a worsened fit is also a possibility. Ref.~\citep{2020arXiv200901728D} discusses the possibility that GW190521 is a PBH binary, concluding that accretion is necessary to reconcile the implied merger rate with existing bounds on $\fpbh$. A full analysis including all sources will shed more light on these intriguing issues.

With the sample size of black hole merger events expected to grow significantly with the conclusion of the third observing run of LIGO-Virgo and the newly online KAGRA facility~\citep{2020arXiv200802921K}, a principled statistical framework for analysing the PBH merger scenario will prove increasingly valuable. In this work we have demonstrated the kind of analysis techniques that will be necessary to constrain the physics of primordial black holes in the coming era of gravitational wave astronomy.

\section*{Acknowledgements}
The authors thanks Nicola Bellomo, Christopher Berry, Gabriele Franciolini, Davide Gerosa, Karsten Jedamzik, Valerio de Luca, Chris Messenger, Paolo Pani, Antonio Riotto and Ville Vaskonen for helpful correspondence. We thank the anonymous referees for helpful suggestions which improved the paper. AH acknowledges support from a Science and Technology Facilities Council Consolidated Grant. AG is funded by a Royal Society Studentship by means of a Royal Society Enhancement Award. CB acknowledges support from the Science and Technology Facilities Council [grant number ST/T000473/1]. We acknowledge use of the software packages \textsc{gwdet}~\citep{2017zndo....889966G}, \textsc{dynesty}~\citep{2020MNRAS.493.3132S}, \textsc{PyCBC}~\citep{2016CQGra..33u5004U} and \textsc{LALSuite}~\citep{lalsuite}.

\bibliography{references}

\begin{thebibliography}{93}
\expandafter\ifx\csname natexlab\endcsname\relax\def\natexlab#1{#1}\fi
\expandafter\ifx\csname bibnamefont\endcsname\relax
  \def\bibnamefont#1{#1}\fi
\expandafter\ifx\csname bibfnamefont\endcsname\relax
  \def\bibfnamefont#1{#1}\fi
\expandafter\ifx\csname citenamefont\endcsname\relax
  \def\citenamefont#1{#1}\fi
\expandafter\ifx\csname url\endcsname\relax
  \def\url#1{\texttt{#1}}\fi
\expandafter\ifx\csname urlprefix\endcsname\relax\def\urlprefix{URL }\fi
\providecommand{\bibinfo}[2]{#2}
\providecommand{\eprint}[2][]{\url{#2}}

\bibitem[{\citenamefont{{Zel'dovich} and
  {Novikov}}(1967)}]{1967SvA....10..602Z}
\bibinfo{author}{\bibfnamefont{Y.~B.} \bibnamefont{{Zel'dovich}}}
  \bibnamefont{and} \bibinfo{author}{\bibfnamefont{I.~D.}
  \bibnamefont{{Novikov}}}, \bibinfo{journal}{Soviet Astronomy}
  \textbf{\bibinfo{volume}{10}}, \bibinfo{pages}{602} (\bibinfo{year}{1967}).

\bibitem[{\citenamefont{{Hawking}}(1971)}]{1971MNRAS.152...75H}
\bibinfo{author}{\bibfnamefont{S.}~\bibnamefont{{Hawking}}},
  \bibinfo{journal}{\mnras} \textbf{\bibinfo{volume}{152}}, \bibinfo{pages}{75}
  (\bibinfo{year}{1971}).

\bibitem[{\citenamefont{{Carr} and {Hawking}}(1974)}]{1974MNRAS.168..399C}
\bibinfo{author}{\bibfnamefont{B.~J.} \bibnamefont{{Carr}}} \bibnamefont{and}
  \bibinfo{author}{\bibfnamefont{S.~W.} \bibnamefont{{Hawking}}},
  \bibinfo{journal}{\mnras} \textbf{\bibinfo{volume}{168}},
  \bibinfo{pages}{399} (\bibinfo{year}{1974}).

\bibitem[{\citenamefont{{Carr}}(1975)}]{1975ApJ...201....1C}
\bibinfo{author}{\bibfnamefont{B.~J.} \bibnamefont{{Carr}}},
  \bibinfo{journal}{\apj} \textbf{\bibinfo{volume}{201}}, \bibinfo{pages}{1}
  (\bibinfo{year}{1975}).

\bibitem[{\citenamefont{{Carr} et~al.}(2016)\citenamefont{{Carr}, {K{\"u}hnel},
  and {Sandstad}}}]{2016PhRvD..94h3504C}
\bibinfo{author}{\bibfnamefont{B.}~\bibnamefont{{Carr}}},
  \bibinfo{author}{\bibfnamefont{F.}~\bibnamefont{{K{\"u}hnel}}},
  \bibnamefont{and}
  \bibinfo{author}{\bibfnamefont{M.}~\bibnamefont{{Sandstad}}},
  \bibinfo{journal}{\prd} \textbf{\bibinfo{volume}{94}}, \bibinfo{eid}{083504}
  (\bibinfo{year}{2016}), \eprint{1607.06077}.

\bibitem[{\citenamefont{{Carr} and {K{\"u}hnel}}(2020)}]{2020ARNPS..7050520C}
\bibinfo{author}{\bibfnamefont{B.}~\bibnamefont{{Carr}}} \bibnamefont{and}
  \bibinfo{author}{\bibfnamefont{F.}~\bibnamefont{{K{\"u}hnel}}},
  \bibinfo{journal}{Annual Review of Nuclear and Particle Science}
  \textbf{\bibinfo{volume}{70}}, \bibinfo{eid}{annurev} (\bibinfo{year}{2020}),
  \eprint{2006.02838}.

\bibitem[{\citenamefont{{Carr} et~al.}(2020)\citenamefont{{Carr}, {Kohri},
  {Sendouda}, and {Yokoyama}}}]{2020arXiv200212778C}
\bibinfo{author}{\bibfnamefont{B.}~\bibnamefont{{Carr}}},
  \bibinfo{author}{\bibfnamefont{K.}~\bibnamefont{{Kohri}}},
  \bibinfo{author}{\bibfnamefont{Y.}~\bibnamefont{{Sendouda}}},
  \bibnamefont{and}
  \bibinfo{author}{\bibfnamefont{J.}~\bibnamefont{{Yokoyama}}},
  \bibinfo{journal}{arXiv e-prints} \bibinfo{eid}{arXiv:2002.12778}
  (\bibinfo{year}{2020}), \eprint{2002.12778}.

\bibitem[{\citenamefont{{Green} and {Kavanagh}}(2020)}]{2020arXiv200710722G}
\bibinfo{author}{\bibfnamefont{A.~M.} \bibnamefont{{Green}}} \bibnamefont{and}
  \bibinfo{author}{\bibfnamefont{B.~J.} \bibnamefont{{Kavanagh}}},
  \bibinfo{journal}{arXiv e-prints} \bibinfo{eid}{arXiv:2007.10722}
  (\bibinfo{year}{2020}), \eprint{2007.10722}.

\bibitem[{\citenamefont{Abbott et~al.}(2016{\natexlab{a}})}]{Abbott:2016blz}
\bibinfo{author}{\bibfnamefont{B.~P.} \bibnamefont{Abbott}}
  \bibnamefont{et~al.} (\bibinfo{collaboration}{LIGO Scientific, Virgo}),
  \bibinfo{journal}{Phys. Rev. Lett.} \textbf{\bibinfo{volume}{116}},
  \bibinfo{pages}{061102} (\bibinfo{year}{2016}{\natexlab{a}}),
  \eprint{1602.03837}.

\bibitem[{\citenamefont{{Bird} et~al.}(2016)\citenamefont{{Bird}, {Cholis},
  {Mu{\~n}oz}, {Ali-Ha{\"\i}moud}, {Kamionkowski}, {Kovetz}, {Raccanelli}, and
  {Riess}}}]{2016PhRvL.116t1301B}
\bibinfo{author}{\bibfnamefont{S.}~\bibnamefont{{Bird}}},
  \bibinfo{author}{\bibfnamefont{I.}~\bibnamefont{{Cholis}}},
  \bibinfo{author}{\bibfnamefont{J.~B.} \bibnamefont{{Mu{\~n}oz}}},
  \bibinfo{author}{\bibfnamefont{Y.}~\bibnamefont{{Ali-Ha{\"\i}moud}}},
  \bibinfo{author}{\bibfnamefont{M.}~\bibnamefont{{Kamionkowski}}},
  \bibinfo{author}{\bibfnamefont{E.~D.} \bibnamefont{{Kovetz}}},
  \bibinfo{author}{\bibfnamefont{A.}~\bibnamefont{{Raccanelli}}},
  \bibnamefont{and} \bibinfo{author}{\bibfnamefont{A.~G.}
  \bibnamefont{{Riess}}}, \bibinfo{journal}{\prl}
  \textbf{\bibinfo{volume}{116}}, \bibinfo{eid}{201301} (\bibinfo{year}{2016}),
  \eprint{1603.00464}.

\bibitem[{\citenamefont{{Sasaki} et~al.}(2016)\citenamefont{{Sasaki}, {Suyama},
  {Tanaka}, and {Yokoyama}}}]{2016PhRvL.117f1101S}
\bibinfo{author}{\bibfnamefont{M.}~\bibnamefont{{Sasaki}}},
  \bibinfo{author}{\bibfnamefont{T.}~\bibnamefont{{Suyama}}},
  \bibinfo{author}{\bibfnamefont{T.}~\bibnamefont{{Tanaka}}}, \bibnamefont{and}
  \bibinfo{author}{\bibfnamefont{S.}~\bibnamefont{{Yokoyama}}},
  \bibinfo{journal}{\prl} \textbf{\bibinfo{volume}{117}}, \bibinfo{eid}{061101}
  (\bibinfo{year}{2016}), \eprint{1603.08338}.

\bibitem[{\citenamefont{{Clesse} and
  {Garc{\'i}a-Bellido}}(2017)}]{2017PDU....15..142C}
\bibinfo{author}{\bibfnamefont{S.}~\bibnamefont{{Clesse}}} \bibnamefont{and}
  \bibinfo{author}{\bibfnamefont{J.}~\bibnamefont{{Garc{\'i}a-Bellido}}},
  \bibinfo{journal}{Physics of the Dark Universe}
  \textbf{\bibinfo{volume}{15}}, \bibinfo{pages}{142} (\bibinfo{year}{2017}),
  \eprint{1603.05234}.

\bibitem[{\citenamefont{Sasaki et~al.}(2018)\citenamefont{Sasaki, Suyama,
  Tanaka, and Yokoyama}}]{Sasaki:2018dmp}
\bibinfo{author}{\bibfnamefont{M.}~\bibnamefont{Sasaki}},
  \bibinfo{author}{\bibfnamefont{T.}~\bibnamefont{Suyama}},
  \bibinfo{author}{\bibfnamefont{T.}~\bibnamefont{Tanaka}}, \bibnamefont{and}
  \bibinfo{author}{\bibfnamefont{S.}~\bibnamefont{Yokoyama}},
  \bibinfo{journal}{Class. Quant. Grav.} \textbf{\bibinfo{volume}{35}},
  \bibinfo{pages}{063001} (\bibinfo{year}{2018}), \eprint{1801.05235}.

\bibitem[{\citenamefont{{De Luca} et~al.}(2020{\natexlab{a}})\citenamefont{{De
  Luca}, {Franciolini}, {Pani}, and {Riotto}}}]{2020JCAP...06..044D}
\bibinfo{author}{\bibfnamefont{V.}~\bibnamefont{{De Luca}}},
  \bibinfo{author}{\bibfnamefont{G.}~\bibnamefont{{Franciolini}}},
  \bibinfo{author}{\bibfnamefont{P.}~\bibnamefont{{Pani}}}, \bibnamefont{and}
  \bibinfo{author}{\bibfnamefont{A.}~\bibnamefont{{Riotto}}},
  \bibinfo{journal}{\jcap} \textbf{\bibinfo{volume}{06}}, \bibinfo{eid}{044}
  (\bibinfo{year}{2020}{\natexlab{a}}), \eprint{2005.05641}.

\bibitem[{\citenamefont{{Jedamzik}}(2020{\natexlab{a}})}]{Jedamzik:2020ypm}
\bibinfo{author}{\bibfnamefont{K.}~\bibnamefont{{Jedamzik}}},
  \bibinfo{journal}{arXiv e-prints} \bibinfo{eid}{arXiv:2006.11172}
  (\bibinfo{year}{2020}{\natexlab{a}}), \eprint{2006.11172}.

\bibitem[{\citenamefont{Abbott
  et~al.}(2019{\natexlab{a}})}]{2019PhRvX...9c1040A}
\bibinfo{author}{\bibfnamefont{B.~P.} \bibnamefont{Abbott}}
  \bibnamefont{et~al.} (\bibinfo{collaboration}{LIGO Scientific, Virgo}),
  \bibinfo{journal}{Physical Review X} \textbf{\bibinfo{volume}{9}},
  \bibinfo{eid}{031040} (\bibinfo{year}{2019}{\natexlab{a}}),
  \eprint{1811.12907}.

\bibitem[{\citenamefont{{Raidal} et~al.}(2017)\citenamefont{{Raidal},
  {Vaskonen}, and {Veerm{\"a}e}}}]{2017JCAP...09..037R}
\bibinfo{author}{\bibfnamefont{M.}~\bibnamefont{{Raidal}}},
  \bibinfo{author}{\bibfnamefont{V.}~\bibnamefont{{Vaskonen}}},
  \bibnamefont{and}
  \bibinfo{author}{\bibfnamefont{H.}~\bibnamefont{{Veerm{\"a}e}}},
  \bibinfo{journal}{\jcap} \textbf{\bibinfo{volume}{09}}, \bibinfo{eid}{037}
  (\bibinfo{year}{2017}), \eprint{1707.01480}.

\bibitem[{\citenamefont{{Chen} and {Huang}}(2018)}]{2018ApJ...864...61C}
\bibinfo{author}{\bibfnamefont{Z.-C.} \bibnamefont{{Chen}}} \bibnamefont{and}
  \bibinfo{author}{\bibfnamefont{Q.-G.} \bibnamefont{{Huang}}},
  \bibinfo{journal}{\apj} \textbf{\bibinfo{volume}{864}}, \bibinfo{eid}{61}
  (\bibinfo{year}{2018}), \eprint{1801.10327}.

\bibitem[{\citenamefont{{Wu}}(2020)}]{2020PhRvD.101h3008W}
\bibinfo{author}{\bibfnamefont{Y.}~\bibnamefont{{Wu}}}, \bibinfo{journal}{\prd}
  \textbf{\bibinfo{volume}{101}}, \bibinfo{eid}{083008} (\bibinfo{year}{2020}),
  \eprint{2001.03833}.

\bibitem[{\citenamefont{{Ali-Ha{\"\i}moud}
  et~al.}(2017)\citenamefont{{Ali-Ha{\"\i}moud}, {Kovetz}, and
  {Kamionkowski}}}]{2017PhRvD..96l3523A}
\bibinfo{author}{\bibfnamefont{Y.}~\bibnamefont{{Ali-Ha{\"\i}moud}}},
  \bibinfo{author}{\bibfnamefont{E.~D.} \bibnamefont{{Kovetz}}},
  \bibnamefont{and}
  \bibinfo{author}{\bibfnamefont{M.}~\bibnamefont{{Kamionkowski}}},
  \bibinfo{journal}{\prd} \textbf{\bibinfo{volume}{96}}, \bibinfo{eid}{123523}
  (\bibinfo{year}{2017}), \eprint{1709.06576}.

\bibitem[{\citenamefont{{Gow} et~al.}(2020{\natexlab{a}})\citenamefont{{Gow},
  {Byrnes}, {Hall}, and {Peacock}}}]{2020JCAP...01..031G}
\bibinfo{author}{\bibfnamefont{A.~D.} \bibnamefont{{Gow}}},
  \bibinfo{author}{\bibfnamefont{C.~T.} \bibnamefont{{Byrnes}}},
  \bibinfo{author}{\bibfnamefont{A.}~\bibnamefont{{Hall}}}, \bibnamefont{and}
  \bibinfo{author}{\bibfnamefont{J.~A.} \bibnamefont{{Peacock}}},
  \bibinfo{journal}{\jcap} \textbf{\bibinfo{volume}{01}}, \bibinfo{eid}{031}
  (\bibinfo{year}{2020}{\natexlab{a}}), \eprint{1911.12685}.

\bibitem[{\citenamefont{Cai et~al.}(2018)\citenamefont{Cai, Tong, Wang, and
  Yan}}]{Cai:2018tuh}
\bibinfo{author}{\bibfnamefont{Y.-F.} \bibnamefont{Cai}},
  \bibinfo{author}{\bibfnamefont{X.}~\bibnamefont{Tong}},
  \bibinfo{author}{\bibfnamefont{D.-G.} \bibnamefont{Wang}}, \bibnamefont{and}
  \bibinfo{author}{\bibfnamefont{S.-F.} \bibnamefont{Yan}},
  \bibinfo{journal}{Phys. Rev. Lett.} \textbf{\bibinfo{volume}{121}},
  \bibinfo{pages}{081306} (\bibinfo{year}{2018}), \eprint{1805.03639}.

\bibitem[{\citenamefont{Byrnes et~al.}(2018)\citenamefont{Byrnes, Hindmarsh,
  Young, and Hawkins}}]{Byrnes:2018clq}
\bibinfo{author}{\bibfnamefont{C.~T.} \bibnamefont{Byrnes}},
  \bibinfo{author}{\bibfnamefont{M.}~\bibnamefont{Hindmarsh}},
  \bibinfo{author}{\bibfnamefont{S.}~\bibnamefont{Young}}, \bibnamefont{and}
  \bibinfo{author}{\bibfnamefont{M.~R.~S.} \bibnamefont{Hawkins}},
  \bibinfo{journal}{JCAP} \textbf{\bibinfo{volume}{08}}, \bibinfo{pages}{041}
  (\bibinfo{year}{2018}), \eprint{1801.06138}.

\bibitem[{\citenamefont{{Vaskonen} and
  {Veerm{\"a}e}}(2020)}]{2020PhRvD.101d3015V}
\bibinfo{author}{\bibfnamefont{V.}~\bibnamefont{{Vaskonen}}} \bibnamefont{and}
  \bibinfo{author}{\bibfnamefont{H.}~\bibnamefont{{Veerm{\"a}e}}},
  \bibinfo{journal}{\prd} \textbf{\bibinfo{volume}{101}}, \bibinfo{eid}{043015}
  (\bibinfo{year}{2020}), \eprint{1908.09752}.

\bibitem[{\citenamefont{{Gow} et~al.}(2020{\natexlab{b}})\citenamefont{{Gow},
  {Byrnes}, {Cole}, and {Young}}}]{2020arXiv200803289G}
\bibinfo{author}{\bibfnamefont{A.~D.} \bibnamefont{{Gow}}},
  \bibinfo{author}{\bibfnamefont{C.~T.} \bibnamefont{{Byrnes}}},
  \bibinfo{author}{\bibfnamefont{P.~S.} \bibnamefont{{Cole}}},
  \bibnamefont{and} \bibinfo{author}{\bibfnamefont{S.}~\bibnamefont{{Young}}},
  \bibinfo{journal}{arXiv e-prints} \bibinfo{eid}{arXiv:2008.03289}
  (\bibinfo{year}{2020}{\natexlab{b}}), \eprint{2008.03289}.

\bibitem[{\citenamefont{{Raidal} et~al.}(2019)\citenamefont{{Raidal},
  {Spethmann}, {Vaskonen}, and {Veerm{\"a}e}}}]{2019JCAP...02..018R}
\bibinfo{author}{\bibfnamefont{M.}~\bibnamefont{{Raidal}}},
  \bibinfo{author}{\bibfnamefont{C.}~\bibnamefont{{Spethmann}}},
  \bibinfo{author}{\bibfnamefont{V.}~\bibnamefont{{Vaskonen}}},
  \bibnamefont{and}
  \bibinfo{author}{\bibfnamefont{H.}~\bibnamefont{{Veerm{\"a}e}}},
  \bibinfo{journal}{\jcap} \textbf{\bibinfo{volume}{02}}, \bibinfo{eid}{018}
  (\bibinfo{year}{2019}), \eprint{1812.01930}.

\bibitem[{\citenamefont{Abbott
  et~al.}(2016{\natexlab{b}})}]{2016PhRvX...6d1015A}
\bibinfo{author}{\bibfnamefont{B.~P.} \bibnamefont{Abbott}}
  \bibnamefont{et~al.} (\bibinfo{collaboration}{LIGO Scientific, Virgo}),
  \bibinfo{journal}{Physical Review X} \textbf{\bibinfo{volume}{6}},
  \bibinfo{eid}{041015} (\bibinfo{year}{2016}{\natexlab{b}}),
  \eprint{1606.04856}.

\bibitem[{\citenamefont{{Fishbach} and {Holz}}(2017)}]{2017ApJ...851L..25F}
\bibinfo{author}{\bibfnamefont{M.}~\bibnamefont{{Fishbach}}} \bibnamefont{and}
  \bibinfo{author}{\bibfnamefont{D.~E.} \bibnamefont{{Holz}}},
  \bibinfo{journal}{\apjl} \textbf{\bibinfo{volume}{851}}, \bibinfo{eid}{L25}
  (\bibinfo{year}{2017}), \eprint{1709.08584}.

\bibitem[{\citenamefont{{Fishbach} et~al.}(2018)\citenamefont{{Fishbach},
  {Holz}, and {Farr}}}]{2018ApJ...863L..41F}
\bibinfo{author}{\bibfnamefont{M.}~\bibnamefont{{Fishbach}}},
  \bibinfo{author}{\bibfnamefont{D.~E.} \bibnamefont{{Holz}}},
  \bibnamefont{and} \bibinfo{author}{\bibfnamefont{W.~M.}
  \bibnamefont{{Farr}}}, \bibinfo{journal}{\apjl}
  \textbf{\bibinfo{volume}{863}}, \bibinfo{eid}{L41} (\bibinfo{year}{2018}),
  \eprint{1805.10270}.

\bibitem[{\citenamefont{{Gerosa} et~al.}(2018)\citenamefont{{Gerosa}, {Berti},
  {O'Shaughnessy}, {Belczynski}, {Kesden}, {Wysocki}, and
  {Gladysz}}}]{2018PhRvD..98h4036G}
\bibinfo{author}{\bibfnamefont{D.}~\bibnamefont{{Gerosa}}},
  \bibinfo{author}{\bibfnamefont{E.}~\bibnamefont{{Berti}}},
  \bibinfo{author}{\bibfnamefont{R.}~\bibnamefont{{O'Shaughnessy}}},
  \bibinfo{author}{\bibfnamefont{K.}~\bibnamefont{{Belczynski}}},
  \bibinfo{author}{\bibfnamefont{M.}~\bibnamefont{{Kesden}}},
  \bibinfo{author}{\bibfnamefont{D.}~\bibnamefont{{Wysocki}}},
  \bibnamefont{and}
  \bibinfo{author}{\bibfnamefont{W.}~\bibnamefont{{Gladysz}}},
  \bibinfo{journal}{\prd} \textbf{\bibinfo{volume}{98}}, \bibinfo{eid}{084036}
  (\bibinfo{year}{2018}), \eprint{1808.02491}.

\bibitem[{\citenamefont{Abbott
  et~al.}(2019{\natexlab{b}})}]{2019ApJ...882L..24A}
\bibinfo{author}{\bibfnamefont{B.~P.} \bibnamefont{Abbott}}
  \bibnamefont{et~al.} (\bibinfo{collaboration}{LIGO Scientific, Virgo}),
  \bibinfo{journal}{\apjl} \textbf{\bibinfo{volume}{882}}, \bibinfo{eid}{L24}
  (\bibinfo{year}{2019}{\natexlab{b}}), \eprint{1811.12940}.

\bibitem[{\citenamefont{{Fernandez} and {Profumo}}(2019)}]{2019JCAP...08..022F}
\bibinfo{author}{\bibfnamefont{N.}~\bibnamefont{{Fernandez}}} \bibnamefont{and}
  \bibinfo{author}{\bibfnamefont{S.}~\bibnamefont{{Profumo}}},
  \bibinfo{journal}{\jcap} \textbf{\bibinfo{volume}{08}}, \bibinfo{eid}{022}
  (\bibinfo{year}{2019}), \eprint{1905.13019}.

\bibitem[{\citenamefont{{Ade} et~al.}(2016)}]{2016A&A...594A..13P}
\bibinfo{author}{\bibfnamefont{P.~A.~R.} \bibnamefont{{Ade}}}
  \bibnamefont{et~al.} (\bibinfo{collaboration}{Planck}),
  \bibinfo{journal}{\aap} \textbf{\bibinfo{volume}{594}}, \bibinfo{eid}{A13}
  (\bibinfo{year}{2016}), \eprint{1502.01589}.

\bibitem[{\citenamefont{{Dolgov} et~al.}(2020)\citenamefont{{Dolgov},
  {Kuranov}, {Mitichkin}, {Porey}, {Postnov}, {Sazhina}, and
  {Simkin}}}]{2020arXiv200500892D}
\bibinfo{author}{\bibfnamefont{A.~D.} \bibnamefont{{Dolgov}}},
  \bibinfo{author}{\bibfnamefont{A.~G.} \bibnamefont{{Kuranov}}},
  \bibinfo{author}{\bibfnamefont{N.~A.} \bibnamefont{{Mitichkin}}},
  \bibinfo{author}{\bibfnamefont{S.}~\bibnamefont{{Porey}}},
  \bibinfo{author}{\bibfnamefont{K.~A.} \bibnamefont{{Postnov}}},
  \bibinfo{author}{\bibfnamefont{O.~S.} \bibnamefont{{Sazhina}}},
  \bibnamefont{and} \bibinfo{author}{\bibfnamefont{I.~V.}
  \bibnamefont{{Simkin}}}, \bibinfo{journal}{arXiv e-prints}
  \bibinfo{eid}{arXiv:2005.00892} (\bibinfo{year}{2020}), \eprint{2005.00892}.

\bibitem[{\citenamefont{{Gerosa}}(2017)}]{2017zndo....889966G}
\bibinfo{author}{\bibfnamefont{D.}~\bibnamefont{{Gerosa}}},
  \emph{\bibinfo{title}{{Dgerosa/Gwdet: V0.1}}} (\bibinfo{year}{2017}).

\bibitem[{\citenamefont{{Finn} and {Chernoff}}(1993)}]{1993PhRvD..47.2198F}
\bibinfo{author}{\bibfnamefont{L.~S.} \bibnamefont{{Finn}}} \bibnamefont{and}
  \bibinfo{author}{\bibfnamefont{D.~F.} \bibnamefont{{Chernoff}}},
  \bibinfo{journal}{\prd} \textbf{\bibinfo{volume}{47}}, \bibinfo{pages}{2198}
  (\bibinfo{year}{1993}), \eprint{gr-qc/9301003}.

\bibitem[{\citenamefont{{Gerosa} et~al.}(2019)\citenamefont{{Gerosa}, {Ma},
  {Wong}, {Berti}, {O'Shaughnessy}, {Chen}, and
  {Belczynski}}}]{2019PhRvD..99j3004G}
\bibinfo{author}{\bibfnamefont{D.}~\bibnamefont{{Gerosa}}},
  \bibinfo{author}{\bibfnamefont{S.}~\bibnamefont{{Ma}}},
  \bibinfo{author}{\bibfnamefont{K.~W.~K.} \bibnamefont{{Wong}}},
  \bibinfo{author}{\bibfnamefont{E.}~\bibnamefont{{Berti}}},
  \bibinfo{author}{\bibfnamefont{R.}~\bibnamefont{{O'Shaughnessy}}},
  \bibinfo{author}{\bibfnamefont{Y.}~\bibnamefont{{Chen}}}, \bibnamefont{and}
  \bibinfo{author}{\bibfnamefont{K.}~\bibnamefont{{Belczynski}}},
  \bibinfo{journal}{\prd} \textbf{\bibinfo{volume}{99}}, \bibinfo{eid}{103004}
  (\bibinfo{year}{2019}), \eprint{1902.00021}.

\bibitem[{\citenamefont{{Usman} et~al.}(2016)\citenamefont{{Usman}, {Nitz},
  {Harry}, {Biwer}, {Brown}, {Cabero}, {Capano}, {Dal Canton}, {Dent},
  {Fairhurst} et~al.}}]{2016CQGra..33u5004U}
\bibinfo{author}{\bibfnamefont{S.~A.} \bibnamefont{{Usman}}},
  \bibinfo{author}{\bibfnamefont{A.~H.} \bibnamefont{{Nitz}}},
  \bibinfo{author}{\bibfnamefont{I.~W.} \bibnamefont{{Harry}}},
  \bibinfo{author}{\bibfnamefont{C.~M.} \bibnamefont{{Biwer}}},
  \bibinfo{author}{\bibfnamefont{D.~A.} \bibnamefont{{Brown}}},
  \bibinfo{author}{\bibfnamefont{M.}~\bibnamefont{{Cabero}}},
  \bibinfo{author}{\bibfnamefont{C.~D.} \bibnamefont{{Capano}}},
  \bibinfo{author}{\bibfnamefont{T.}~\bibnamefont{{Dal Canton}}},
  \bibinfo{author}{\bibfnamefont{T.}~\bibnamefont{{Dent}}},
  \bibinfo{author}{\bibfnamefont{S.}~\bibnamefont{{Fairhurst}}},
  \bibnamefont{et~al.}, \bibinfo{journal}{Class. Quant. Grav.}
  \textbf{\bibinfo{volume}{33}}, \bibinfo{eid}{215004} (\bibinfo{year}{2016}),
  \eprint{1508.02357}.

\bibitem[{\citenamefont{{De Luca} et~al.}(2019)\citenamefont{{De Luca},
  {Desjacques}, {Franciolini}, {Malhotra}, and {Riotto}}}]{2019JCAP...05..018D}
\bibinfo{author}{\bibfnamefont{V.}~\bibnamefont{{De Luca}}},
  \bibinfo{author}{\bibfnamefont{V.}~\bibnamefont{{Desjacques}}},
  \bibinfo{author}{\bibfnamefont{G.}~\bibnamefont{{Franciolini}}},
  \bibinfo{author}{\bibfnamefont{A.}~\bibnamefont{{Malhotra}}},
  \bibnamefont{and} \bibinfo{author}{\bibfnamefont{A.}~\bibnamefont{{Riotto}}},
  \bibinfo{journal}{\jcap} \textbf{\bibinfo{volume}{05}}, \bibinfo{eid}{018}
  (\bibinfo{year}{2019}), \eprint{1903.01179}.

\bibitem[{\citenamefont{{Mirbabayi} et~al.}(2020)\citenamefont{{Mirbabayi},
  {Gruzinov}, and {Nore{\~n}a}}}]{2020JCAP...03..017M}
\bibinfo{author}{\bibfnamefont{M.}~\bibnamefont{{Mirbabayi}}},
  \bibinfo{author}{\bibfnamefont{A.}~\bibnamefont{{Gruzinov}}},
  \bibnamefont{and}
  \bibinfo{author}{\bibfnamefont{J.}~\bibnamefont{{Nore{\~n}a}}},
  \bibinfo{journal}{\jcap} \textbf{\bibinfo{volume}{03}}, \bibinfo{eid}{017}
  (\bibinfo{year}{2020}), \eprint{1901.05963}.

\bibitem[{\citenamefont{{De Luca} et~al.}(2020{\natexlab{b}})\citenamefont{{De
  Luca}, {Franciolini}, {Pani}, and {Riotto}}}]{2020JCAP...04..052D}
\bibinfo{author}{\bibfnamefont{V.}~\bibnamefont{{De Luca}}},
  \bibinfo{author}{\bibfnamefont{G.}~\bibnamefont{{Franciolini}}},
  \bibinfo{author}{\bibfnamefont{P.}~\bibnamefont{{Pani}}}, \bibnamefont{and}
  \bibinfo{author}{\bibfnamefont{A.}~\bibnamefont{{Riotto}}},
  \bibinfo{journal}{\jcap} \textbf{\bibinfo{volume}{04}}, \bibinfo{eid}{052}
  (\bibinfo{year}{2020}{\natexlab{b}}), \eprint{2003.02778}.

\bibitem[{\citenamefont{Abbott
  et~al.}(2016{\natexlab{c}})}]{2016LRR....19....1A}
\bibinfo{author}{\bibfnamefont{B.~P.} \bibnamefont{Abbott}}
  \bibnamefont{et~al.} (\bibinfo{collaboration}{LIGO Scientific, Virgo}),
  \bibinfo{journal}{Living Reviews in Relativity}
  \textbf{\bibinfo{volume}{19}}, \bibinfo{eid}{1}
  (\bibinfo{year}{2016}{\natexlab{c}}), \eprint{1304.0670v3}.

\bibitem[{\citenamefont{{Wysocki} et~al.}(2019)\citenamefont{{Wysocki},
  {Lange}, and {O'Shaughnessy}}}]{2019PhRvD.100d3012W}
\bibinfo{author}{\bibfnamefont{D.}~\bibnamefont{{Wysocki}}},
  \bibinfo{author}{\bibfnamefont{J.}~\bibnamefont{{Lange}}}, \bibnamefont{and}
  \bibinfo{author}{\bibfnamefont{R.}~\bibnamefont{{O'Shaughnessy}}},
  \bibinfo{journal}{\prd} \textbf{\bibinfo{volume}{100}}, \bibinfo{eid}{043012}
  (\bibinfo{year}{2019}), \eprint{1805.06442}.

\bibitem[{\citenamefont{Abbott
  et~al.}(2020{\natexlab{a}})}]{2020arXiv200408342T}
\bibinfo{author}{\bibfnamefont{R.}~\bibnamefont{Abbott}} \bibnamefont{et~al.}
  (\bibinfo{collaboration}{LIGO Scientific Collaboration and Virgo
  Collaboration}), \bibinfo{journal}{Phys. Rev. D}
  \textbf{\bibinfo{volume}{102}}, \bibinfo{pages}{043015}
  (\bibinfo{year}{2020}{\natexlab{a}}), \eprint{2004.08342}.

\bibitem[{\citenamefont{Abbott et~al.}(2020{\natexlab{b}})}]{Abbott:2020khf}
\bibinfo{author}{\bibfnamefont{R.}~\bibnamefont{Abbott}} \bibnamefont{et~al.}
  (\bibinfo{collaboration}{LIGO Scientific, Virgo}), \bibinfo{journal}{\apjl}
  \textbf{\bibinfo{volume}{896}}, \bibinfo{pages}{L44}
  (\bibinfo{year}{2020}{\natexlab{b}}), \eprint{2006.12611}.

\bibitem[{\citenamefont{{Roulet} and
  {Zaldarriaga}}(2019)}]{2019MNRAS.484.4216R}
\bibinfo{author}{\bibfnamefont{J.}~\bibnamefont{{Roulet}}} \bibnamefont{and}
  \bibinfo{author}{\bibfnamefont{M.}~\bibnamefont{{Zaldarriaga}}},
  \bibinfo{journal}{\mnras} \textbf{\bibinfo{volume}{484}},
  \bibinfo{pages}{4216} (\bibinfo{year}{2019}), \eprint{1806.10610}.

\bibitem[{\citenamefont{{Nakama} et~al.}(2017)\citenamefont{{Nakama}, {Silk},
  and {Kamionkowski}}}]{2017PhRvD..95d3511N}
\bibinfo{author}{\bibfnamefont{T.}~\bibnamefont{{Nakama}}},
  \bibinfo{author}{\bibfnamefont{J.}~\bibnamefont{{Silk}}}, \bibnamefont{and}
  \bibinfo{author}{\bibfnamefont{M.}~\bibnamefont{{Kamionkowski}}},
  \bibinfo{journal}{\prd} \textbf{\bibinfo{volume}{95}}, \bibinfo{eid}{043511}
  (\bibinfo{year}{2017}), \eprint{1612.06264}.

\bibitem[{\citenamefont{{Kavanagh} et~al.}(2018)\citenamefont{{Kavanagh},
  {Gaggero}, and {Bertone}}}]{2018PhRvD..98b3536K}
\bibinfo{author}{\bibfnamefont{B.~J.} \bibnamefont{{Kavanagh}}},
  \bibinfo{author}{\bibfnamefont{D.}~\bibnamefont{{Gaggero}}},
  \bibnamefont{and}
  \bibinfo{author}{\bibfnamefont{G.}~\bibnamefont{{Bertone}}},
  \bibinfo{journal}{\prd} \textbf{\bibinfo{volume}{98}}, \bibinfo{eid}{023536}
  (\bibinfo{year}{2018}), \eprint{1805.09034}.

\bibitem[{\citenamefont{Peters}(1964)}]{PhysRev.136.B1224}
\bibinfo{author}{\bibfnamefont{P.~C.} \bibnamefont{Peters}},
  \bibinfo{journal}{Phys. Rev.} \textbf{\bibinfo{volume}{136}},
  \bibinfo{pages}{B1224} (\bibinfo{year}{1964}),
  \urlprefix\url{https://link.aps.org/doi/10.1103/PhysRev.136.B1224}.

\bibitem[{\citenamefont{Eroshenko}(2018)}]{Eroshenko:2016hmn}
\bibinfo{author}{\bibfnamefont{Y.~N.} \bibnamefont{Eroshenko}},
  \bibinfo{journal}{J. Phys. Conf. Ser.} \textbf{\bibinfo{volume}{1051}},
  \bibinfo{pages}{012010} (\bibinfo{year}{2018}), \eprint{1604.04932}.

\bibitem[{\citenamefont{{Ioka} et~al.}(1998)\citenamefont{{Ioka}, {Chiba},
  {Tanaka}, and {Nakamura}}}]{1998PhRvD..58f3003I}
\bibinfo{author}{\bibfnamefont{K.}~\bibnamefont{{Ioka}}},
  \bibinfo{author}{\bibfnamefont{T.}~\bibnamefont{{Chiba}}},
  \bibinfo{author}{\bibfnamefont{T.}~\bibnamefont{{Tanaka}}}, \bibnamefont{and}
  \bibinfo{author}{\bibfnamefont{T.}~\bibnamefont{{Nakamura}}},
  \bibinfo{journal}{\prd} \textbf{\bibinfo{volume}{58}}, \bibinfo{eid}{063003}
  (\bibinfo{year}{1998}), \eprint{astro-ph/9807018}.

\bibitem[{\citenamefont{{Inman} and
  {Ali-Ha{\"\i}moud}}(2019)}]{2019PhRvD.100h3528I}
\bibinfo{author}{\bibfnamefont{D.}~\bibnamefont{{Inman}}} \bibnamefont{and}
  \bibinfo{author}{\bibfnamefont{Y.}~\bibnamefont{{Ali-Ha{\"\i}moud}}},
  \bibinfo{journal}{\prd} \textbf{\bibinfo{volume}{100}}, \bibinfo{eid}{083528}
  (\bibinfo{year}{2019}), \eprint{1907.08129}.

\bibitem[{\citenamefont{{Jedamzik}}(2020{\natexlab{b}})}]{Jedamzik:2020omx}
\bibinfo{author}{\bibfnamefont{K.}~\bibnamefont{{Jedamzik}}},
  \bibinfo{journal}{arXiv e-prints} \bibinfo{eid}{arXiv:2007.03565}
  (\bibinfo{year}{2020}{\natexlab{b}}), \eprint{2007.03565}.

\bibitem[{\citenamefont{{Young} and {Hamers}}(2020)}]{Young:2020scc}
\bibinfo{author}{\bibfnamefont{S.}~\bibnamefont{{Young}}} \bibnamefont{and}
  \bibinfo{author}{\bibfnamefont{A.~S.} \bibnamefont{{Hamers}}},
  \bibinfo{journal}{arXiv e-prints} \bibinfo{eid}{arXiv:2006.15023}
  (\bibinfo{year}{2020}), \eprint{2006.15023}.

\bibitem[{\citenamefont{{Trashorras} et~al.}(2020)\citenamefont{{Trashorras},
  {Garc{\'i}a-Bellido}, and {Nesseris}}}]{Trashorras:2020mwn}
\bibinfo{author}{\bibfnamefont{M.}~\bibnamefont{{Trashorras}}},
  \bibinfo{author}{\bibfnamefont{J.}~\bibnamefont{{Garc{\'i}a-Bellido}}},
  \bibnamefont{and}
  \bibinfo{author}{\bibfnamefont{S.}~\bibnamefont{{Nesseris}}},
  \bibinfo{journal}{arXiv e-prints} \bibinfo{eid}{arXiv:2006.15018}
  (\bibinfo{year}{2020}), \eprint{2006.15018}.

\bibitem[{\citenamefont{Liu et~al.}(2019)\citenamefont{Liu, Guo, and
  Cai}}]{Liu:2019rnx}
\bibinfo{author}{\bibfnamefont{L.}~\bibnamefont{Liu}},
  \bibinfo{author}{\bibfnamefont{Z.-K.} \bibnamefont{Guo}}, \bibnamefont{and}
  \bibinfo{author}{\bibfnamefont{R.-G.} \bibnamefont{Cai}},
  \bibinfo{journal}{Eur. Phys. J. C} \textbf{\bibinfo{volume}{79}},
  \bibinfo{pages}{717} (\bibinfo{year}{2019}), \eprint{1901.07672}.

\bibitem[{\citenamefont{Adamek et~al.}(2019)\citenamefont{Adamek, Byrnes,
  Gosenca, and Hotchkiss}}]{Adamek:2019gns}
\bibinfo{author}{\bibfnamefont{J.}~\bibnamefont{Adamek}},
  \bibinfo{author}{\bibfnamefont{C.~T.} \bibnamefont{Byrnes}},
  \bibinfo{author}{\bibfnamefont{M.}~\bibnamefont{Gosenca}}, \bibnamefont{and}
  \bibinfo{author}{\bibfnamefont{S.}~\bibnamefont{Hotchkiss}},
  \bibinfo{journal}{Phys. Rev. D} \textbf{\bibinfo{volume}{100}},
  \bibinfo{pages}{023506} (\bibinfo{year}{2019}), \eprint{1901.08528}.

\bibitem[{\citenamefont{{Bosch-Ramon} and
  {Bellomo}}(2020)}]{2020A&A...638A.132B}
\bibinfo{author}{\bibfnamefont{V.}~\bibnamefont{{Bosch-Ramon}}}
  \bibnamefont{and}
  \bibinfo{author}{\bibfnamefont{N.}~\bibnamefont{{Bellomo}}},
  \bibinfo{journal}{\aap} \textbf{\bibinfo{volume}{638}}, \bibinfo{eid}{A132}
  (\bibinfo{year}{2020}), \eprint{2004.11224}.

\bibitem[{\citenamefont{Sobrinho and Augusto}(2020)}]{Sobrinho:2020cco}
\bibinfo{author}{\bibfnamefont{J.~L.~G.} \bibnamefont{Sobrinho}}
  \bibnamefont{and} \bibinfo{author}{\bibfnamefont{P.}~\bibnamefont{Augusto}},
  \bibinfo{journal}{Mon. Not. Roy. Astron. Soc.}
  \textbf{\bibinfo{volume}{496}}, \bibinfo{pages}{60} (\bibinfo{year}{2020}),
  \eprint{2005.10037}.

\bibitem[{\citenamefont{Dolgov and Silk}(1993)}]{PhysRevD.47.4244}
\bibinfo{author}{\bibfnamefont{A.}~\bibnamefont{Dolgov}} \bibnamefont{and}
  \bibinfo{author}{\bibfnamefont{J.}~\bibnamefont{Silk}},
  \bibinfo{journal}{Phys. Rev. D} \textbf{\bibinfo{volume}{47}},
  \bibinfo{pages}{4244} (\bibinfo{year}{1993}).

\bibitem[{\citenamefont{Carr et~al.}(2017)\citenamefont{Carr, Raidal, Tenkanen,
  Vaskonen, and Veerm\"ae}}]{PhysRevD.96.023514}
\bibinfo{author}{\bibfnamefont{B.}~\bibnamefont{Carr}},
  \bibinfo{author}{\bibfnamefont{M.}~\bibnamefont{Raidal}},
  \bibinfo{author}{\bibfnamefont{T.}~\bibnamefont{Tenkanen}},
  \bibinfo{author}{\bibfnamefont{V.}~\bibnamefont{Vaskonen}}, \bibnamefont{and}
  \bibinfo{author}{\bibfnamefont{H.}~\bibnamefont{Veerm\"ae}},
  \bibinfo{journal}{Phys. Rev. D} \textbf{\bibinfo{volume}{96}},
  \bibinfo{pages}{023514} (\bibinfo{year}{2017}), \eprint{1705.05567}.

\bibitem[{\citenamefont{{Gow} et~al.}(2020{\natexlab{c}})\citenamefont{{Gow},
  {Byrnes}, and {Hall}}}]{2020arXiv200903204G}
\bibinfo{author}{\bibfnamefont{A.~D.} \bibnamefont{{Gow}}},
  \bibinfo{author}{\bibfnamefont{C.~T.} \bibnamefont{{Byrnes}}},
  \bibnamefont{and} \bibinfo{author}{\bibfnamefont{A.}~\bibnamefont{{Hall}}},
  \bibinfo{journal}{arXiv e-prints} \bibinfo{eid}{arXiv:2009.03204}
  (\bibinfo{year}{2020}{\natexlab{c}}), \eprint{2009.03204}.

\bibitem[{\citenamefont{Abbott et~al.}(2017)}]{2017PhRvL.118v1101A}
\bibinfo{author}{\bibfnamefont{B.~P.} \bibnamefont{Abbott}}
  \bibnamefont{et~al.} (\bibinfo{collaboration}{LIGO Scientific, Virgo}),
  \bibinfo{journal}{\prl} \textbf{\bibinfo{volume}{118}}, \bibinfo{eid}{221101}
  (\bibinfo{year}{2017}), \eprint{1706.01812}.

\bibitem[{\citenamefont{{Kovetz} et~al.}(2017)\citenamefont{{Kovetz}, {Cholis},
  {Breysse}, and {Kamionkowski}}}]{2017PhRvD..95j3010K}
\bibinfo{author}{\bibfnamefont{E.~D.} \bibnamefont{{Kovetz}}},
  \bibinfo{author}{\bibfnamefont{I.}~\bibnamefont{{Cholis}}},
  \bibinfo{author}{\bibfnamefont{P.~C.} \bibnamefont{{Breysse}}},
  \bibnamefont{and}
  \bibinfo{author}{\bibfnamefont{M.}~\bibnamefont{{Kamionkowski}}},
  \bibinfo{journal}{\prd} \textbf{\bibinfo{volume}{95}}, \bibinfo{eid}{103010}
  (\bibinfo{year}{2017}), \eprint{1611.01157}.

\bibitem[{\citenamefont{{Talbot} and {Thrane}}(2018)}]{2018ApJ...856..173T}
\bibinfo{author}{\bibfnamefont{C.}~\bibnamefont{{Talbot}}} \bibnamefont{and}
  \bibinfo{author}{\bibfnamefont{E.}~\bibnamefont{{Thrane}}},
  \bibinfo{journal}{\apj} \textbf{\bibinfo{volume}{856}}, \bibinfo{eid}{173}
  (\bibinfo{year}{2018}), \eprint{1801.02699}.

\bibitem[{\citenamefont{{{\"O}zel} et~al.}(2010)\citenamefont{{{\"O}zel},
  {Psaltis}, {Narayan}, and {McClintock}}}]{2010ApJ...725.1918O}
\bibinfo{author}{\bibfnamefont{F.}~\bibnamefont{{{\"O}zel}}},
  \bibinfo{author}{\bibfnamefont{D.}~\bibnamefont{{Psaltis}}},
  \bibinfo{author}{\bibfnamefont{R.}~\bibnamefont{{Narayan}}},
  \bibnamefont{and} \bibinfo{author}{\bibfnamefont{J.~E.}
  \bibnamefont{{McClintock}}}, \bibinfo{journal}{\apj}
  \textbf{\bibinfo{volume}{725}}, \bibinfo{pages}{1918} (\bibinfo{year}{2010}),
  \eprint{1006.2834}.

\bibitem[{\citenamefont{{Talbot} and {Thrane}}(2017)}]{2017PhRvD..96b3012T}
\bibinfo{author}{\bibfnamefont{C.}~\bibnamefont{{Talbot}}} \bibnamefont{and}
  \bibinfo{author}{\bibfnamefont{E.}~\bibnamefont{{Thrane}}},
  \bibinfo{journal}{\prd} \textbf{\bibinfo{volume}{96}}, \bibinfo{eid}{023012}
  (\bibinfo{year}{2017}), \eprint{1704.08370}.

\bibitem[{\citenamefont{{Garcia-Bellido}
  et~al.}(2020)\citenamefont{{Garcia-Bellido}, {Nu{\~n}o Siles}, and {Ruiz
  Morales}}}]{2020arXiv201013811G}
\bibinfo{author}{\bibfnamefont{J.}~\bibnamefont{{Garcia-Bellido}}},
  \bibinfo{author}{\bibfnamefont{J.~F.} \bibnamefont{{Nu{\~n}o Siles}}},
  \bibnamefont{and} \bibinfo{author}{\bibfnamefont{E.}~\bibnamefont{{Ruiz
  Morales}}}, \bibinfo{journal}{arXiv e-prints} \bibinfo{eid}{arXiv:2010.13811}
  (\bibinfo{year}{2020}), \eprint{2010.13811}.

\bibitem[{\citenamefont{{Kimball} et~al.}(2020)\citenamefont{{Kimball},
  {Talbot}, {Berry}, {Carney}, {Zevin}, {Thrane}, and
  {Kalogera}}}]{2020arXiv200500023K}
\bibinfo{author}{\bibfnamefont{C.}~\bibnamefont{{Kimball}}},
  \bibinfo{author}{\bibfnamefont{C.}~\bibnamefont{{Talbot}}},
  \bibinfo{author}{\bibfnamefont{C.~P.~L.} \bibnamefont{{Berry}}},
  \bibinfo{author}{\bibfnamefont{M.}~\bibnamefont{{Carney}}},
  \bibinfo{author}{\bibfnamefont{M.}~\bibnamefont{{Zevin}}},
  \bibinfo{author}{\bibfnamefont{E.}~\bibnamefont{{Thrane}}}, \bibnamefont{and}
  \bibinfo{author}{\bibfnamefont{V.}~\bibnamefont{{Kalogera}}},
  \bibinfo{journal}{arXiv e-prints} \bibinfo{eid}{arXiv:2005.00023}
  (\bibinfo{year}{2020}), \eprint{2005.00023}.

\bibitem[{\citenamefont{{Loredo}}(2004)}]{2004AIPC..735..195L}
\bibinfo{author}{\bibfnamefont{T.~J.} \bibnamefont{{Loredo}}}, in
  \emph{\bibinfo{booktitle}{American Institute of Physics Conference Series}},
  edited by \bibinfo{editor}{\bibfnamefont{R.}~\bibnamefont{{Fischer}}},
  \bibinfo{editor}{\bibfnamefont{R.}~\bibnamefont{{Preuss}}}, \bibnamefont{and}
  \bibinfo{editor}{\bibfnamefont{U.~V.} \bibnamefont{{Toussaint}}}
  (\bibinfo{year}{2004}), vol. \bibinfo{volume}{735} of
  \emph{\bibinfo{series}{American Institute of Physics Conference Series}}, pp.
  \bibinfo{pages}{195--206}, \eprint{astro-ph/0409387}.

\bibitem[{\citenamefont{{Mandel} et~al.}(2019)\citenamefont{{Mandel}, {Farr},
  and {Gair}}}]{2019MNRAS.486.1086M}
\bibinfo{author}{\bibfnamefont{I.}~\bibnamefont{{Mandel}}},
  \bibinfo{author}{\bibfnamefont{W.~M.} \bibnamefont{{Farr}}},
  \bibnamefont{and} \bibinfo{author}{\bibfnamefont{J.~R.}
  \bibnamefont{{Gair}}}, \bibinfo{journal}{\mnras}
  \textbf{\bibinfo{volume}{486}}, \bibinfo{pages}{1086} (\bibinfo{year}{2019}),
  \eprint{1809.02063}.

\bibitem[{\citenamefont{{Vattis} et~al.}(2020)\citenamefont{{Vattis},
  {Goldstein}, and {Koushiappas}}}]{Vattis:2020iuz}
\bibinfo{author}{\bibfnamefont{K.}~\bibnamefont{{Vattis}}},
  \bibinfo{author}{\bibfnamefont{I.~S.} \bibnamefont{{Goldstein}}},
  \bibnamefont{and} \bibinfo{author}{\bibfnamefont{S.~M.}
  \bibnamefont{{Koushiappas}}}, \bibinfo{journal}{arXiv e-prints}
  \bibinfo{eid}{arXiv:2006.15675} (\bibinfo{year}{2020}), \eprint{2006.15675}.

\bibitem[{\citenamefont{{Tsai} et~al.}(2020)\citenamefont{{Tsai}, {Palmese},
  {Profumo}, and {Jeltema}}}]{Tsai:2020hpi}
\bibinfo{author}{\bibfnamefont{Y.-D.} \bibnamefont{{Tsai}}},
  \bibinfo{author}{\bibfnamefont{A.}~\bibnamefont{{Palmese}}},
  \bibinfo{author}{\bibfnamefont{S.}~\bibnamefont{{Profumo}}},
  \bibnamefont{and}
  \bibinfo{author}{\bibfnamefont{T.}~\bibnamefont{{Jeltema}}},
  \bibinfo{journal}{arXiv e-prints} \bibinfo{eid}{arXiv:2007.03686}
  (\bibinfo{year}{2020}), \eprint{2007.03686}.

\bibitem[{\citenamefont{{Vitale} et~al.}(2017)\citenamefont{{Vitale}, {Lynch},
  {Sturani}, and {Graff}}}]{2017CQGra..34cLT01V}
\bibinfo{author}{\bibfnamefont{S.}~\bibnamefont{{Vitale}}},
  \bibinfo{author}{\bibfnamefont{R.}~\bibnamefont{{Lynch}}},
  \bibinfo{author}{\bibfnamefont{R.}~\bibnamefont{{Sturani}}},
  \bibnamefont{and} \bibinfo{author}{\bibfnamefont{P.}~\bibnamefont{{Graff}}},
  \bibinfo{journal}{Class. Quant. Grav.} \textbf{\bibinfo{volume}{34}},
  \bibinfo{eid}{03LT01} (\bibinfo{year}{2017}), \eprint{1503.04307}.

\bibitem[{\citenamefont{{Skilling}}(2004)}]{2004AIPC..735..395S}
\bibinfo{author}{\bibfnamefont{J.}~\bibnamefont{{Skilling}}}, in
  \emph{\bibinfo{booktitle}{American Institute of Physics Conference Series}}
  (\bibinfo{year}{2004}), vol. \bibinfo{volume}{735} of
  \emph{\bibinfo{series}{American Institute of Physics Conference Series}}, pp.
  \bibinfo{pages}{395--405}.

\bibitem[{\citenamefont{Skilling}(2006)}]{skilling2006}
\bibinfo{author}{\bibfnamefont{J.}~\bibnamefont{Skilling}},
  \bibinfo{journal}{Bayesian Anal.} \textbf{\bibinfo{volume}{1}},
  \bibinfo{pages}{833} (\bibinfo{year}{2006}),
  \urlprefix\url{https://doi.org/10.1214/06-BA127}.

\bibitem[{\citenamefont{{Feroz} et~al.}(2009)\citenamefont{{Feroz}, {Hobson},
  and {Bridges}}}]{2009MNRAS.398.1601F}
\bibinfo{author}{\bibfnamefont{F.}~\bibnamefont{{Feroz}}},
  \bibinfo{author}{\bibfnamefont{M.~P.} \bibnamefont{{Hobson}}},
  \bibnamefont{and}
  \bibinfo{author}{\bibfnamefont{M.}~\bibnamefont{{Bridges}}},
  \bibinfo{journal}{\mnras} \textbf{\bibinfo{volume}{398}},
  \bibinfo{pages}{1601} (\bibinfo{year}{2009}), \eprint{0809.3437}.

\bibitem[{\citenamefont{{Speagle}}(2020)}]{2020MNRAS.493.3132S}
\bibinfo{author}{\bibfnamefont{J.~S.} \bibnamefont{{Speagle}}},
  \bibinfo{journal}{\mnras} \textbf{\bibinfo{volume}{493}},
  \bibinfo{pages}{3132} (\bibinfo{year}{2020}), \eprint{1904.02180}.

\bibitem[{\citenamefont{{Trotta}}(2007)}]{2007MNRAS.378...72T}
\bibinfo{author}{\bibfnamefont{R.}~\bibnamefont{{Trotta}}},
  \bibinfo{journal}{\mnras} \textbf{\bibinfo{volume}{378}}, \bibinfo{pages}{72}
  (\bibinfo{year}{2007}), \eprint{astro-ph/0504022}.

\bibitem[{\citenamefont{MacKay}(2002)}]{10.5555/971143}
\bibinfo{author}{\bibfnamefont{D.~J.~C.} \bibnamefont{MacKay}},
  \emph{\bibinfo{title}{Information Theory, Inference \& Learning Algorithms}}
  (\bibinfo{publisher}{Cambridge University Press}, \bibinfo{address}{USA},
  \bibinfo{year}{2002}), ISBN \bibinfo{isbn}{0521642981}.

\bibitem[{\citenamefont{Gelman et~al.}(1996)\citenamefont{Gelman, li~Meng, and
  Stern}}]{Gelman96posteriorpredictive}
\bibinfo{author}{\bibfnamefont{A.}~\bibnamefont{Gelman}},
  \bibinfo{author}{\bibfnamefont{X.}~\bibnamefont{li~Meng}}, \bibnamefont{and}
  \bibinfo{author}{\bibfnamefont{H.}~\bibnamefont{Stern}},
  \bibinfo{journal}{Statistica Sinica} pp. \bibinfo{pages}{733--807}
  (\bibinfo{year}{1996}).

\bibitem[{\citenamefont{{Fishbach} et~al.}(2020)\citenamefont{{Fishbach},
  {Farr}, and {Holz}}}]{2020ApJ...891L..31F}
\bibinfo{author}{\bibfnamefont{M.}~\bibnamefont{{Fishbach}}},
  \bibinfo{author}{\bibfnamefont{W.~M.} \bibnamefont{{Farr}}},
  \bibnamefont{and} \bibinfo{author}{\bibfnamefont{D.~E.}
  \bibnamefont{{Holz}}}, \bibinfo{journal}{\apjl}
  \textbf{\bibinfo{volume}{891}}, \bibinfo{eid}{L31} (\bibinfo{year}{2020}),
  \eprint{1911.05882}.

\bibitem[{\citenamefont{Carr and K{\"u}hnel}(2019)}]{Carr:2018poi}
\bibinfo{author}{\bibfnamefont{B.}~\bibnamefont{Carr}} \bibnamefont{and}
  \bibinfo{author}{\bibfnamefont{F.}~\bibnamefont{K{\"u}hnel}},
  \bibinfo{journal}{Phys. Rev. D} \textbf{\bibinfo{volume}{99}},
  \bibinfo{pages}{103535} (\bibinfo{year}{2019}), \eprint{1811.06532}.

\bibitem[{\citenamefont{{Clesse} and
  {Garc\'ia-Bellido}}(2020)}]{2020arXiv200706481C}
\bibinfo{author}{\bibfnamefont{S.}~\bibnamefont{{Clesse}}} \bibnamefont{and}
  \bibinfo{author}{\bibfnamefont{J.}~\bibnamefont{{Garc\'ia-Bellido}}},
  \bibinfo{journal}{arXiv e-prints} \bibinfo{eid}{arXiv:2007.06481}
  (\bibinfo{year}{2020}), \eprint{2007.06481}.

\bibitem[{\citenamefont{{Quinlan} and {Shapiro}}(1989)}]{1989ApJ...343..725Q}
\bibinfo{author}{\bibfnamefont{G.~D.} \bibnamefont{{Quinlan}}}
  \bibnamefont{and} \bibinfo{author}{\bibfnamefont{S.~L.}
  \bibnamefont{{Shapiro}}}, \bibinfo{journal}{\apj}
  \textbf{\bibinfo{volume}{343}}, \bibinfo{pages}{725} (\bibinfo{year}{1989}).

\bibitem[{\citenamefont{{Mouri} and {Taniguchi}}(2002)}]{2002ApJ...566L..17M}
\bibinfo{author}{\bibfnamefont{H.}~\bibnamefont{{Mouri}}} \bibnamefont{and}
  \bibinfo{author}{\bibfnamefont{Y.}~\bibnamefont{{Taniguchi}}},
  \bibinfo{journal}{\apjl} \textbf{\bibinfo{volume}{566}}, \bibinfo{pages}{L17}
  (\bibinfo{year}{2002}), \eprint{astro-ph/0201102}.

\bibitem[{\citenamefont{{Ricotti}}(2007)}]{2007ApJ...662...53R}
\bibinfo{author}{\bibfnamefont{M.}~\bibnamefont{{Ricotti}}},
  \bibinfo{journal}{\apj} \textbf{\bibinfo{volume}{662}}, \bibinfo{pages}{53}
  (\bibinfo{year}{2007}), \eprint{0706.0864}.

\bibitem[{\citenamefont{Abbott et~al.}(2020{\natexlab{c}})}]{Abbott:2020tfl}
\bibinfo{author}{\bibfnamefont{R.}~\bibnamefont{Abbott}} \bibnamefont{et~al.}
  (\bibinfo{collaboration}{LIGO Scientific, Virgo}), \bibinfo{journal}{Phys.
  Rev. Lett.} \textbf{\bibinfo{volume}{125}}, \bibinfo{pages}{101102}
  (\bibinfo{year}{2020}{\natexlab{c}}), \eprint{2009.01075}.

\bibitem[{\citenamefont{{Bhagwat} et~al.}(2020)\citenamefont{{Bhagwat}, {De
  Luca}, {Franciolini}, {Pani}, and {Riotto}}}]{2020arXiv200812320B}
\bibinfo{author}{\bibfnamefont{S.}~\bibnamefont{{Bhagwat}}},
  \bibinfo{author}{\bibfnamefont{V.}~\bibnamefont{{De Luca}}},
  \bibinfo{author}{\bibfnamefont{G.}~\bibnamefont{{Franciolini}}},
  \bibinfo{author}{\bibfnamefont{P.}~\bibnamefont{{Pani}}}, \bibnamefont{and}
  \bibinfo{author}{\bibfnamefont{A.}~\bibnamefont{{Riotto}}},
  \bibinfo{journal}{arXiv e-prints} \bibinfo{eid}{arXiv:2008.12320}
  (\bibinfo{year}{2020}), \eprint{2008.12320}.

\bibitem[{\citenamefont{{De Luca} et~al.}(2020{\natexlab{c}})\citenamefont{{De
  Luca}, {Desjacques}, {Franciolini}, {Pani}, and
  {Riotto}}}]{2020arXiv200901728D}
\bibinfo{author}{\bibfnamefont{V.}~\bibnamefont{{De Luca}}},
  \bibinfo{author}{\bibfnamefont{V.}~\bibnamefont{{Desjacques}}},
  \bibinfo{author}{\bibfnamefont{G.}~\bibnamefont{{Franciolini}}},
  \bibinfo{author}{\bibfnamefont{P.}~\bibnamefont{{Pani}}}, \bibnamefont{and}
  \bibinfo{author}{\bibfnamefont{A.}~\bibnamefont{{Riotto}}},
  \bibinfo{journal}{arXiv e-prints} \bibinfo{eid}{arXiv:2009.01728}
  (\bibinfo{year}{2020}{\natexlab{c}}), \eprint{2009.01728}.

\bibitem[{\citenamefont{{Akutsu} et~al.}(2020)}]{2020arXiv200802921K}
\bibinfo{author}{\bibfnamefont{T.}~\bibnamefont{{Akutsu}}} \bibnamefont{et~al.}
  (\bibinfo{collaboration}{KAGRA}), \bibinfo{journal}{arXiv e-prints}
  \bibinfo{eid}{arXiv:2008.02921} (\bibinfo{year}{2020}), \eprint{2008.02921}.

\bibitem[{\citenamefont{{LIGO Scientific Collaboration}}(2018)}]{lalsuite}
\bibinfo{author}{\bibnamefont{{LIGO Scientific Collaboration}}},
  \emph{\bibinfo{title}{{LIGO} {A}lgorithm {L}ibrary - {LALS}uite}},
  \bibinfo{howpublished}{free software (GPL)} (\bibinfo{year}{2018}).

\bibitem[{\citenamefont{Wolfram}(2001 (accessed July 14, 2020))}]{wolfram_page}
\bibinfo{author}{\bibnamefont{Wolfram}},
  \emph{\bibinfo{title}{HypergeometricPFQ}} (\bibinfo{year}{2001 (accessed July
  14, 2020)}),
  \urlprefix\url{https://functions.wolfram.com/HypergeometricFunctions/Hypergeometric1F2/06/02/03/0001/}.

\end{thebibliography}

\appendix

\section{Fast numerical implementation of the suppression factor}
\label{app:numerics}

Bayesian inference of the posterior for PBH mergers involves many computations of the likelihood, and fast implementations are therefore crucial. The computational bottleneck in the likelihood evaluation step is the computation of the suppression factor, given in Equation~\eqref{eq:sfac} as
\begin{equation}
  S = \frac{e^{-\bar{N}(y)}}{\Gamma(21/37)}\int_0^{\infty} \ud v \, v^{-\frac{16}{37}} \exp{\left[-\bar{N}(y) \langle m \rangle \int_0^{\infty} \frac{\ud m}{m} \psi(m) F\left(\frac{m}{\langle m \rangle} \frac{v}{\bar{N}(y)}\right) - \frac{3\sigma^2_M v^2}{10 \fpbh^2} \right]}.
  \label{eq:appsfac}
\end{equation}
where we remind the reader that $F(x) = {}_1F_2(-1/2; 3/4, 5/4;-9x^2/16) - 1$ with ${}_1F_{2}$ a generalized hypergeometric function.

Fast evaluation of Equation~\eqref{eq:appsfac} is numerically challenging for general mass functions due to the double integral and potential for large dynamic range in $\psi(m)$\footnote{Additionally at the time of writing there was no public implementation of the ${}_1F_2$ hypergeometric function available in Python.}. We can make progress however by using the Taylor expansion of $F(x)$ at both high and low values of its argument.

\subsection{Switching approximation to $F(x)$}

Around $z=0$ we can truncate the definition of the generalized hypergeometric function to obtain the Taylor series of ${}_1F_2$ to order $n_{\mathrm{max}}$
\begin{equation}
  {}_1F_2(a_1; b_2, b_2; z) \approx \sum_{n=0}^{n_{\mathrm{max}}} \frac{(a_1)_n}{(b_1)_n (b_2)_n}\frac{z^n}{n!},
  \label{eq:HGlo}
\end{equation}
where $(\alpha)_n$ is a Pochhammer symbol given by
\begin{equation}
  (\alpha)_n = \alpha(\alpha+1)(\alpha+2) \ldots (\alpha+n-1)
\end{equation}
for $n \geq 1$, with $(\alpha)_0 = 1$. We choose $n_{\mathrm{max}} = 3$, additionally computing the next order $n=4$ term to quantify the perturbative error.

At large values of $z$, we use the asymptotic expansion available from Ref.~\citep{wolfram_page}, given by
\begin{align}
  &{}_1F_{2}(a_1; b_1, b_2; z) \approx \frac{\Gamma(b_1)\Gamma(b_2)}{\Gamma(b_1-a_1)\Gamma(b_2-a_2)}(-z)^{-a_1} \nonumber \\
    &\times  \left\{1 + \frac{a_1(a_1 - b_1 + 1)(a_1 - b_2 + 1)}{z} + \frac{1}{2z^2}[a_1(a_1+1)(a_1-b_1+1)(a_1-b_1+2)(a_1-b_2+2)(a_1-b_2+1)] + \ldots \right\} \nonumber \\
    & + \frac{\Gamma(b_1)\Gamma(b_2)}{2\sqrt{\pi}\Gamma(a_1)}(-z)^{\frac{1}{2}\left(a_1-b_1-b_2+\frac{1}{2}\right)} \nonumber \\
  &\times \left\{e^{-i\left[\frac{1}{2}\pi\left(a_1-b_1-b_2+\frac{1}{2}\right) + 2\sqrt{-z}\right]}\left(1+\frac{d_1}{\sqrt{-z}} + \frac{d_2}{z} + \ldots\right) + e^{i\left[\frac{1}{2}\pi\left(a_1-b_1-b_2+\frac{1}{2}\right) + 2\sqrt{-z}\right]}\left(1 - \frac{d_1}{\sqrt{-z}} + \frac{d_2}{z} + \ldots\right) \right\},
  \label{eq:HGhi}
\end{align}
where
\begin{align}
  d_1 &= \frac{i}{16}[-3 + 12a_1^2 - 4b_1^2 + 8b_2 - 4b_2^2 + 8b_1(1+b_2) - 8a_1(1 + b_1 + b_2)], \\
  d_2 &= \frac{1}{512}\{-15 + 144a_1^4 + 16b_1^4 + 16b_2 + 56b_2^2 - 64b_2^3 + 16b_2^4 - 64b_1^3(1+b_2) - 64a_1^3(7+3b_1+3b_2) \nonumber \\
  & + 8b_1^2(7+8b_2+12b_2^2) + 16b_1(1+25b_2 + 4b_2^2 - 4b_2^3) - 8a_1^2[-43 + 4b_1^2 - 72b_2 + 4b_2^2 - 8b_1(9+5b_2)] \nonumber \\
  &+ 16a_1[-1 + 4b_1^3 - 25b_2 - 4b_2^2 + 4b_2^3 - 4b_1^2(1+b_2) - b_1(25 + 40b_2 + 4b_2^2)] \}.
\end{align}
The complex exponentials in this expansion give rise to oscillating terms in $F(x)$ at order $\mathcal{O}(x^{-n})$ for $n \geq 2$. Neglecting these terms, $F(x)$ has the asymptotic expansion for $x \gg 1$
\begin{equation}
  F(x) \approx x - 1 + \frac{1}{6x} + \ldots
  \label{eq:Fapprox}
\end{equation}

At small values of $x$ we use Equation~\eqref{eq:HGlo}, keeping terms in ${}_1F_2$ up to order $z^3$ (i.e.~terms of order $x^6$ in $F(x)$). We additionally compute the error arising from neglect of the $x^8$ term. At some value $x_*$ we switch to the asymptotic expansion of Equation~\eqref{eq:HGhi}, keeping terms up to order $x^{-3}$ and using the neglected $x^{-4}$ term to approximate the error (with oscillating terms set to their maximal value). The switching value $x_*$ is then chosen to minimize the relative error of the approximation. This yields $x_* = 2.72$ in the case when all $x^{-4}$ terms are used and $x_*=2.74$ when the approximation Equation~\eqref{eq:Fapprox} is used.
\begin{figure}
\centering
\includegraphics[width=0.5\textwidth]{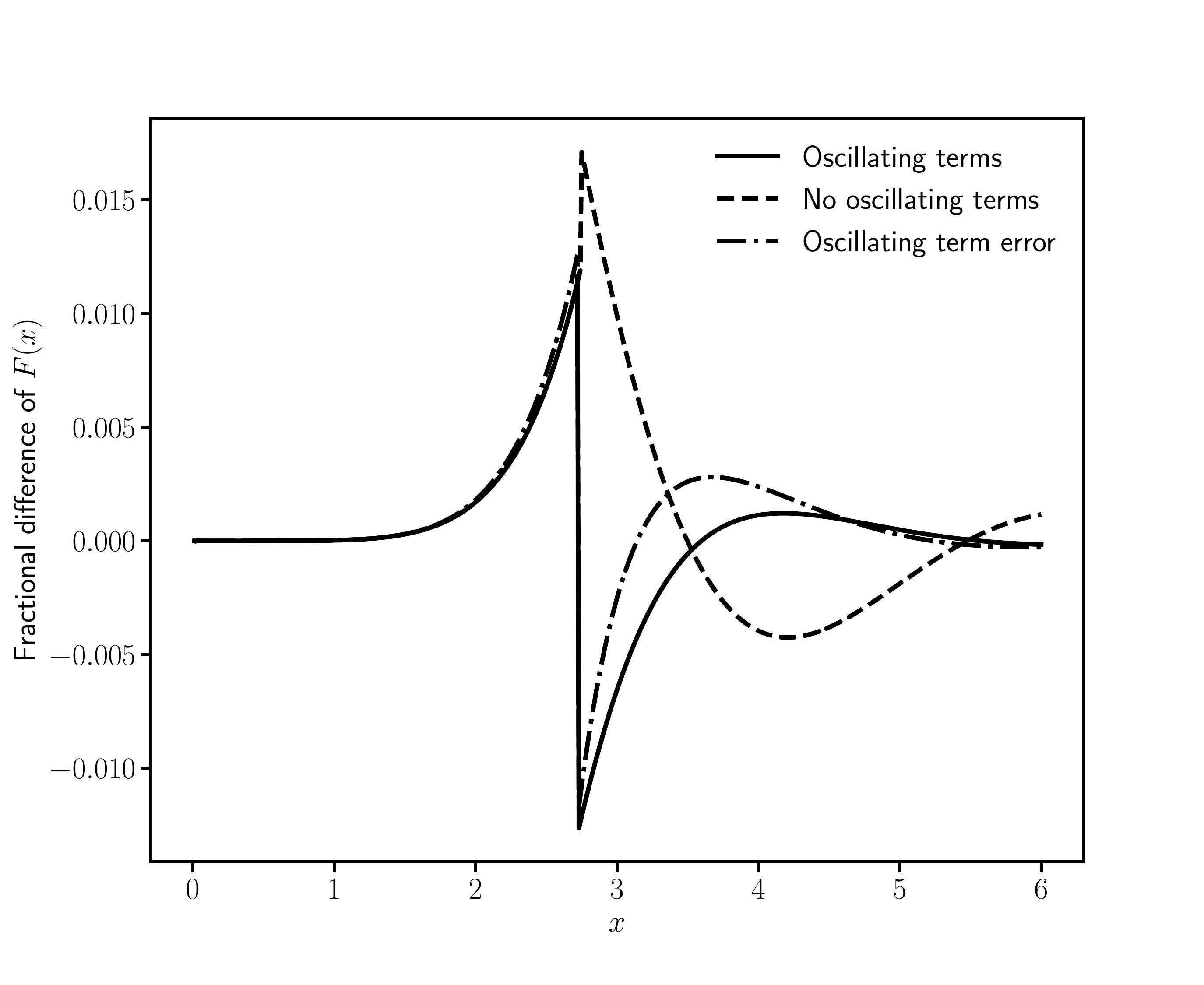}
\caption{Fractional difference of the approximation described in the text to the function $F(x)$, defined after Equation~\eqref{eq:appsfac}, and its true value with (solid) and without (dashed) oscillating terms. We also show the estimate of the error of the approximation based on the neglected higher-order terms (dot-dashed). We achieve better than 2\% accuracy across all values of the argument, and additionally can accurately predict the error of the approximation. Note that all curves are indistinguishable for small values of $x$.}
\label{fig:HG_validation}
\end{figure}

In Figure~\ref{fig:HG_validation} we show the relative error of our approximation scheme compared to an exact calculation implemented in MATLAB. The errors are generally sub-percent, reaching maximal values of $\lesssim 2\%$ around $x_*$. The error estimate derived from neglected higher-order terms is generally an accurate approximation to the true error, and we also see that oscillating terms are generally negligible with Equation~\eqref{eq:Fapprox} sufficient for $x \gtrsim x_*$. The maximum absolute error of our approximation is also $\sim \mathcal{O}(1\%)$.

The results of this section may be summarised as follows: in our baseline PBH likelihood analysis we use the following approximation for $F(x)$, accurate at the percent level:
\begin{equation}
  F(x) \approx 
  \begin{cases}
    \frac{3}{10}x^2 - \frac{3}{280}x^4 + \frac{27}{80080}x^6 \quad & x \leq 2.74 \\
    x - 1 + \frac{1}{6x} \quad & x > 2.74
  \end{cases}
\end{equation}

\subsection{Exact integration for lognormal mass functions}

A polynomial expansion for $F(x)$ is particularly useful when $\psi(m)$ is a lognormal distribution, since the inner-most integral in Equation~\eqref{eq:appsfac} can be done analytically term by term. For this we need to neglect the small oscillating terms in the large-$x$ expansion of $F(x)$, which by Figure~\ref{fig:HG_validation} are negligible at the percent level. We also need the results
\begin{align}
  I^+_p(m_c, \sigma; m_*) &\equiv \int_0^{m_*} \ud m \frac{1}{\sqrt{2\pi\sigma^2}} \exp\left[-\frac{\ln^2(m/m_c)}{2\sigma^2}\right] m^{p-2} \nonumber \\
  & = m_c^{p-1} e^{\frac{(1-p)^2\sigma^2}{2}}\Phi\left[\frac{1}{\sigma}\ln(m_*/m_c) + (1-p)\sigma\right], \\
  I^-_p(m_c, \sigma; m_*) &\equiv \int_{m_*}^\infty \ud m \frac{1}{\sqrt{2\pi\sigma^2}} \exp\left[-\frac{\ln^2(m/m_c)}{2\sigma^2}\right] m^{p-2} \nonumber \\
  & = m_c^{p-1} e^{\frac{(1-p)^2\sigma^2}{2}}\left\{ 1 - \Phi\left[\frac{1}{\sigma}\ln(m_*/m_c) + (1-p)\sigma\right]\right\},
  \label{eq:Ip}
\end{align}
where $\Phi(x)$ is the cumulative distribution function of the normal distribution (expressible in terms of the error function). Substituting in the polynomial expansions for $F(x)$ at small $x$ and large $x$, we can use Equation~\eqref{eq:Ip} to leave only the outermost integral in Equation~\eqref{eq:appsfac}. We perform this integral using numerical quadrature, which results in a fast and accurate (to roughly 1\%) approximation to the suppression factor which can be used in likelihood evaluations.

Using that $\langle m \rangle = m_c e^{-\sigma^2/2}$ for a lognormal mass function\footnote{We remind the reader that angle brackets denote expectation values over $\ud n /\ud m \propto \psi/m$, i.e.~$\langle m \rangle \equiv \left[\int \ud m \, m \psi(m)/m\right]/\int \ud m \psi(m)/m$.}, it is straightforward to verify that at fixed $\bar{N}(y)$ the suppression factor is independent of the  absolute mass scale $m_c$, and depends only on the width $\sigma$. Physically this is due to the assumption that the PBHs are distributed in space in a way that is independent of their mass. In reality this assumption might be broken if PBHs cluster significantly due to mass segregation, a complication which we neglect here.

It is useful to consider a few limiting cases of the suppression factor. Firstly, for any mass function, in the limit that $F(x)$ is dominated by its leading order quadratic part at small $x$ the distribution of angular momentum tends to a Gaussian and the suppression factor is~\citep{2017JCAP...09..037R}
\begin{equation}
  S_{\mathrm{min}} = \frac{\pi}{\Gamma(29/37)}\left(\frac{\sigma_j}{j_0}\right)^{-\frac{21}{37}} e^{-\bar{N}(y)},
  \label{eq:Smin}
\end{equation}
where
\begin{equation}
  \frac{\sigma^2_j}{j_0^2} = \frac{6}{5}\left[\frac{1 + \sigma_m^2/\langle m \rangle^2}{\bar{N}(y)} + \frac{\sigma_M^2}{\fpbh^2} \right],
\end{equation}
and $\sigma_m \equiv \sqrt{\langle m^2 \rangle - \langle m \rangle^2}$ is the width of the mass function. It may be shown that $S \geq S_{\mathrm{min}}$. Physically this limit is realised when $\bar{N}(y) \rightarrow \infty$, i.e.~the spatial density of PBHs becomes sufficiently large that the central limit theorem Gaussianizes the distribution of torquing angular momenta (which has already been assumed to have happened for the dark matter component). The variance of this Gaussian is $\sigma_j^2$, which has contributions from the PBHs and dark matter adding in quadrature. The model of Ref.~\citep{2017JCAP...09..037R} imposes an exponential suppression in this regime, realised via the $e^{-\bar{N}(y)}$ term in Equation~\eqref{eq:Smin}, which accounts for the fact that a high spatial density of PBHs will increase the likelihood of a PBH being sufficiently close to the binary as to prevent its formation due to three-body effects. In our implementation we simply set the suppression factor to zero for $\bar{N}(y) > 23$, where the exponential suppression ensures that $S$ is practically zero.

In the opposite limit the term linear in $x$ dominates in $F(x)$, and the suppression factor tends to
\begin{equation}
  S_{\mathrm{max}} = \left(\frac{5\fpbh^2}{6 \sigma_M^2} \right)^{\frac{21}{74}}U \left(\frac{21}{74}, \frac{1}{2}, \frac{5 \fpbh^2}{6\sigma_M^2} \right),
  \label{eq:Smax}
\end{equation}
where $U$ is a confluent hypergeometric function. One can show that $S \leq S_{\mathrm{max}}$. This limit is realised when $\bar{N}(y) \rightarrow 0$, although how quickly the limit is reached depends on $\fpbh/\sigma_M$. Ref.~\citep{2017JCAP...09..037R} advocates $\bar{N}(y) \ll \fpbh^2/\sigma_M^2$. We find that Equation~\eqref{eq:Smax} is accurate to $\lesssim 2\%$ for all $\fpbh > 10^{-6}$ when $\bar{N}(y) < 0.01$ and for all $\sigma$ we consider. We adopt this threshold for $\bar{N}(y)$ when switching to the asymptotic limit Equation~\eqref{eq:Smax}. This gives percent-level accuracy for the merger rate while maintaining a reasonable run time.

\section{Suppression-induced parameter degeneracies in the PBH model}
\label{app:degs}

When sampling from the posterior in the PBH model we find a pronounced degeneracy tail in each of the three projected two-dimensional planes defined by $\fpbh$, $m_c$, and $\sigma$, see Figure~\ref{fig:PBH_posterior}. The feature allows for large values of all three parameters, and is not present when the model is analysed fixing the suppression factor to unity. In this section we discuss the origin of the degeneracy feature.

\begin{figure}
\centering
\includegraphics[width=0.85\textwidth]{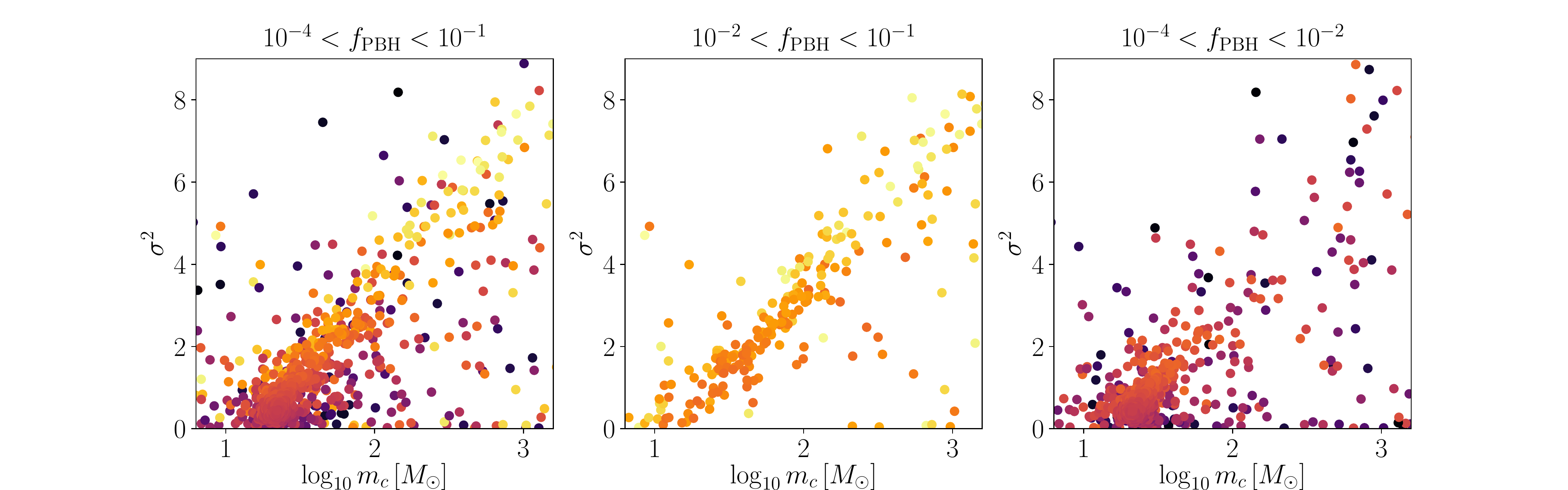}
\caption{Posterior samples in the $(\log_{10}{m_c}, \sigma^2)$ plane for the suppression factor PBH model and GWTC-1 data, colour-coded by $\fpbh$ (warmer colour corresponding to larger values). \emph{Left panel}: Samples having $10^{-4} < \fpbh < 10^{-1}$, thinned by a factor of 4 for visual clarity. \emph{Middle panel}: Samples with higher values of $\fpbh$ in the range  $10^{-2} < \fpbh < 10^{-1}$. \emph{Right panel}: Samples with lower values of $\fpbh$ in the range $10^{-4} < \fpbh < 10^{-2}$. Note that the colour coding is the same across the three panels.}
\label{fig:PBH_posterior_fpbh_split}
\end{figure}

In Figure~\ref{fig:PBH_posterior_fpbh_split} we show posterior samples in the parameter space $(\log_{10}{m_c}, \sigma^2)$ colour-coded by the value of $\fpbh$. The degeneracy feature appears linear in this space, and roughly corresponds to fixing $\langle m \rangle = m_ce^{-\sigma^2/2} \approx 20 \, M_\odot $, i.e.~fixing the average PBH mass to the observed mass scale of the LIGO sources.

In the middle and right panels of Figure~\ref{fig:PBH_posterior_fpbh_split} we split the samples into a set with $10^{-2} < \fpbh < 10^{-1}$ and a set with $10^{-4} < \fpbh < 10^{-2}$ respectively. Samples in the degeneracy tail typically have higher $\fpbh$, which indicates that we are looking at the projection of a three-parameter degeneracy.

\begin{figure}
\centering
\includegraphics[width=0.85\textwidth]{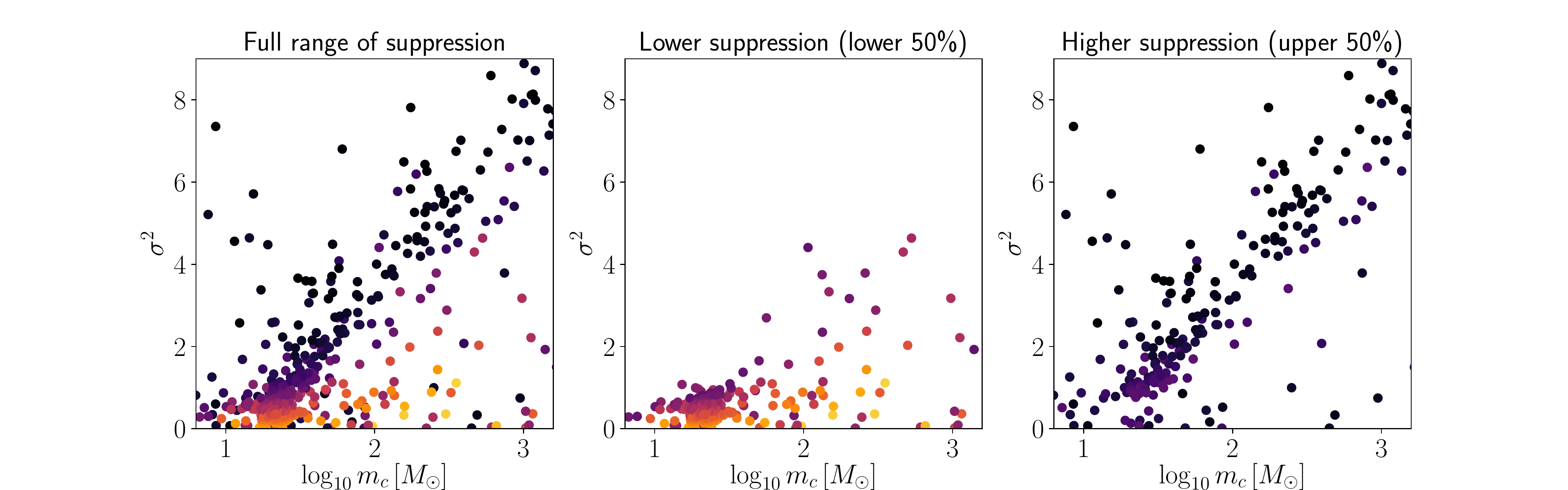}
\caption{Posterior samples in the $(\log_{10}{m_c}, \sigma^2)$ plane for the suppression factor PBH model and GWTC-1 data, colour-coded by suppression factor, defined as $\beta(S=1)/\beta(S)$ (warmer colour corresponding to lower suppression). \emph{Left panel}: All samples, thinned by a factor of 8 for visual clarity. \emph{Middle panel}: Samples with suppression in the lower 50\% quantile. \emph{Right panel}: Samples with suppression in the upper 50\% quantile. Note that the colour coding is the same across the three panels.}
\label{fig:PBH_posterior_S_split}
\end{figure}

In Figure~\ref{fig:PBH_posterior_S_split} we split the samples by the degree of suppression, defined as the ratio between the total detectable number of mergers $\beta$ in a model with suppression factor set to unity to its value in a model with suppression set by Equation~\eqref{eq:sfac} and Equation~\eqref{eq:Eq3p5}. The degeneracy tail is clearly characterised by higher degrees of suppression, which compensate for the overproduction of mergers that large values of $\fpbh$ would otherwise imply.

These plots suggest that the degeneracy tail is caused by the dependence of the suppression factor on the mass function parameters via Equation~\eqref{eq:Eq3p5}, namely
\begin{equation}
  \bar{N}(y) = \frac{M}{\langle m \rangle} \frac{\fpbh}{\fpbh + \sigma_M},
  \label{eq:appEq3p5}
\end{equation}
where $\sigma_M \approx 0.006$ and $M$ is the total binary mass. We remind the reader that $\bar{N}(y)$ is the expected number of PBHs in a spherical region of comoving radius $y$ which contains no other black holes except the binary pair, and that the suppression factor has a pre-factor of $e^{-\bar{N}(y)}$, introduced in Ref.~\citep{2017JCAP...09..037R} to ensure that no other PBH gets close enough to the binary to disrupt it prior to the merger event.

The suppression factor enters the likelihood in two places: in the average over source parameter MCMC samples (the term in square brackets in Equation~\eqref{eq:likeMC}) and in the $e^{-\beta}$ factor entering via the Poisson probability of seeing $N_{\mathrm{obs}}$ events given $\beta$ were expected. The former essentially fixes the total mass in Equation~\eqref{eq:appEq3p5} to the LIGO mass scale, which implies that $\langle m \rangle $ is fixed to keep the source likelihoods high and unsuppressed; note that in the degeneracy tail $\fpbh \gg \sigma_M$ such that the second term in Equation~\eqref{eq:appEq3p5} is irrelevant. Models with high $\sigma$ have contributions to $\beta$ from a much broader range of masses; the typical total masses contributing to $\beta$ (Equation~\eqref{eq:dN}) are $\sim 2m_c$, meaning $M \gg \langle m \rangle$ when $\sigma$ is large. This implies that $\bar{N}(y) \gg 1$ for much of the integration range in $\beta$, which in turn implies a high degree of suppression in the total number of mergers and compensates for the high $\fpbh$.

We note that the degeneracy tail skews the one-dimensional posteriors on the PBH model parameters, but has a relatively minor impact on their median and best-fit values. Nevertheless, what we have uncovered is a mechanism for generating high values of $\fpbh$ which can still fit the LIGO data. Equation~\eqref{eq:appEq3p5} was arrived at in Ref.~\citep{2017JCAP...09..037R} by a combination of simulation results and analytic arguments, but since it was not tested for the extreme mass function parameters favoured in the degeneracy feature the model could well be unreliable in this regime. Further investigations with $N$-body simulations will be required to rigorously test the degeneracy feature seen in Figure~\ref{fig:PBH_posterior}.

\section{The posterior predictive distribution}
\label{app:ppd}

In this section we present a derivation of the posterior predictive distribution (PPD). This quantity is sufficient to understand the differences in Bayesian evidences between our BBH merger models.

Each model has the flexibility to fit the total number of observed mergers by adjusting an amplitude-like parameter ($\fpbh$ or $R_0$), so the best fitting $\beta$ is roughly 10 for all models. The factor of $e^{-\beta}$ in the likelihood Equation~\eqref{eq:like} thus cancels in the likelihood ratio and evidence ratio, and we can approximate $D$ with $\mathbf{d}_{\mathrm{new}}$, the new GW strain data. Formally the likelihood for $\mathbf{d}$ is that used to produce the posterior samples for the source parameters. We will instead simplify the analysis by compressing the data to a set of estimators for the source parameters which best constrain the population models. For both the PBH and LIGO models, the relevant source parameters are the total mass $M$, the mass ratio $q$, and the redshift $z$, or combinations of these. We have seen from Figure~\ref{fig:dbetas} that the redshift distributions of the observable mergers in both models are indistinguishable, so we can anticipate that the likelihood in $M$ and $q$ will be sufficient to explain the differing maximum likelihood values of the models. Since it is the detector-frame chirp mass $\mathcal{M}_z$ which is most closely related to the signal measured in the data (as discussed in Section~\ref{sec:data}), our two estimators will be $\widehat{\mathcal{M}}_z$ and $\hat{q}$, where the precise expression of these in terms of the data is left unspecified for now.

Although the preferred $q$-distributions are quite different (middle panel of Figure~\ref{fig:dbetas}) the typical errors on $q$ are large for the GWTC-1 sources, so it is unclear how powerful the mass ratio distribution is in discriminating between models. To test this, we generated new source parameter posteriors for each merger event by randomizing the $q$ values at each sample point. Specifically, for each sample $\boldsymbol{\lambda}_i$, we recorded the chirp mass $\mathcal{M}_{\mathrm{chirp}}$, replaced $q$ with a random sample from a uniform distribution between 0 and 1, then set $m_1$ and $m_2$ using the saved $\mathcal{M}_{\mathrm{chirp}}$ and the new $q$. We then re-ran the inference of the PBH $S=1$ model and the LIGO models A and B using these new source posteriors. This procedure destroys all information on mass ratio in each source, preserving that on redshift and detector-frame chirp mass. The (natural log) evidence ratio of the PBH model to Model B with this new data is $-7.30 \pm 0.25$. For the PBH model to Model A the evidence ratio is $-5.72 \pm 0.23$, and for Model A to Model B it is $-1.59 \pm 0.23$. The $S=1$ PBH model is now slightly less disfavoured compared with Model A and Model B, but is still heavily disfavoured compared with both these models. There is no significant change in the evidence ratio between the two LIGO models. This test strongly suggests that the distribution of chirp masses preferred by the models is the key discriminator between them. The mass ratio uncertainties are too large in the GWTC-1 catalogue for $q$ to be effective at constraining the space of allowed models.

These arguments strongly suggest that the source likelihood $p(\mathbf{d} \vert \boldsymbol{\theta}, M)$ needed for the PPD should be the probability of an `observed chirp mass' $\widehat{\mathcal{M}}_z$ given source parameters. An expression for $p(\mathbf{d} \vert \boldsymbol{\theta}, M)$ is given in Ref.~\citep{2019MNRAS.486.1086M}. For a single source with \emph{detectable} GW strain $\mathbf{d}$ and source parameters $\boldsymbol{\lambda}$, the likelihood is 
\begin{equation}
  p(\mathbf{d} \vert \boldsymbol{\theta}) = \frac{I(\mathbf{d}) \int \ud \boldsymbol{\lambda} \, p(\mathbf{d} \vert \boldsymbol{\lambda}) p(\boldsymbol{\lambda} \vert \boldsymbol{\theta})}{\int \ud \boldsymbol{\lambda} \, p_{\mathrm{det}}(\boldsymbol{\lambda}) p(\boldsymbol{\lambda} \vert \boldsymbol{\theta})},
  \label{eq:MFGlike}
\end{equation}
where $I(\mathbf{d})$ is unity if the data pass the detection threshold and zero if it does not. We have also defined the detection probability over the complete set of source parameters $p_{\mathrm{det}}(\boldsymbol{\lambda})$ as
\begin{equation}
  p_{\mathrm{det}}(\boldsymbol{\lambda}) = \int \ud \mathbf{d} \, I(\mathbf{d}) p(\mathbf{d} \vert \boldsymbol{\lambda}),
\end{equation}
where $p(\mathbf{d} \vert \boldsymbol{\lambda})$ is the probability of \emph{any} data set, not just those observable by the detector. The joint likelihood of $N$ sources is simply the product of individual likelihoods each given by Equation~\eqref{eq:MFGlike}. The prior distribution of source parameters given the population model $p(\boldsymbol{\lambda} \vert \boldsymbol{\theta})$ is simply proportional to the merger rate $\ud N/\ud m_1 \ud m_2 \ud z$ appearing in Equation~\eqref{eq:like}. Note that the overall normalisation of the merger rate drops out of Equation~\eqref{eq:MFGlike}.

Equation~\eqref{eq:MFGlike} is the likelihood for GW strain data, and we wish to re-write it as a one-dimensional likelihood for observed chirp mass $\widehat{\mathcal{M}}_z$. This will allow us to study the origin of the large likelihood ratios between models, which we have argued is primarily due to the relative ability of models to fit the observed distribution of chirp masses. It will also allow an assessment of the absolute quality of model fits via the PPD. To do this, we will make a series of well-motivated approximations to construct a likelihood for the compressed data $\widehat{\mathcal{M}}_z$.

Firstly we will assume that $I(\mathbf{d}) = \Theta(\hat{\rho} - \rho_*)$ where $\Theta$ is the Heaviside step function, $\hat{\rho}$ is the $S/N$ of the observed waveform, and $\rho_*$ is a threshold $S/N$, i.e.~we assume the merger is detectable if its $S/N$ is above a sharp threshold. Secondly, we assume that $\hat{\rho}$ is unaffected by noise fluctuations, such that for source parameters $\boldsymbol{\lambda}$ we have $p(\hat{\rho} \vert \boldsymbol{\lambda}) = \delta^D[\hat{\rho} - \rho(\boldsymbol{\lambda})]$ where $\delta^D$ is the Dirac delta function and $\rho(\boldsymbol{\lambda})$ is the $S/N$ of a model template with source parameters $\boldsymbol{\lambda}$. This is a reasonable approximation since $\hat{\rho}$ is a stack across the whole waveform, and is typically well constrained. These two approximations imply that $p_{\mathrm{det}}(\boldsymbol{\lambda}) = \Theta[\rho(\boldsymbol{\lambda}) - \rho_*]$. Since $\hat{\rho}$ is assumed to be non-stochastic, we can integrate it out of Equation~\eqref{eq:MFGlike} and redefine the data vector $\mathbf{d}$ as having the overall amplitude projected out.

Next, we assume that the $S/N$ can be written in terms of the orientation parameter $\omega$ introduced in Section~\ref{sec:data} as $\rho(\boldsymbol{\lambda}) = \omega \rho_{\mathrm{opt}}(\tilde{\boldsymbol{\lambda}})$, where $\tilde{\boldsymbol{\lambda}}$ are the source parameters excluding the orientation and angular position parameters (which have been combined into $\omega$), and $\rho_{\mathrm{opt}}$ is the $S/N$ of an optimally oriented binary. For isotropic sources we have $p(\boldsymbol{\lambda} \vert \boldsymbol{\theta}) = p(\tilde{\boldsymbol{\lambda}} \vert \boldsymbol{\theta})p(\omega)$, where $p(\omega)$ is the distribution discussed in Section~\ref{sec:data} and produced by \textsc{gwdet}. With these approximations, Equation~\eqref{eq:MFGlike} becomes
\begin{align}
  p(\mathbf{d} \vert \boldsymbol{\theta}) &\propto \int \ud \tilde{\boldsymbol{\lambda}} \ud \omega \, \Theta[\omega \rho_{\mathrm{opt}}(\tilde{\boldsymbol{\lambda}}) - \rho_*] p(\mathbf{d} \vert  \tilde{\boldsymbol{\lambda}}, \omega) p(\omega) p(\tilde{\boldsymbol{\lambda}} \vert \boldsymbol{\theta}) \nonumber \\
  &\propto  \int \ud \tilde{\boldsymbol{\lambda}} \,  p(\tilde{\boldsymbol{\lambda}} \vert \boldsymbol{\theta}) \int^1_{\frac{\rho_*}{\rho_{\mathrm{opt}}(\tilde{\boldsymbol{\lambda}})}} \ud \omega \, p(\omega) p(\mathbf{d} \vert  \tilde{\boldsymbol{\lambda}}, \omega) .
  \label{eq:MFGlikeapprox}
\end{align}

Now, we will assume that an estimator $\widehat{\mathcal{M}}_z$ for the chirp mass can be constructed from the data using some form of massive data compression. Writing $\mathbf{d} = (\widehat{\mathcal{M}}_z, \tilde{\mathbf{d}})$ we can integrate out all other `modes' of the data $\tilde{\mathbf{d}}$ which keep $\widehat{\mathcal{M}}_z$ fixed. This simply amounts to replacing $\mathbf{d}$ with $\widehat{\mathcal{M}}_z$ everywhere in Equation~\eqref{eq:MFGlikeapprox}.

Our next approximation sets $p(\widehat{\mathcal{M}}_z \vert  \tilde{\boldsymbol{\lambda}}, \omega) \approx  p(\widehat{\mathcal{M}}_z \vert  \tilde{\boldsymbol{\lambda}})$, i.e.~we assume the estimated chirp mass on its own provides no information on the orientation or angular position of the binary. With this we have
\begin{equation}
  p(\widehat{\mathcal{M}}_z \vert \boldsymbol{\theta}) \propto \int \ud \tilde{\boldsymbol{\lambda}} \, p_{\mathrm{det}}(\tilde{\boldsymbol{\lambda}}) p(\widehat{\mathcal{M}}_z \vert \tilde{\boldsymbol{\lambda}}) p(\tilde{\boldsymbol{\lambda}} \vert \boldsymbol{\theta})
  \label{eq:likeapprox2}
\end{equation}
where we used Equation~\eqref{eq:pdet} to substitute for the angle-averaged detection probability.

We now assume that both $p_{\mathrm{det}}(\tilde{\boldsymbol{\lambda}})$ and $p(\widehat{\mathcal{M}}_z \vert \tilde{\boldsymbol{\lambda}})$ depend only upon $\mathcal{M}_z$, $q$, and $z$, i.e.~we neglect any spin dependence in the $S/N$. This allows us to integrate out all other source parameters from Equation~\eqref{eq:likeapprox2}. This just leaves $p(\widehat{\mathcal{M}}_z \vert \mathcal{M}_z, q, z)$, the likelihood for the observed chirp mass, to be specified.

At this point, we must make a choice for the form of $p(\widehat{\mathcal{M}}_z \vert \mathcal{M}_z, q, z)$, since so far we have only used it as a function of model parameters and not of `observed' parameters. Since the data compression producing $\widehat{\mathcal{M}}_z$ is massive we will assume that the sampling distribution of the estimator is Gaussian with mean $\mu$ and variance $\sigma^2$. An unbiased estimator for the chirp mass would have $\mu = \mathcal{M}_z$, but the variance can have a general dependence on $\mathcal{M}_z$, $q$, and $z$. This could in principle be inferred from the source parameter posterior given priors on the source parameters, but we will make the ansatz that $\sigma^2$ is roughly constant and can be set equal to the marginalised \emph{posterior} variances on chirp mass from the MCMC samples -- this is justified if $p(\widehat{\mathcal{M}}_z \vert \mathcal{M}_z)$ really is Gaussian and the prior on chirp mass is uniform (although recall that the chirp mass is measured with $\sim 15\%$ precision with little sensitivity to the choice of prior).

With these assumptions we can write
\begin{align}
  p(\widehat{\mathcal{M}}_z \vert \boldsymbol{\theta}) &= \int \ud \mathcal{M}_z \, p(\widehat{\mathcal{M}}_z \vert  \mathcal{M}_z) p(\mathcal{M}_z \vert \boldsymbol{\theta}), \label{eq:convolvedlike}\\
  p(\mathcal{M}_z \vert \boldsymbol{\theta}) &= \frac{\int \ud q \ud z \, p_{\mathrm{det}}(\mathcal{M}_z, q, z) p(\mathcal{M}_z, q, z \vert \boldsymbol{\theta})}{\int \ud \mathcal{M}_z \ud q \ud z \, p_{\mathrm{det}}(\mathcal{M}_z, q, z) p(\mathcal{M}_z, q, z \vert \boldsymbol{\theta})}. \label{eq:dbetanorm}
\end{align}

Equation~\eqref{eq:convolvedlike} is simply the predicted distribution of chirp mass averaged over all other source parameters \emph{convolved} with the observational uncertainty specified by $p(\widehat{\mathcal{M}}_z \vert  \mathcal{M}_z)$. We immediately recognise Equation~\eqref{eq:dbetanorm} as a normalised version of the differential detectable merger rate, i.e.~$\partial \ln \beta/\partial \mathcal{M}_z$. Recall that in Figure~\ref{fig:dbetas} we plotted $\partial \beta/ \partial M$, $\partial \beta/ \partial q$, and $\partial \beta/ \partial z$.

Finally, we can average over the posterior of the model parameters $\boldsymbol{\theta}$ for each source to get the PPD. This gives, for the full catalogue of sources,
\begin{align}
  p(\{ \widehat{\mathcal{M}}_z \} \vert \{ \mathbf{d} \}) &= \prod_{i=1}^{N_{\mathrm{obs}}} \int \ud \mathcal{M}_z \, p(\widehat{\mathcal{M}}^i_z \vert  \mathcal{M}_z) p(\mathcal{M}_z \vert \{ \mathbf{d} \}), \label{eq:convlike_2} \\
  p(\mathcal{M}_z \vert \{ \mathbf{d} \}) &= \int \ud \boldsymbol{\theta} \, p(\mathcal{M}_z \vert \boldsymbol{\theta}) p(\boldsymbol{\theta} \vert \{ \mathbf{d} \}). \label{eq:PPD_2}
\end{align}
In a slight abuse of terminology we will also refer to $p(\mathcal{M}_z \vert \{ \mathbf{d} \})$ as the PPD. This object quantifies the probability distribution of the detector-frame chirp mass given the data, and can be convolved with the observational errors according to Equation~\eqref{eq:convlike_2} to give an equivalent predictive distribution for the observed chirp mass.

It is straightforward to show that Equation~\eqref{eq:convolvedlike} evaluated at the locations of the observed strain data gives a likelihood function equivalent to that specified in Equation~\eqref{eq:like} and used throughout this work for inference. Indeed, the above derivation shows how this simplified likelihood may be obtained from first principles, and makes all approximations transparent. The only difference with Equation~\eqref{eq:like} is the Poisson probability of observing $N_{\mathrm{obs}}$ sources when $\beta$ were expected -- the likelihood Equation~\eqref{eq:convolvedlike} only accounts for the relative merger rate, which is sufficient to understand the disparity in likelihood ratios since $\beta \approx 10$ at the best-fit point for all models.

\end{document}